%% file: main.tex
\documentclass[twocolumn,twocolappendix]{aastex631}
\usepackage{amsmath,amssymb,amsfonts}
\usepackage{graphicx}
\usepackage{multirow,tabularx,booktabs,makecell,threeparttable,array}
\usepackage{placeins}
\usepackage{xcolor}

\usepackage{color}

\newcommand{\Msun}{$M_{\odot}$}

\def\astrid{\texttt{ASTRID}}

\usepackage{hyperref}

\shorttitle{DESI Dual and Offset AGN}
\shortauthors{Dadiani et al.}

\begin{document}

\title{Cosmic Pairs: A DESI Census of Dual and Offset AGN as Precursors to Massive Black Hole Binaries}

\author[0000-0001-6586-4297]{Ekaterine Dadiani}
\affiliation{McWilliams Center for Cosmology and Astrophysics, Department of Physics, Carnegie Mellon University, Pittsburgh, PA 15213, USA}

\author[0000-0002-6011-0530]{Antonella Palmese}
\affiliation{McWilliams Center for Cosmology and Astrophysics, Department of Physics, Carnegie Mellon University, Pittsburgh, PA 15213, USA}

\author[0000-0002-8828-8461]{Yihao Zhou}
\affiliation{McWilliams Center for Cosmology and Astrophysics, Department of Physics, Carnegie Mellon University, Pittsburgh, PA 15213, USA}

\author[0000-0001-6627-2533]{Nianyi Chen}
\affiliation{Max-Planck-Institut f\"ur Astrophysik, Karl-Schwarzschild-Str 1, D-85748 Garching, Germany}
\affiliation{School of Natural Sciences, Institute for Advanced Study, Princeton, NJ 08540, USA}

\author[0000-0002-6462-5734]{Tiziana Di Matteo}
\affiliation{McWilliams Center for Cosmology and Astrophysics, Department of Physics, Carnegie Mellon University, Pittsburgh, PA 15213, USA}

\author[0000-0002-6729-2373]{Alejandro Er\'ostegui}
\affiliation{Institute of Space Sciences, ICE-CSIC, Campus UAB, Carrer de Can Magrans s/n, 08913 Bellaterra, Barcelona, Spain}

\author[0000-0003-4440-259X]{Mar Mezcua}
\affiliation{Institute of Space Sciences, ICE-CSIC, Campus UAB, Carrer de Can Magrans s/n, 08913 Bellaterra, Barcelona, Spain}
\affiliation{Institut d'Estudis Espacials de Catalunya (IEEC), c/ Esteve Terradas 1, Edifici RDIT, Campus PMT-UPC, 08860 Castelldefels, Spain}


\author{Jessica Nicole Aguilar}
\affiliation{Lawrence Berkeley National Laboratory, 1 Cyclotron Road, Berkeley, CA 94720, USA}

\author[0000-0001-6098-7247]{Steven Ahlen}
\affiliation{Department of Physics, Boston University, 590 Commonwealth Avenue, Boston, MA 02215, USA}

\author[0000-0003-4162-6619]{Stephen Bailey}
\affiliation{Lawrence Berkeley National Laboratory, 1 Cyclotron Road, Berkeley, CA 94720, USA}

\author[0000-0003-0467-5438]{Florian Beutler}
\affiliation{Institute for Astronomy, University of Edinburgh, Royal Observatory, Blackford Hill, Edinburgh EH9 3HJ, UK}

\author[0000-0001-9712-0006]{Davide Bianchi}
\affiliation{Dipartimento di Fisica ``Aldo Pontremoli'', Universit\`a degli Studi di Milano, Via Celoria 16, I-20133 Milano, Italy}
\affiliation{INAF-Osservatorio Astronomico di Brera, Via Brera 28, 20122 Milano, Italy}

\author{David Brooks}
\affiliation{Department of Physics \& Astronomy, University College London, Gower Street, London, WC1E 6BT, UK}

\author{Todd Claybaugh}
\affiliation{Lawrence Berkeley National Laboratory, 1 Cyclotron Road, Berkeley, CA 94720, USA}

\author[0000-0002-1769-1640]{Axel de la Macorra}
\affiliation{Instituto de F\'{\i}sica, Universidad Nacional Aut\'onoma de M\'exico, Circuito de la Investigaci\'on Cient\'{\i}fica, Ciudad Universitaria, Cd. de M\'exico, C.~P.~04510, M\'exico}

\author[0000-0002-4928-4003]{Arjun Dey}
\affiliation{NSF NOIRLab, 950 N. Cherry Ave., Tucson, AZ 85719, USA}

\author[0000-0002-5665-7912]{Biprateep Dey}
\affiliation{Department of Astronomy \& Astrophysics, University of Toronto, Toronto, ON M5S 3H4, Canada}
\affiliation{Department of Physics \& Astronomy and Pittsburgh Particle Physics, Astrophysics, and Cosmology Center (PITT PACC), University of Pittsburgh, 3941 O'Hara Street, Pittsburgh, PA 15260, USA}

\author{Peter Doel}
\affiliation{Department of Physics \& Astronomy, University College London, Gower Street, London, WC1E 6BT, UK}

\author[0000-0003-1251-532X]{Victoria A. Fawcett}
\affiliation{European Southern Observatory, Karl-Schwarzschild-Stra\ss e 2, 85748 Garching bei M\"unchen, Germany}

\author[0000-0003-4175-571X]{Benjamin Floyd}
\affiliation{Institute of Cosmology and Gravitation, University of Portsmouth, Dennis Sciama Building, Portsmouth, PO1 3FX, UK}

\author[0000-0002-3033-7312]{Andreu Font-Ribera}
\affiliation{Instituci\'o Catalana de Recerca i Estudis Avan\c{c}ats, Passeig de Llu\'{\i}s Companys, 23, 08010 Barcelona, Spain}
\affiliation{Institut de F\'{\i}sica d'Altes Energies (IFAE), The Barcelona Institute of Science and Technology, Edifici Cn, Campus UAB, 08193, Bellaterra (Barcelona), Spain}

\author[0000-0002-2890-3725]{Jaime E. Forero-Romero}
\affiliation{Departamento de F\'isica, Universidad de los Andes, Cra. 1 No. 18A-10, Edificio Ip, CP 111711, Bogot\'a, Colombia}
\affiliation{Observatorio Astron\'omico, Universidad de los Andes, Cra. 1 No. 18A-10, Edificio H, CP 111711, Bogot\'a, Colombia}

\author[0000-0001-9632-0815]{Enrique Gazta\~naga}
\affiliation{Institut d'Estudis Espacials de Catalunya (IEEC), c/ Esteve Terradas 1, Edifici RDIT, Campus PMT-UPC, 08860 Castelldefels, Spain}
\affiliation{Institute of Cosmology and Gravitation, University of Portsmouth, Dennis Sciama Building, Portsmouth, PO1 3FX, UK}
\affiliation{Institute of Space Sciences, ICE-CSIC, Campus UAB, Carrer de Can Magrans s/n, 08913 Bellaterra, Barcelona, Spain}

\author[0000-0003-3142-233X]{Satya Gontcho A Gontcho}
\affiliation{University of Virginia, Department of Astronomy, Charlottesville, VA 22904, USA}

\author{Gaston Gutierrez}
\affiliation{Fermi National Accelerator Laboratory, PO Box 500, Batavia, IL 60510, USA}

\author[0000-0002-6024-466X]{Mustapha Ishak}
\affiliation{Department of Physics, The University of Texas at Dallas, 800 W. Campbell Rd., Richardson, TX 75080, USA}

\author{Robert Kehoe}
\affiliation{Department of Physics, Southern Methodist University, 3215 Daniel Avenue, Dallas, TX 75275, USA}

\author[0000-0001-6356-7424]{Anthony Kremin}
\affiliation{Lawrence Berkeley National Laboratory, 1 Cyclotron Road, Berkeley, CA 94720, USA}

\author[0000-0002-1134-9035]{Ofer Lahav}
\affiliation{Department of Physics \& Astronomy, University College London, Gower Street, London, WC1E 6BT, UK}

\author{Andrew Lambert}
\affiliation{Lawrence Berkeley National Laboratory, 1 Cyclotron Road, Berkeley, CA 94720, USA}

\author[0000-0003-1838-8528]{Martin Landriau}
\affiliation{Lawrence Berkeley National Laboratory, 1 Cyclotron Road, Berkeley, CA 94720, USA}

\author[0000-0001-7178-8868]{Laurent Le Guillou}
\affiliation{Sorbonne Universit\'e, CNRS/IN2P3, Laboratoire de Physique Nucl\'eaire et de Hautes Energies (LPNHE), FR-75005 Paris, France}

\author[0000-0003-4962-8934]{Marc Manera}
\affiliation{Departament de F\'{\i}sica, Serra H\'{u}nter, Universitat Aut\`{o}noma de Barcelona, 08193 Bellaterra (Barcelona), Spain}
\affiliation{Institut de F\'{\i}sica d'Altes Energies (IFAE), The Barcelona Institute of Science and Technology, Edifici Cn, Campus UAB, 08193, Bellaterra (Barcelona), Spain}

\author[0000-0002-1125-7384]{Aaron Meisner}
\affiliation{NSF NOIRLab, 950 N. Cherry Ave., Tucson, AZ 85719, USA}

\author{Ramon Miquel}
\affiliation{Instituci\'o Catalana de Recerca i Estudis Avan\c{c}ats, Passeig de Llu\'{\i}s Companys, 23, 08010 Barcelona, Spain}
\affiliation{Institut de F\'{\i}sica d'Altes Energies (IFAE), The Barcelona Institute of Science and Technology, Edifici Cn, Campus UAB, 08193, Bellaterra (Barcelona), Spain}

\author[0000-0002-2733-4559]{John Moustakas}
\affiliation{Department of Physics and Astronomy, Siena University, 515 Loudon Road, Loudonville, NY 12211, USA}

\author[0000-0001-9070-3102]{Seshadri Nadathur}
\affiliation{Institute of Cosmology and Gravitation, University of Portsmouth, Dennis Sciama Building, Portsmouth, PO1 3FX, UK}

\author[0000-0002-4637-2868]{Enrique Paillas}
\affiliation{Instituto de Estudios Astrof\'isicos, Facultad de Ingenier\'ia y Ciencias, Universidad Diego Portales, Av. Ej\'ercito Libertador 441, Santiago, Chile}
\affiliation{Steward Observatory, University of Arizona, 933 N. Cherry Avenue, Tucson, AZ 85721, USA}

\author[0000-0002-0644-5727]{Will J. Percival}
\affiliation{Department of Physics and Astronomy, University of Waterloo, 200 University Ave W, Waterloo, ON N2L 3G1, Canada}
\affiliation{Perimeter Institute for Theoretical Physics, 31 Caroline St. North, Waterloo, ON N2L 2Y5, Canada}
\affiliation{Waterloo Centre for Astrophysics, University of Waterloo, 200 University Ave W, Waterloo, ON N2L 3G1, Canada}

\author[0000-0001-7145-8674]{Francisco Prada}
\affiliation{Instituto de Astrof\'{\i}sica de Andaluc\'{\i}a (CSIC), Glorieta de la Astronom\'{\i}a, s/n, E-18008 Granada, Spain}

\author[0000-0001-6979-0125]{Ignasi P\'erez-R\`afols}
\affiliation{Departament de F\'isica, EEBE, Universitat Polit\`ecnica de Catalunya, c/Eduard Maristany 10, 08930 Barcelona, Spain}

\author[0000-0002-4940-3009]{Ragadeepika Pucha}
\affiliation{Steward Observatory, University of Arizona, 933 N. Cherry Avenue, Tucson, AZ 85721, USA}
\affiliation{Department of Physics and Astronomy, The University of Utah, 115 South 1400 East, Salt Lake City, UT 84112, USA}

\author[0000-0002-3500-6635]{Corentin Ravoux}
\affiliation{Universit\'e Clermont-Auvergne, CNRS, LPCA, 63000 Clermont-Ferrand, France}

\author{Graziano Rossi}
\affiliation{Department of Physics and Astronomy, Sejong University, 209 Neungdong-ro, Gwangjin-gu, Seoul 05006, Republic of Korea}

\author[0000-0002-0394-0896]{Rossana Ruggeri}
\affiliation{Queensland University of Technology, School of Chemistry \& Physics, George St, Brisbane 4001, Australia}

\author[0000-0002-9646-8198]{Eusebio Sanchez}
\affiliation{CIEMAT, Avenida Complutense 40, E-28040 Madrid, Spain}

\author[0000-0002-0408-5633]{Christoph Saulder}
\affiliation{Max Planck Institute for Extraterrestrial Physics, Gie\ss enbachstra\ss e 1, 85748 Garching, Germany}

\author{David Schlegel}
\affiliation{Lawrence Berkeley National Laboratory, 1 Cyclotron Road, Berkeley, CA 94720, USA}

\author{Michael Schubnell}
\affiliation{Department of Physics, University of Michigan, 450 Church Street, Ann Arbor, MI 48109, USA}
\affiliation{University of Michigan, 500 S. State Street, Ann Arbor, MI 48109, USA}

\author[0000-0002-6588-3508]{Hee-Jong Seo}
\affiliation{Department of Physics \& Astronomy, Ohio University, 139 University Terrace, Athens, OH 45701, USA}

\author[0000-0002-3461-0320]{Joseph Harry Silber}
\affiliation{Lawrence Berkeley National Laboratory, 1 Cyclotron Road, Berkeley, CA 94720, USA}

\author[0000-0002-2949-2155]{Ma{\l}gorzata Siudek}
\affiliation{Institute of Space Sciences, ICE-CSIC, Campus UAB, Carrer de Can Magrans s/n, 08913 Bellaterra, Barcelona, Spain}
\affiliation{Instituto de Astrof\'{\i}sica de Canarias, C/ V\'{\i}a L\'{a}ctea, s/n, E-38205 La Laguna, Tenerife, Spain}

\author[0000-0003-1704-0781]{Gregory Tarl\'e}
\affiliation{University of Michigan, 500 S. State Street, Ann Arbor, MI 48109, USA}

\author{Benjamin Alan Weaver}
\affiliation{NSF NOIRLab, 950 N. Cherry Ave., Tucson, AZ 85719, USA}

\author{DESI Collaboration}
\noaffiliation

\correspondingauthor{Ekaterine Dadiani}
\email{edadiani@andrew.cmu.edu}

\begin{abstract}
We present a systematic census of dual and offset active galactic nuclei (AGN) using spectroscopic data from the first data release (DR1) of the Dark Energy Spectroscopic Instrument (DESI).
After correcting for observational systematics, our final sample contains $>7,000$ dual AGN and 27,000 galaxy pairs containing one AGN over the redshift range $0 \lesssim z \lesssim 3.6$. This sample expands the known dual AGN  sample by $\sim1$--$2$ orders of magnitude at $0.2 \lesssim z \lesssim 0.4$, includes ${\sim}50$ dwarf dual AGN candidates in a regime where only a handful were previously known, and triples the census at $z>2$.
Dual AGN are preferentially found at small separations, consistent with merger-driven triggering of AGN activity. 
The two members of a pair differ in their star formation response: the more massive (primary) host changes little with separation, while the less massive (secondary) lies ${\sim}0.3$ dex above matched inactive and one-AGN companions in main-sequence offset.
Using \texttt{ASTRID} simulations, we predict that the fraction of DESI dual AGN whose central black holes will merge by $z\sim 0$ increases with redshift, reaching $\sim76\%$ by $z \sim 2$, while the fraction producing LISA-detectable mergers peaks at $\sim37\%$ near $z \sim 0.9$. 
These results provide the largest uniformly selected spectroscopic sample of kpc-scale dual and offset AGN candidates from a single survey, connecting their host-galaxy and AGN demographics to the progenitor population of massive black hole mergers detectable by LISA.
\end{abstract}


\input{Sec1_Intro}
\input{Sec2_Methods}

\input{Sec3_Data}
\input{Sec4_Result}
\input{Sec5_Conclusion}

\section*{Acknowledgements}
This material is based upon work supported by the National Aeronautics and Space Administration under Grant No. 22-LPS22-0025.

This material is based upon work supported by the U.S. Department of Energy (DOE), Office of Science, Office of High-Energy Physics, under Contract No. DE--AC02--05CH11231, and by the National Energy Research Scientific Computing Center, a DOE Office of Science User Facility under the same contract. Additional support for DESI was provided by the U.S. National Science Foundation (NSF), Division of Astronomical Sciences under Contract No. AST-0950945 to the NSF's National Optical-Infrared Astronomy Research Laboratory; the Science and Technology Facilities Council of the United Kingdom; the Gordon and Betty Moore Foundation; the Heising-Simons Foundation; the French Alternative Energies and Atomic Energy Commission (CEA); the National Council of Humanities, Science and Technology of Mexico (CONAHCYT); the Ministry of Science, Innovation and Universities of Spain (MICIU/AEI/10.13039/501100011033), and by the DESI Member Institutions: \url{https://www.desi.lbl.gov/collaborating-institutions}.

The DESI Legacy Imaging Surveys consist of three individual and complementary projects: the Dark Energy Camera Legacy Survey (DECaLS), the Beijing-Arizona Sky Survey (BASS), and the Mayall z-band Legacy Survey (MzLS). DECaLS, BASS and MzLS together include data obtained, respectively, at the Blanco telescope, Cerro Tololo Inter-American Observatory, NSF's NOIRLab; the Bok telescope, Steward Observatory, University of Arizona; and the Mayall telescope, Kitt Peak National Observatory, NOIRLab. NOIRLab is operated by the Association of Universities for Research in Astronomy (AURA) under a cooperative agreement with the National Science Foundation. Pipeline processing and analyses of the data were supported by NOIRLab and the Lawrence Berkeley National Laboratory. Legacy Surveys also uses data products from the Near-Earth Object Wide-field Infrared Survey Explorer (NEOWISE), a project of the Jet Propulsion Laboratory/California Institute of Technology, funded by the National Aeronautics and Space Administration. Legacy Surveys was supported by: the Director, Office of Science, Office of High Energy Physics of the U.S. Department of Energy; the National Energy Research Scientific Computing Center, a DOE Office of Science User Facility; the U.S. National Science Foundation, Division of Astronomical Sciences; the National Astronomical Observatories of China, the Chinese Academy of Sciences and the Chinese National Natural Science Foundation. LBNL is managed by the Regents of the University of California under contract to the U.S. Department of Energy. The complete acknowledgments can be found at \url{https://www.legacysurvey.org/}.

Any opinions, findings, and conclusions or recommendations expressed in this material are those of the author(s) and do not necessarily reflect the views of the U.S. National Science Foundation, the U.S. Department of Energy, or any of the listed funding agencies.

The authors are honored to be permitted to conduct scientific research on I'oligam Du'ag (Kitt Peak), a mountain with particular significance to the Tohono O'odham Nation.

The authors used generative AI tools, including ChatGPT and Claude, for language editing and clarity. All scientific content and final wording were reviewed and verified by the authors, who take full responsibility for the content of the manuscript.

\appendix
\input{Sec6_Appendix}

\bibliographystyle{aasjournal}
\bibliography{main, desi_papers, obs, duals_lit,astrid}
\end{document}

%% file: Sec1_Intro.tex
\section{Introduction}

Observational evidence indicates that nearly all massive galaxies host supermassive black holes (SMBHs) at their centers, with masses ranging from $10^6$ to $10^{10}\,M_\odot$ \citep{1982MNRAS.200..115S, 1998AJ....115.2285M, 2002MNRAS.336L..61H, 2013ARA&A..51..511K}. Under the hierarchical framework of galaxy formation, galaxies grow through repeated mergers, which inevitably lead to the coalescence of their central SMBHs \citep{2003ApJ...582..559V, 2006ApJ...652..864H}. Galaxy mergers are believed to be the primary drivers of SMBH growth. The resulting gravitational torques trigger massive gas inflows toward the galactic centers, fueling rapid accretion and igniting active galactic nucleus (AGN) activity \citep{DiMatteo2005_BH_model, 2008ApJS..175..356H, 2012ApJ...746L..22K}.

Theoretical models describe this evolution as a sequence from wide galaxy pairs to bound SMBH binaries through three main stages \citep{1980Natur.287..307B, 2001ApJ...563...34M, 2003ApJ...596..860M}. Initially, SMBHs reside at galactic-scale separations of tens of kiloparsecs within merging galaxies. Through dynamical friction, they lose orbital energy and form dual SMBHs at separations of $\sim$tens of kiloparsecs down to tens of parsecs \citep{2013degn.book.....M, 2019NewAR..8601525D}. Over several hundred million years, these dual SMBHs evolve into gravitationally bound binaries at parsec scales, eventually emitting gravitational waves and merging into a single SMBH \citep{2009ApJ...696L..89C, 2021ApJ...911..132X}. Gravitational wave emission from such inspirals and mergers is a key target for pulsar timing arrays \citep[PTA;][]{2016ApJ...821...13A, 2020ApJ...905L..34A, 2023arXiv230900693T} and future space-based observatories such as the Laser Interferometer Space Antenna (LISA; \citealt{2017arXiv170200786A, 2023LRR....26....2A, 2022hgwa.bookE..17A}).

A critical observational signature of this process occurs during the gas-rich galaxy merger phase. When both SMBHs are simultaneously accreting gas, the system appears as a dual AGN, providing the most direct electromagnetic evidence of galaxy merger-driven growth and allowing constraints on models of galaxy assembly \citep{2016MNRAS.458.1013S, 2019NewAR..8601525D, 2022MNRAS.514..640V}. If only one nucleus is active, the system may instead appear as an ``offset AGN'' (referred to here as a one-AGN pair)  \citep{2012ApJ...746L..22K, 2023ApJ...943...38S}. Despite their theoretical importance, a statistical census of dual AGNs at cosmological distances remains elusive.

{Identifying dual AGNs remains observationally challenging because resolving kiloparsec-scale separations requires high angular resolution, particularly at intermediate and high redshifts ($z\gtrsim1$; see \citealt{2019NewAR..8601525D} for a review and \citealt{2025ApJS..281...25P} for a recent census). High-resolution X-ray and radio observations provide some of the most secure confirmations by directly resolving two accreting nuclei, but their sensitivity and angular-resolution requirements limit large statistical samples, especially beyond the local Universe \citep{2003ApJ...582L..15K, 2008MNRAS.386..105B, 2006ApJ...646...49R, 2011ApJ...735L..42K, 2020ApJ...892...29F, 2021ApJ...907...71F}. HST and JWST imaging reveals double nuclei, disturbed merger morphologies, and obscured high-redshift candidates, but are limited by survey area and often require spectroscopic or X-ray confirmation \citep{2011ApJ...737..101L, 2012ApJ...746L..22K, 2023ApJ...950L...5Y, 2025A&A...696A..59P, 2025ApJ...986..101L}. Optical spectroscopy has identified candidates through double-peaked narrow emission lines, spatially offset emission-line regions, and projected quasar pairs, although blending, finite spatial resolution, and contamination from outflows or rotating disks can complicate interpretation \citep{2006AJ....131....1H, 2008ApJ...678..635M, 2009ApJ...705L..76W, 2010ApJ...708..427L, 2010ApJ...716..866S, 2010ApJ...719.1672H, 2012ApJS..201...31G, 2013ApJ...777...64C, 2020ApJ...899..154S}. IFU observations help distinguish true dual AGNs from complex gas kinematics \citep{2012ApJ...745...67F}, while \textit{Gaia} astrometry enables searches for unresolved luminous quasar pairs through photocenter variability and multi-component point-spread-function fitting, though such candidates require follow-up to confirm their nature \citep{2019ApJ...885L...4S, 2020ApJ...888...73H, 2022ApJ...925..162C, 2022NatAs...6.1185M}.}

Taken together, these techniques have assembled a rich but heterogeneous literature of fewer than $\sim$2000 spatially resolved dual AGN systems at $r_\mathrm{proj} < 50$\,kpc \citep{2025ApJS..281...25P}, the majority concentrated at $z \lesssim 0.3$. The intermediate and high-redshift regimes are covered mainly by deep, narrow, or targeted programs and remain poorly sampled, leaving no uniform wide-area census of kpc-scale dual AGNs across cosmic time.
The Dark Energy Spectroscopic Instrument \citep[DESI;][]{DESI2016b.Instr}, with precise spectroscopic redshifts for tens of millions of galaxies and quasars across $\sim$9000\,deg$^2$, allows us to identify close galaxy pairs and classify their nuclear activity homogeneously over a broad redshift baseline.
Combining DESI fiber spectroscopy with mid-infrared photometry from the Wide-field Infrared Survey Explorer \citep[WISE;][]{2010AJ....140.1868W, 2011ApJ...735..112J,2012MNRAS.426.3271M,2012ApJ...753...30S,2018ApJS..234...23A,2020ApJ...903...91Y,2022AJ....163..224H}
for AGN identification, we construct large, homogeneous samples of dual and offset AGN candidates spanning a wide range of redshifts and projected separations. DESI particularly expands the census at intermediate redshifts ($z \sim 0.2$--$0.4$), where earlier samples were sparse, and substantially increases the number of known systems at $z \sim 0.7$--$1$.

Placing the observed population in a theoretical context requires a simulation matched to the survey selection. In a related study, the large-volume cosmological hydrodynamical simulation \texttt{ASTRID} \citep[described in][]{Bird2022, Ni-Astrid, Chen2022, Chen2023, DiMatteo2023, Ni2024, Zhou2025} is used to construct a DESI-like mock dual AGN catalog by applying the same selection functions and observational constraints as used here \citep{2025arXiv251216844C}. The mock sample reproduces the observed dual fraction, separation distribution, and host galaxy and AGN properties, validating \texttt{ASTRID} as a reliable forward-modeling tool for the observed population.

DESI dual AGN catalog presented here provides a large, uniform sample of electromagnetically identified SMBH pairs at kiloparsec separations. By forward modeling these systems in \texttt{ASTRID}, we connect observable properties, such as stellar mass, star formation activity, and AGN luminosity, to the likelihood of subsequent black hole mergers, including the subset expected to enter the sensitivity regimes of LISA and PTAs. Although individual systems cannot be uniquely associated with future gravitational wave events, the DESI dual AGN population provides empirical constraints on massive black hole merger demographics and the progenitors of the low-frequency gravitational wave background.

In this paper, we use DESI spectroscopy complemented by WISE mid-infrared AGN selection to systematically search for dual and offset AGN candidates, and we use the \textsc{Astrid} mock to translate the resulting census into predictions for massive black hole mergers and their detectability with LISA.
This paper is organized as follows. 
In Section~\ref{methods}, we describe the DESI survey and the \texttt{ASTRID} simulation framework. 
Section~\ref{data} outlines our AGN identification, dual candidate selection, and data cleaning procedures, including a detailed treatment of fiber-assignment incompleteness for close pairs. 
In Section~\ref{results}, we present the demographics of dual AGNs and one-AGN pairs, and examine how their host galaxy and AGN properties depend on redshift and projected separation, relative to the full AGN population. 
Section~\ref{simulation} compares these observational results with simulation predictions, focusing on dual AGN fractions, merger rates, and implications for gravitational wave observables. 
Section~\ref{Case Studies} highlights systems of particular interest, including high redshift dual AGNs and dwarf galaxy pairs that may serve as potential LISA sources. 
Finally, Section~\ref{conclusions} summarizes our main findings and discussions.

%% file: Sec2_Methods.tex
\section{Data and Simulations}
\label{methods}
\subsection{DESI}
\label{sec:DESI}

The Dark Energy Spectroscopic Instrument is a highly multiplexed, robotic, fiber-fed spectroscopic survey operating on the Mayall 4m telescope at Kitt Peak National Observatory \citep{DESI2022.KP1.Instr, DESI2016b.Instr}. Its focal plane hosts 5000 robotic fibers over a $\sim3^\circ$ diameter field of view \citep{FocalPlane.Silber.2023, Corrector.Miller.2023}, feeding ten spectrographs that cover $3600$--$9800~\AA$ at resolution $R\sim2000$--$5500$ \citep{FiberSystem.Poppett.2024}. Designed as a Stage-IV dark energy experiment \citep{Snowmass2013.Levi, DESI2016a.Science}, the main survey began in May 2021 and will ultimately obtain redshifts for ${\sim}63$ million galaxies and quasars over ${\sim}17{,}000$\,deg$^2$ \citep{SurveyOps.Schlafly.2023}. In this study, we use the first public data release (DR1), comprising the first 13 months of main-survey observations together with uniformly reprocessed survey-validation data \citep{DESI2024.I.DR1}. DR1 provides high-confidence redshifts for ${\sim}18.7$ million unique targets over more than $9000$\,deg$^2$, including ${\sim}13.1$ million galaxies and ${\sim}1.6$ million quasars \citep{DESI2024.I.DR1}. 
These data have already enabled precision cosmological constraints from baryon acoustic oscillations and clustering analyses \citep{DESI2024.VII.KP7B}.
Subsequent releases, including DR2 and later datasets, continue to expand the statistical power of DESI measurements \citep{DESI.DR2.DR2,Y3.clust-s1.Andrade.2025,Y3.lya-s1.Casas.2025}. 

Spectra are reduced with the DESI pipeline \citep{Spectro.Pipeline.Guy.2023}, with redshifts and classifications from the \texttt{Redrock} template-fitting framework\footnote{\url{https://github.com/desihub/redrock/releases/tag/0.15.4}} \citep{Redrock.Bailey.2024, RedrockQSO.Brodzeller.2023}; the supporting survey infrastructure and value-added products are described in \citet{FBA.Raichoor.2024, Expcalc.Kirkby.2024, AstrometricCalib.Kent.2023, SGA.Moustakas.2023, LSSCatalogs.Ross.2025}.

DESI targets four extragalactic tracer classes relevant to this work, spanning a broad redshift range: the Bright Galaxy Survey (BGS) \citep{BGSPrelim.RuizMacias.2020, BGS.TS.Hahn.2023, 2025AJ....169..157J}, Luminous Red Galaxies (LRG) \citep{LGRPrelim.Zhou.2020, LRG.TS.Zhou.2023}, Emission Line Galaxies (ELG) \citep{ELGPrelim.Raichoor.2020, ELG.TS.Raichoor.2023}, and quasars (QSO) \citep{QSOPrelim.Yeche.2020}, alongside the Milky Way Survey \citep{MWSPrelim.AllendePrieto.2020,
MWS.TS.Cooper.2023}. Secondary target programs (SCND) use spare fibers to observe objects beyond the primary selection; those most relevant here include \texttt{LOW\_MASS\_AGN}, targeting faint low-redshift AGN in dwarf galaxies, and \texttt{QSO\_RED} and \texttt{WISE\_VAR\_QSO}, targeting additional quasar populations (see Appendix~B of \citealt{DESI2023b.KP1.EDR}; also \citealt{2023ApJ...954..149D, 2023MNRAS.525.5575F, hall2026desitransients}). Target selection is based primarily on optical and mid-infrared photometry from Data Release~9 of the DESI Legacy Imaging Surveys\footnote{\url{https://www.legacysurvey.org/dr9/}} \citep{LS.Overview.Dey.2019, BASS.Zou.2017, TS.Pipeline.Myers.2023, LS.dr9.Schegel.2024}, with \textit{Gaia} information used mainly for stellar targeting and star--galaxy separation \citep{2016A&A...595A...1G}. Survey validation and visual inspection confirmed the reliability of the spectral classifications and redshifts \citep{DESI2023a.KP1.SV, VIGalaxies.Lan.2023, 2023AJ....165..124A}.

Although DESI is designed for precision cosmology, its combination of wide area, dense sampling, and uniform spectroscopy makes it possible to assemble statistical samples of rare systems. This work exploits the DR1 spectroscopic dataset to identify and characterize candidate dual and offset AGN systems.

\subsection{\texttt{ASTRID}}

Hosting $0.33$ trillion particles in a simulation box of $250\,{\rm Mpc}/h$ per side, \astrid\ \citep{Bird2022,Ni2025, Zhou2025} is one of the largest hydrodynamical simulations evolved to $z=0$. 
Its mass resolution is $m_{\rm DM}=6.74\times 10^{6}\,h^{-1}$~\Msun\ for dark matter (DM) and $m_{\rm gas}=1.27\times 10^{6}\,h^{-1}$~\Msun\ for gas in the initial conditions. 
The gravitational softening length is $\epsilon_{\rm g}=1.5\,h^{-1}$ kpc for both DM and gas particles. The adopted cosmological parameters are from \citet{Planck}.
With its large cosmological volume and relatively high resolution, \astrid\ hosts a large massive black hole (MBH) population covering a mass range of $10^{4.5}-10^{11}\ M_{\odot}$ \citep{Zhou2025}. 
The statistical power of \astrid\ allows us to study AGN across a wide range of luminosities and galactic environments, which is crucial for constructing mock catalogs of dual AGN systems for large surveys like DESI. \astrid\ has previously been used to characterize the dual AGN population and its host galaxies across cosmic time \citep{Chen2023, 2024OJAp....7E..84D}.

\astrid\ includes a full-physics sub-grid treatment for galaxy formation, black holes, stellar and AGN feedback, and inhomogeneous reionization. In the following we briefly describe the black hole model in \astrid\, for more details on the simulation see \citet{Ni2025}.
In \astrid, black holes are seeded in halos with mass above $5\times 10^{9}\,h^{-1}$~\Msun\ with initial seed masses drawn from a power law between $3\times 10^{4}\,h^{-1}$~\Msun and $3\times 10^{5}\,h^{-1}$~\Msun. 
The black hole accretion rate $\dot{M}$ is computed using the Bondi-Hoyle-Lyttleton formula \citep{BondiHoyle1944, DiMatteo2005_BH_model}, and is capped at twice the Eddington limit.
We estimate the AGN bolometric luminosity as $L_{\rm bol}=\eta\dot{M}_{\rm BH}c^2$, where $\eta=0.1$ is the mass-to-energy efficiency \citep{Shakura1973}.
AGN feedback is implemented in two modes: a quasar mode for high accretion rates and a radio mode for low accretion rates \citep{Weinberger2017_tng}.
Uniquely among large-scale cosmological simulations, \astrid\ adopts a subgrid model for MBH dynamical friction following the prescription of \citet{Tremmel2017,Chen2022_DF}, which gives a more realistic treatment of black hole orbital decay during galaxy mergers, allowing them to sink toward the centers of their host galaxies and merge on physical timescales. 
Two black holes are merged when their separation becomes smaller than twice the gravitational softening length and they are gravitationally bound.

As the DESI survey targets different tracers across redshift, we select dual AGN based on their host galaxy properties to match the DESI tracer population. 
To match the QSO tracer in DESI, we select the AGN in \astrid\ whose UV emission outshines that of their host galaxy. 
We model each star particle as a simple stellar population (SSP) with its birth time, metallicity, and mass extracted from the simulation. We use the \textsc{FSPS} stellar population synthesis code \citep{Conroy2009, Conroy2010} with the PARSEC isochrones \citep{Bressan2012_parsec}, MILES stellar library \citep{Sanchez-Blazquez2006_miles}, assuming a Chabrier initial mass function \citep{Chabrier2003}. 
The luminosity for an individual galaxy is then obtained by summing the emission from all star particles in that galaxy.  
For the ELG tracer, we follow \citet{Yuan2025MNRAS.538.1216Y} and select AGN hosted by galaxies with star formation rate (SFR) $>1\,M_\odot/{\rm yr}$ and stellar mass $10^{8.5}\,M_\odot <M_{\rm gal} < 10^{10.5}\,M_\odot$.
For the LRG tracer, we follow \citet{Yuan2022MNRAS.512.5793Y} and select AGN hosted by galaxies with stellar mass $M_{\rm gal} > 10^{11}$ $M_\odot$.
We construct the dual AGN sample from different mock tracer populations across redshift because the dominant DESI tracer varies with redshift.
Specifically, we divide the sample into four redshift intervals.
At $z\geq 1.6$, dual AGN are selected from the QSO population in \astrid. At $0.8\leq z<1.6$, we select dual AGN from both the ELG-like and QSO samples. At $0.5\leq z<0.8$, the dual AGN population is identified from LRG and QSO.
Finally, at $z<0.5$, the DESI AGN population contains a mixture of BGS sources and secondary targets. In this regime, we directly match the single AGN luminosity distribution and select duals accordingly. 
For more details of constructing the DESI-mock dual AGN catalog from \astrid\ simulation, we refer readers to \cite{2025arXiv251216844C}.

To identify dual AGN that are likely to merge within a Hubble time and contribute to LISA detections, we trace the evolution of dual AGN in \astrid\ from their initial identification as duals to their eventual merger. We calculate the LISA signal-to-noise ratio (SNR) for each merger event 
using the \texttt{gwsnrcalc} package \citep{Katz2019}, assuming circular orbits. We sum the signals over the final four years before coalescence. 
If a merger event has ${\rm SNR}>10$, we label its progenitor dual AGN as a LISA-detectable dual AGN.

%% file: Sec3_Data.tex
\section{Dual and offset AGN Selection} \label{data}

We begin our candidate selection using the DESI AGN Host Galaxies Physical Properties value-added catalog (VAC)\footnote{\url{https://data.desi.lbl.gov/doc/releases/dr1/vac/cigale/}} \citep{2024A&A...691A.308S}, which is derived through spectral energy distribution (SED) fitting using the \texttt{Code Investigating GALaxy Emission} (\texttt{CIGALE} v22.1; \citealt{2019A&A...622A.103B}). 
This catalog provides detailed physical parameters for DESI DR1 galaxies, obtained by simultaneously fitting galaxy and AGN emission components to multi-wavelength photometric data (optical bands $g$, $r$, $z$, and mid-infrared WISE bands W1, W2, W3, W4;  \citealt{BASS.Zou.2017,LS.Overview.Dey.2019,LS.dr9.Schegel.2024}). 
The \texttt{CIGALE} model incorporates a delayed star formation history with an optional exponential burst, \citet{2003MNRAS.344.1000B} stellar population synthesis with a \cite{Chabrier2003} initial mass function, nebular emission modeling from \citet{2011MNRAS.415.2920I}, dust attenuation following \citet{2000ApJ...533..682C}, dust emission modeling from \citet{2014ApJ...784...83D}, and AGN emission modeling according to \citet{2006MNRAS.366..767F}. The quality of the fit is evaluated through reduced $\chi^2$ values, and estimates and uncertainties of the physical properties represent the means and standard deviations weighted by the likelihood of their probability distribution functions.

The DESI DR1 parent sample contains $\sim17$ million \texttt{GALAXY} and \texttt{QSO} spectra after applying stringent quality criteria described below. Spectra are required to satisfy $\texttt{COADD\_FIBERSTATUS}=0$ to ensure the absence of known fiber or observational issues. Reliable redshift measurements are selected based on results from the DESI spectroscopic reduction pipeline and the \texttt{Redrock} redshift-fitting framework \citep{Spectro.Pipeline.Guy.2023, Redrock.Bailey.2024}, requiring either $\texttt{ZWARN}=0$ (secure redshift) or $\texttt{ZWARN}=4$ that flags spectra with a high fraction of outlier pixels relative to the best-fit template, which is commonly triggered in quasar spectra due to their intrinsic complexity and is not indicative of redshift failure \citep{RedrockQSO.Brodzeller.2023}. Only primary observations ($\texttt{ZCAT\_PRIMARY}=\mathrm{True}$) are retained to ensure a single spectrum per target. Duplicate entries are removed through cross-matching with the Legacy Surveys Tractor photometric catalog, which provides uniform multi-band photometry derived from forward modeling of the imaging data \citep{2016ascl.soft04008L,TS.Pipeline.Myers.2023}.

We systematically search within  HEALPix pixels and their neighboring pixels for galaxy/QSO pairs separated by projected distances $\Delta r < 50$ kpc and line-of-sight velocity difference $\Delta v < 600~$  km~s$^{-1}$. 
{The $\Delta r$ cut selects galaxy scale pairs in the regime where interaction-driven star formation and AGN enhancement have been observed, while remaining more conservative than the broader $\sim100$ kpc-scale limits sometimes adopted in literature dual AGN and QSO-pair compilations \citep[e.g.,][]{2025ApJS..281...25P, 2026ApJ..1000..311J}.
The $\Delta v$ cut removes chance superpositions of galaxies at different redshifts.  
When both objects are QSOs at $z>0.5$, we allow $\Delta v < 2000~{\rm km~s^{-1}}$, following high redshift quasar pair studies where broad UV emission line redshifts can have systematic uncertainties of order $\sim1000~{\rm km~s^{-1}}$ and where peculiar velocities may further broaden the observed velocity difference \citep{2006AJ....131....1H, 2010ApJ...719.1672H}.}

We apply additional quality cuts to ensure reliable estimates of physical properties for both members of each pair. Both objects in the pair are required to have $\log (M_\star/M_\odot) \geq 6$ to exclude poorly constrained stellar-mass estimates, a reduced chi-square of  $\chi^2 \leq 10$ from \texttt{CIGALE} SED fitting to remove sources with unreliable model photometry,
and a minimum signal-to-noise ratio of $\mathrm{SNR} \geq 5$ in each of the $g$-, $r$-, and $z$-bands to ensure robust flux measurements across all optical bands.

\subsection{Removing False Pairs and Photometric Artifacts}
We apply additional cleaning filters to eliminate likely false pairs and problematic cases. 
First, we remove any candidate with an angular separation $\Delta\theta<0.05''$. Such extremely small separations are far below the $\sim1''$ typical seeing and the $\sim0.07''$ astrometric precision of DESI fiber positioning. Objects this close together are likely false detections of a single source, such as a galaxy observed twice or a deblending error. In practice, virtually no truly distinct galaxies can be resolved at $<0.05''$ even with HST, so we safely discard these cases. 

Additionally, we set aside pairs with separations in the range $0.05''\leq\Delta\theta\leq 1.6''$ for special consideration. The scale of $\sim1.6''$ corresponds to the diameter of a DESI fiber in the sky, so pairs closer than this threshold would fall into a single fiber if observed on the same tile, or at least significantly overlap within the fiber’s aperture. Although DESI uses multiple passes and a tiling strategy to target close pairs, any objects within $1.6''$ are inherently difficult to deblend spectroscopically. 
Moreover, the Legacy Surveys imaging (with $\sim 1''$ PSF) starts to blur sources at these scales, so the photometry of such close pairs is also suspect. 
For our main sample, we use a conservative minimum separation of $\sim1.6''$ to avoid fiber overlap issues, while acknowledging that true dual AGNs can exist at smaller separations (we will treat those with specialized methods in a follow-up study).

We also exclude any systems where a galaxy has multiple companions satisfying the pair criteria (triplets or larger groups) from the main catalog. This exclusion is to avoid double counting and because the dynamics of such systems are more complex and difficult to interpret. {More broadly, this work does not include a dedicated environmental analysis; assessing the role of group and cluster environments including higher-order systems will be an important direction for future work.}

Another systematic affecting close galaxy pairs is photometric shredding in the DESI Legacy Imaging Surveys. Shredding occurs when a single extended or clumpy galaxy is fragmented into multiple catalog entries by the Tractor source-modeling pipeline \citep{2016ascl.soft04008L}, particularly for nearby, irregular, or low-surface-brightness systems with multi-component structure. This can lead to underestimated fluxes, biased stellar masses, and spurious faint companions.
A common mitigation is to apply \texttt{FRACFLUX} cuts in the \textit{grz} bands, typically requiring $\texttt{FRACFLUX}\lesssim0.2$--$0.35$, where high values indicate significant flux contamination from nearby sources \citep{2023ApJ...954..149D, 2025ApJ...982...10P}. 
However, such cuts are problematic for genuine close pairs, where elevated \texttt{FRACFLUX} is expected by construction and may reflect real companions rather than artifacts. They may also preferentially remove nearby, clumpy, star-forming dwarfs, which are scientifically important as possible low-mass massive black hole merger progenitors.
For example, Manwadkar et al. (in prep.) address this limitation with a dedicated DESI DR1 pipeline that identifies shredded sources, reconstructs parent galaxies using image segmentation and color-based pixel association, and remeasures total fluxes with curve-of-growth photometry. This approach recovers genuine dwarf galaxies that simple \texttt{FRACFLUX} cuts would discard, but is optimized for low-redshift dwarf systems ($M_\star \lesssim 3\times10^9\,M_\odot$, $z\lesssim0.1$) and therefore does not cover the full redshift range of our AGN pair sample.
For our sample, we visually inspect dwarf pairs with $\texttt{FRACFLUX}>0.35$ in the \textit{grz} bands, using the Manwadkar et al. (in prep.) shredded-source catalog as a guide where available, to distinguish genuine companions from likely shredding artifacts. 
{
This cleaning is subject to visual assessment and is therefore imperfect. Systems with small galaxy companions may look very similar to single galaxies with clumpy or pronounced substructure, making them difficult to distinguish by eye without additional information, such as velocity fields from integral-field spectroscopy or much higher spatial-resolution imaging. Some true companions may therefore be removed as suspected shreds, while residual artifacts may remain.
}

We further apply Legacy Survey DR9 \texttt{MASKBITS}\footnote{\url{https://www.legacysurvey.org/dr9/bitmasks/}} cuts to remove sources affected by severe imaging artifacts or contamination. We exclude objects near very bright stars (\texttt{MASKBITS}=1), bad or masked pixels in the $g$, $r$, or $z$ bands (\texttt{MASKBITS}=5,6,7), and globular-cluster regions (\texttt{MASKBITS}=13). We also visually inspect, rather than automatically remove, objects flagged by medium-bright Gaia star masks or SGA-2020 large-galaxy footprints (\texttt{MASKBITS}=11,12; \citealt{SGA.Moustakas.2023}), since genuine dual or offset AGN candidates may lie near bright stars or extended galaxies.

\subsection{Fiber assignment incompleteness for close duals}
\label{weights}

Close angular pairs in DESI suffer from spectroscopic incompleteness due to the finite patrol radius and physical size of fiber positioners, which prevent both members of very close pairs from being observed simultaneously within a single exposure. This leads to a scale-dependent loss of close pairs and must be corrected when interpreting the incidence of small-separation dual AGN.

To obtain unbiased small-separation statistics while preserving consistency on large scales, we adopt the DESI DR1 fiber-assignment mitigation scheme presented by \citet{2025JCAP...04..074B}. This approach combines pairwise inverse-probability (PIP) weights with a small-scale angular upweighting correction and has been shown to recover the true clustering signal across all scales in DESI DR1. Throughout this work, these corrections are applied only when comparing our measurements to theoretical predictions \citep[e.g.\ \texttt{ASTRID} dual/offset AGN models][]{2025arXiv251216844C}, while raw counts are used for descriptive statistics.

The PIP method estimates the probability that each galaxy pair is observed given the DESI fiber-assignment process. This is done using multiple realizations of the targeting algorithm, encoded in the DESI bitweights. 
If $K$ is the number of realizations and $c_i$ is the number of times galaxy $i$ is successfully observed, the efficient estimator for the individual inverse-probability (IIP) weight is \citep[eq.~5.8 from][]{2025JCAP...04..074B}
\begin{equation}
w_{\rm eff}(c_i)=\frac{K+1}{c_i+1}.
\end{equation}
For a pair of galaxies $(i,j)$, the pairwise inverse-probability weight is then constructed as \citep[eq.~5.11 from][]{2025JCAP...04..074B}
\begin{equation}
w^{\rm PIP}_{ij}=w_{\rm eff}(c_i)\,w_{\rm eff}(c_j)\,
\frac{w_{\rm eff}(c_{ij})}{g_K(c_i,c_j)},
\end{equation}
where $c_{ij}$ is the number of realizations in which both galaxies are observed together and $g_K(c_i,c_j)$ is a normalization term that removes finite-sampling biases. This implementation follows the DESI DR1 formalism and improves upon earlier SDSS/eBOSS fiber-collision corrections by providing an unbiased estimator even for highly incomplete samples and limited numbers of assignment realizations \citep{2020MNRAS.498..128M}.

Intuitively, PIP weighting corrects for the fact that some close galaxy pairs are less likely to be observed because two nearby targets cannot always be assigned fibers simultaneously. Observed pairs are therefore upweighted according to the probability that both members were successfully assigned fibers across multiple realizations of the fiber-assignment algorithm. However, a small fraction of pairs have zero joint assignment probability and therefore cannot be recovered by PIP weighting alone.

To account for these cases, we apply an additional angular upweighting factor $U(\theta)$ following \citet{2025JCAP...04..074B}. This factor rescales the weighted pair counts so that the angular clustering of the spectroscopically observed sample (denoted by ``obs'' in the equation below) matches that of the full photometric target sample prior to fiber assignment (denoted as ``full''). Here, the full sample corresponds to all photometric targets of a given DESI tracer class satisfying the survey selection criteria, regardless of whether they were spectroscopically observed. $U(\theta)$ is computed following:
\begin{equation}
U(\theta)=
\frac{DD^{\rm (full)}(\theta)/p_{\rm full}}
     {DD^{\rm (obs)}(\theta)/p_{\rm obs}},
\end{equation}
where $DD(\theta)$ represents raw angular pair counts and $p$ is the total number of pairs in each sample. 
We compute $DD(\theta)$ in logarithmically spaced angular bins from $1.6''$ to $900''$, using 12 bins per dex and assign corrections using interpolation in $\log\theta$. The correction is significant only below the fiber-collision scale and approaches unity at larger separations. This correction is therefore particularly important for close galaxy pairs, where fiber-assignment constraints are strongest.

Because fiber-assignment corrections are computed independently for each DESI tracer population, residual incompleteness may remain for pairs involving different tracer classes. For such cross-tracer pairs, pairwise inverse-probability weights are not always directly available, so we approximate the pair weight as the product of the individual galaxy weights, $w^{\rm IIP}_{ij}=w^{\rm IIP}_i w^{\rm IIP}_j$. This approximation becomes exact when the selection probabilities of the two galaxies are independent, which is typically expected for pairs separated by more than the fiber-collision scale \citep{2025JCAP...04..074B}. 

We note that not all galaxies in the AGN and galaxy catalogs have a corresponding entry in the DESI LSS catalogs. For such objects, no bitweight information is available and we assign a completeness weight of unity ($w = 1$).

\begin{figure}
\centering
\includegraphics[width=0.39\textwidth]{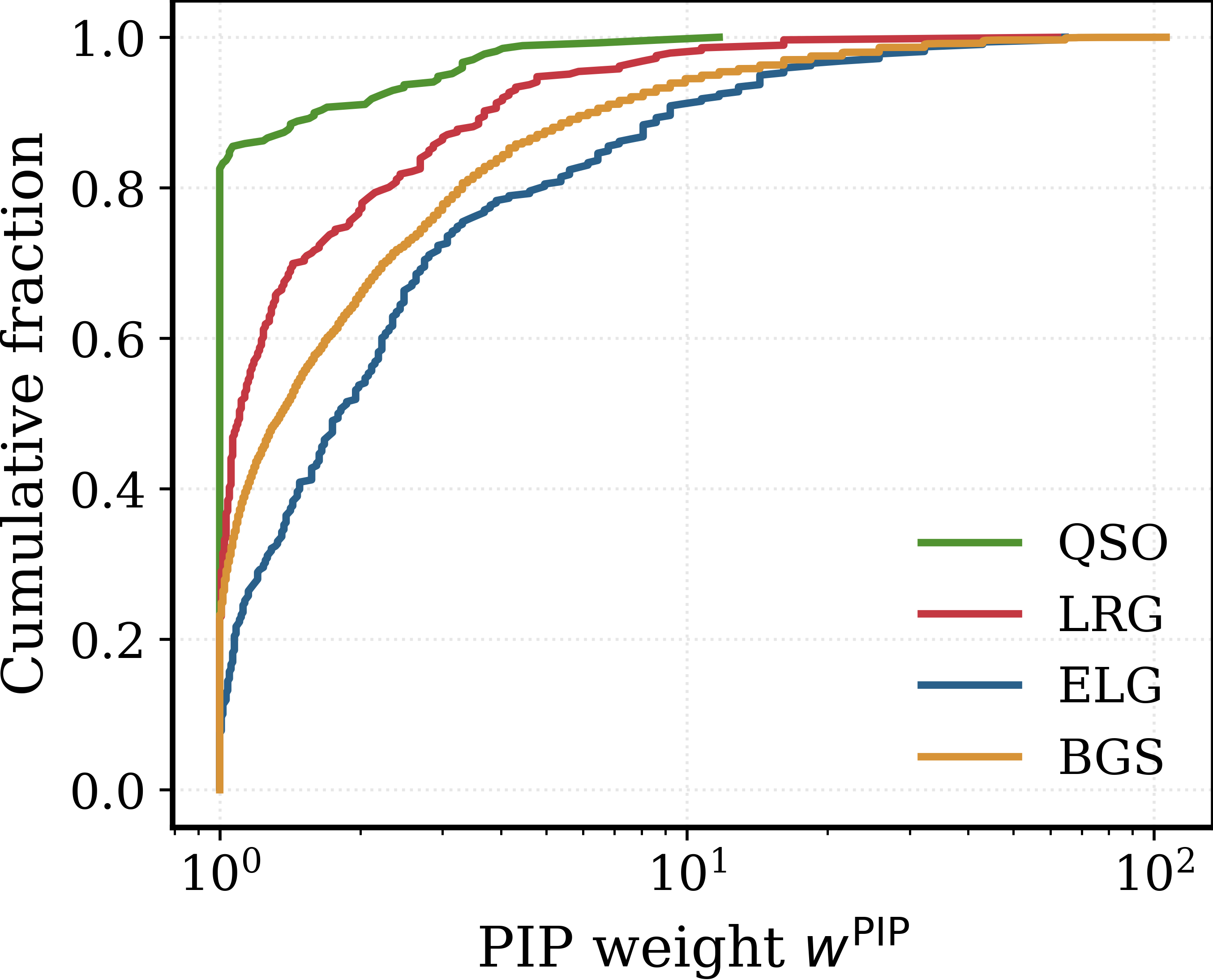}
\caption{Cumulative distribution functions (CDFs) of the pair-inverse-probability (PIP) weights, $w^{\rm PIP}$, for dual AGN candidate pairs, shown separately for the different tracers. 
}
\label{fig:w_pip}
\end{figure}
It is worth noting that the magnitude of these corrections varies with target tracers. 
Figure \ref{fig:w_pip} shows the cumulative distributions of PIP weights, $w^{\rm PIP}$, for close dual AGN candidate pairs separated by tracer.
The ELG sample suffers significant fiber collisions, only about one-third of the close ELG-ELG pairs are observed in some regions ($\sim35\%$ completeness), necessitating large PIP weights and $U(\theta)$ factors to restore missing pairs. 
In contrast, the much sparser QSO sample is $\sim88\%$  complete, so fiber collisions have a minimal effect on QSO–QSO pair counts (hence $w^{\rm PIP}\approx1$ and $U(\theta)\approx1$ for most quasar pairs). 
Intermediate tracers like LRGs and bright galaxies (BGS) fall in between, with moderate incompleteness (on the order of $30$--$40\%$ missing pairs in high-density regions). 

DESI DR1 analyses have verified that PIP with angular upweighting recovers the true clustering signal in both configuration and Fourier space, even for the most fiber-collision-impacted samples \citep{2025JCAP...04..074B}. Thus, by including these corrections, we ensure that the observed incidence of close dual AGN is not biased by spectroscopic targeting limitations, allowing us to robustly interpret the small-separation behavior and compare it directly with model predictions.

\subsection{AGN Identification and Classification}
\begin{figure*}
\centering
\includegraphics[width=\textwidth]{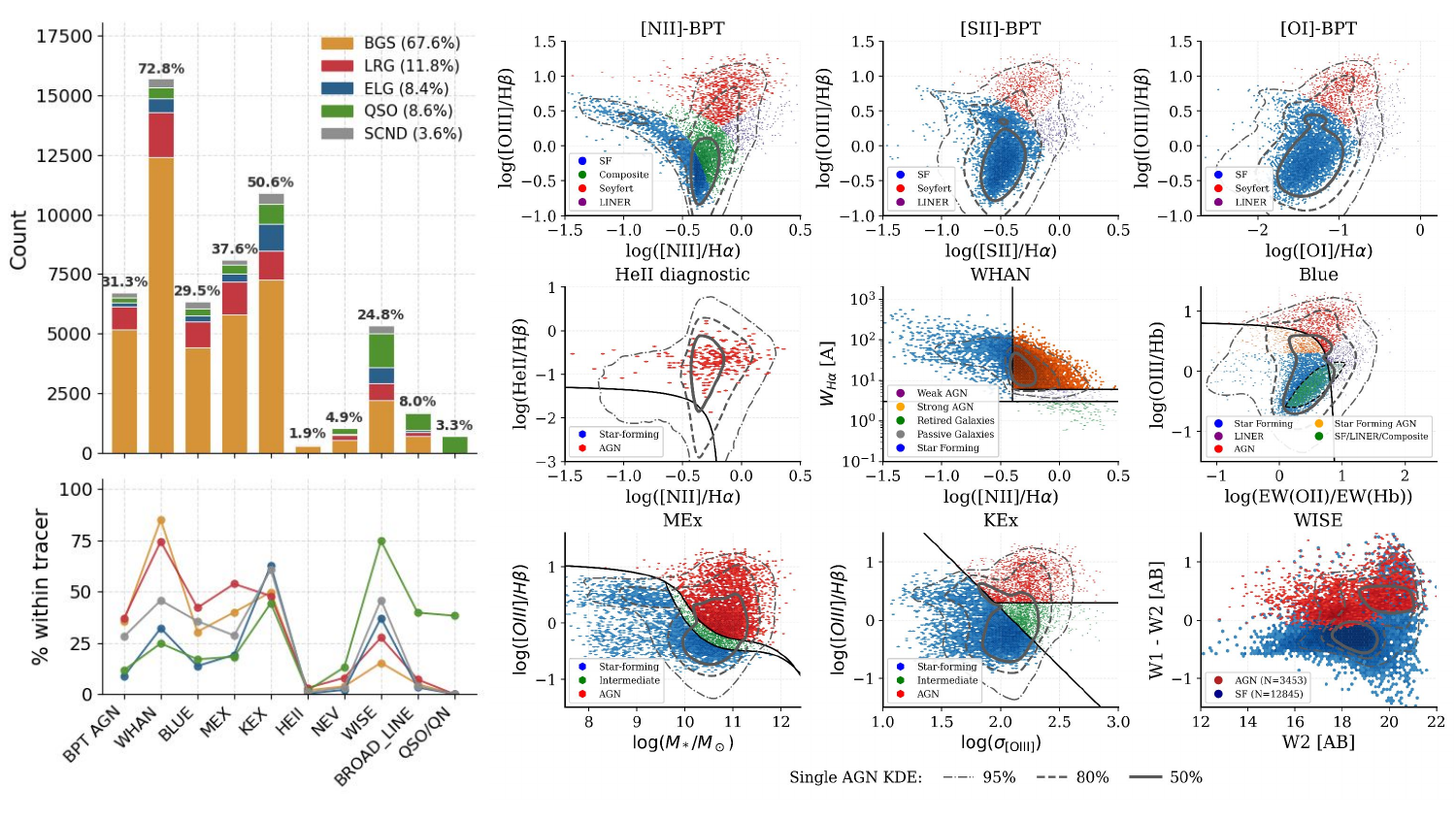}
\caption{
\textbf{Left:} Distribution of AGN selection methods across the sample of candidate dual AGN systems, categorized by their DESI target class. 
The top panel shows the total number of spectroscopic fibers (primary and secondary combined) that satisfy each AGN selection criterion, with percentage labels indicating each method's contribution to the full AGN sample. 
Bar colors denote the original DESI target class. 
The bottom panel shows the fractional contribution of each diagnostic within individual DESI tracer populations.
\textbf{Right:} Emission-line and photometric diagnostic diagrams used in the AGN classification. 
Dual AGN candidates are shown as colored points, with colors indicating their classification within each diagnostic diagram. The background distribution of isolated AGNs from the VAC is shown as kernel-density contours enclosing $50\%$, $80\%$, and $95\%$ of the population. 
Standard demarcation curves separating star-forming galaxies, composite systems, and AGNs are shown for each diagnostic diagram, including classical BPT line-ratio diagrams, the WHAN diagram, BLUE diagram, He\,II diagnostic, Mass–Excitation (MEx), Kinematics–Excitation (KEx), and mid-infrared WISE color selection.
}
\label{fig:AGN_maskbit}
\end{figure*}

For AGN classification, we use the DESI DR1 AGN/Galaxy Classification Value-Added Catalog (VAC)\footnote{\url{https://data.desi.lbl.gov/doc/releases/dr1/vac/agngal/}}, which provides a homogeneous AGN classification for all spectra across the main DESI target classes (Juneau et al., in preparation). 
This catalog starts from all extragalactic spectra ($z > 0.001$) and combines the outputs of the DESI spectral pipeline and several classification algorithms. The redshift and initial type of each spectrum come from the \texttt{Redrock} pipeline \citep[][]{Redrock.Bailey.2024}, which assigns SPECTYPE = GALAXY, QSO, or STAR along with the redshift. For quasars, the machine-learning tool \texttt{QuasarNet} and a Mg II emission-line “afterburner” are applied to refine the redshift and identify broad-line QSOs missed by \texttt{Redrock} \citep[see][]{2023ApJ...944..107C, 2023AJ....165..124A}. 
AGN classifications in the VAC are based on both optical spectroscopic diagnostics and mid-infrared photometry. Emission-line measurements are obtained with the \texttt{FastSpecFit} pipeline \citep{2023ascl.soft08005M}, which fits the continuum and emission lines for each spectrum. These measurements are used to construct several classical optical diagnostic diagrams, including the [N\,II], [S\,II], and [O\,I] BPT diagrams \citep{1981PASP...93....5B,2001ApJ...556..121K,2003MNRAS.346.1055K,2006MNRAS.372..961K,2007MNRAS.382.1415S,2021ApJ...915...35L, 1987ApJS...63..295V}, which distinguish star-forming galaxies from AGN and further separate Seyfert and LINER populations. Additional optical diagnostics include the WHAN diagram \citep{2011MNRAS.413.1687C}, the BLUE diagram \citep{2004MNRAS.350..396L,2010A&A...509A..53L}, the He\,II diagram \citep{2012MNRAS.421.1043S}, and the Mass–Excitation (MEx) and Kinematics–Excitation (KEx) diagrams \citep{2014ApJ...788...88J, 2018ApJ...856..171Z}. 
For the MEx classification, the stellar masses used in the diagnostic are taken from \texttt{FastSpecFit}, not from the \texttt{CIGALE} VAC used later for host galaxy property analysis.
The presence of high-ionization emission such as [Ne\,V]$\lambda3426$ also serves as a direct indicator of AGN activity \citep{1998A&A...329..495S}. Complementary AGN classifications are obtained from mid-infrared photometry using WISE colors \citep{2011ApJ...735..112J,2012MNRAS.426.3271M,2012ApJ...753...30S,2018ApJS..234...23A,2020ApJ...903...91Y,2022AJ....163..224H}, which identify obscured AGN through the characteristic emission of the dusty torus. 
For sources identified exclusively through optical diagnostics, we require \texttt{FLAG\_OPTICAL} from CIGALE VAC $\geq 2$, corresponding to $\mathrm{SNR}\geq 10$ in at least two of the $g$, $r$, and $z$ bands, to reduce contamination from low-quality spectra.

Using the above diagnostics, each galaxy in a pair is classified as AGN or non-AGN. Dual AGN candidates are pairs in which both components independently meet one or more AGN criteria (optical, infrared (IR), or broad-line). If only one member is flagged as an AGN and the other is inactive, it is referred to as a one-AGN pair. Pairs where neither galaxy shows any AGN signature form the inactive pairs sample (these serve as a control for analyzing galaxy pairs without nuclear activity). Note, we do not require both AGNs to be of the same type (for instance, one could be an IR-selected obscured AGN while the other is an optical Seyfert). We also compile the entire set of isolated AGNs from the VAC (i.e. all galaxies classified as AGN, regardless of companionship) as a comparison sample of isolated AGNs. This allows us to compare the properties and incidence of AGNs in pairs with those in the general field.

The limited angular resolution of WISE introduces an additional cleanup step for dual AGN identification. Because the WISE W1/W2 PSF has a FWHM of $\sim6''$, close pairs can be strongly blended, so a WISE-only AGN classification may reflect the combined mid-infrared emission of the pair rather than two distinct AGNs. We therefore flag candidates in which both components are classified as AGNs only by WISE and have $\Delta\theta<6''$, consistent with previous cautions against interpreting WISE-selected AGNs in merging systems below the WISE beam scale \citep{2017ApJ...848..126S, 2017MNRAS.464.3882W, 2018MNRAS.478.3056B}. These likely blended systems are excluded from the main catalog but retained as a systematic uncertainty estimate.

{
We also caution that for systems near the DESI fiber scale, the fiber aperture, seeing, and extended emission can cause the spectrum of one component to include light from its companion, and optical emission line diagnostics can be less reliable. Among pairs with $1.6'' \leq \Delta\theta < 2.0''$ ($\sim5.2\%$) $\sim80\%$ have \texttt{FRACFLUX}$>0.5$ in all three $grz$ bands for at least one component, indicating substantial imaging blending. We also find that $\sim45\%$  of this subset have more than one AGN evidence flag for both components. We retain this subset in the catalog, but caution that component-wise dual AGN classifications at these separations should be treated as candidates pending spatially resolved follow-up.}

Figure~\ref{fig:AGN_maskbit} (left) breaks down the AGN classifications in the dual sample by diagnostic and by DESI target class. Both primary and secondary targets are included, and because individual objects can carry multiple tracer labels, the percentages sum to more than $100\%$.

Optical AGN identifications based on non-BPT diagnostics (``OPT\_OTHER\_AGN'') dominate the sample, followed by classical BPT and mid-infrared (WISE) selections. 
Among individual diagnostics, WHAN classifications represent the largest category ($\sim73\%$ of all AGNs), followed by KEx ($\sim51\%$) and MEx ($\sim38\%$).
{However, only $\sim15\%$ of dual AGN sample have both nuclei classified exclusively by WHAN, while $\sim37\%$ have exactly one WHAN-only classified nucleus.}
Classical [NII], [SII], and [OI] BPT diagrams account for $\sim31\%$ of the sample, while mid-infrared–selected AGNs identified via WISE comprise $\sim25\%$. 
Broad-line AGNs contribute $\sim8\%$, of which $\sim83\%$ are classified as \texttt{QSO} by \texttt{Redrock} and the remainder carry additional AGN diagnostics.
Pipeline quasar identifications (QSO+QN, combining \texttt{Redrock}, \texttt{QuasarNet}, and Mg\,II afterburner recoveries) account for $\sim3\%$, with He\,II and [Ne\,V] adding $\sim2\%$ and $\sim5\%$.
{Additionally, $\sim31\%$ of dual AGN have both nuclei supported by two or more AGN diagnostic families (WHAN + BPT, WHAN + MEx, BPT + WISE, KEx + MEx); while $\sim6.7\%$ have both AGNs supported by multiple channel diagnostics: optical narrow-line diagnostics, WISE, broad-line/QSO. The latter is the most conservative subset.}

These fractions largely trace the target mix. BGS galaxies make up $\sim68\%$ of all AGNs (orange bars), which is why optical diagnostics dominate globally. QSO targets are only $\sim9\%$ of the sample (green bars) but contribute most of the broad-line, WISE, and QSO+QN identifications. Within tracers (bottom-left panel), BGS and LRG AGNs are identified overwhelmingly through optical emission lines (WHAN in $\sim86\%$ and $\sim72\%$, respectively, with KEx/MEx at $\sim40$--$60\%$ and BPT near $40\%$), ELGs split between KEx ($\sim55\%$) and WISE ($\sim39\%$), and QSO targets rely on luminous AGN indicators (QSO+QN and WISE colors in $\sim75\%$ each, broad lines in $\sim56\%$).

The right panels of Figure~\ref{fig:AGN_maskbit} place the dual AGN candidates on the diagnostic diagrams themselves, against kernel-density contours of the isolated AGN population enclosing $50\%$, $80\%$, and $95\%$. Dual AGNs occupy the same broad loci as isolated AGNs rather than forming a distinct spectral class.
Two-sample Kolmogorov--Smirnov (KS) and Anderson--Darling (AD) tests applied to each plotted quantity quantify this. Because the isolated sample is far larger, we use the KS distance, $D_{\rm KS}$, as the effect size. The BPT diagrams show negligible differences ($D_{\rm KS}=0.040$--$0.047$), WHAN, BLUE, MEx, KEx, and HeII show moderate differences ($D_{\rm KS}\simeq0.13$--$0.15$, although the HeII comparison uses only $N_{\rm dual}=283$ systems), and the only substantial offset is in WISE color ($D_{\rm KS}=0.29$).

%% file: Sec4_Result.tex
\section{Results and Discussions} \label{results}

\subsection{Pair Separations and Distance} \label{pair_sep}

\begin{figure*}
\centering
\includegraphics[width=0.99\textwidth]{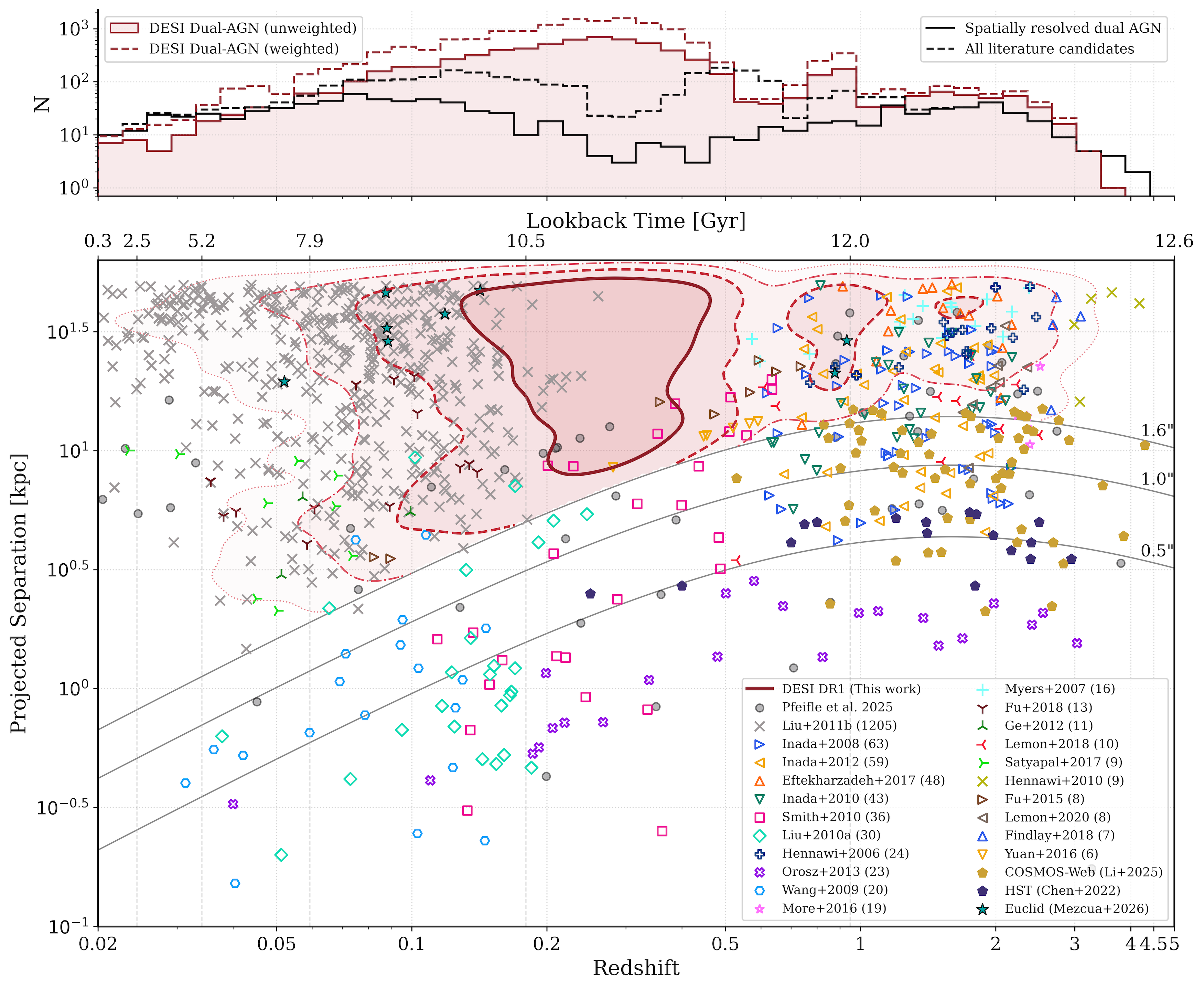}
\caption{
Projected physical separation as a function of redshift  for the dual AGN candidate sample. The red shaded regions represent the Kernel Density Estimates of the DESI-selected candidates, with contours indicating the $50\%$, $80\%$, $95\%$, and $99\%$ enclosed-probability levels. The gray lines indicate fixed angular separations of $0.5''$, $1''$, and $1.6''$; the latter corresponds to the DESI fiber diameter and marks the effective resolution limit for spectroscopic pairs. For comparison, colored symbols show confirmed dual AGNs from the literature, with the corresponding reference and sample size listed in the legend. 
The literature data are primarily drawn from the Big Multi-AGN Catalog \citep{2025ApJS..281...25P}, which includes major surveys by \citet{2011ApJ...737..101L, 2012ApJS..201...31G, 2016MNRAS.462.1603Y, 2014ApJ...789..112C, 2012AJ....143..119I}. 
We also include recent high-redshift results from \citet{2022ApJ...925..162C, 2025ApJ...986..101L, 2025A&A...696A..59P}, and dwarf dual AGN sample from \citet{2026arXiv260413170M}.
The upper panel shows the redshift distribution of the DESI dual AGN candidates compared to literature samples. The red-filled histogram represents the DESI dual AGN candidates from this work, while the red-dashed histogram shows the same sample after applying observational completeness weights. The black solid histogram shows spatially resolved dual AGN reported in the literature. The black dashed histogram shows the full literature sample, including systems selected through unresolved spectroscopic diagnostics (e.g., double-peaked emission lines), which may represent dual AGN candidates but do not necessarily correspond to spatially resolved pairs.
}
\label{fig:dr_z}
\end{figure*}

In Figure~\ref{fig:dr_z}, we present the distribution of projected separations between dual AGNs as a function of redshift for the candidates identified in this work, shown by the red kernel density estimate (KDE) contours. Projected separations are computed as $\Delta r_{\rm proj}=D_A(z_{\rm mean})\Delta\theta$, where $\Delta\theta$ is the angular separation between the two nuclei and $z_{\rm mean}$ is their mean redshift. The diagonal gray curves mark fixed angular separations. The $1.6''$ line corresponds to the DESI fiber diameter and represents the effective resolution floor for resolving distinct DESI sources, while the $0.5''$ line approximately marks the regime accessible to high-resolution space-based imaging such as {\it HST}.

To contextualize the DESI sample and identify the parameter space newly probed by this survey, we compare our candidates with literature dual AGNs, shown as colored symbols, restricted to $r_{\rm proj}<50$ kpc to match our selection. The comparison sample is drawn primarily from the Big Multi-AGN Catalog \citep[BigMAC;][]{2025ApJS..281...25P}, which compiles candidates reported up to $\sim2020$ from more than 150 studies, spanning projected separations of $\sim0.03$--$110$ kpc and including major surveys such as \cite{2011ApJ...737..101L} and \cite{2012AJ....143..119I}. 
We supplement BigMAC with recent close and high-redshift samples from COSMOS-Web, VODKA, and GA-NIFS, identified with deep {\it HST}/{\it JWST} imaging, Gaia varstrometry plus {\it HST} confirmation, and {\it JWST}/NIRSpec IFU spectroscopy, respectively \citep{2025ApJ...986..101L,2022ApJ...925..162C,2025A&A...696A..59P}, and with dual AGN candidates in low-mass galaxies ($M_\star < 10^{10}\,M_\odot$, $r_{\rm proj} \sim 20$--51 kpc) from the Euclid Quick Data Release \citep{2026arXiv260413170M}.
For the main panel, we include only spatially resolved literature systems with reliable projected separations, yielding 1,754 systems from 87 studies. Candidates identified only through unresolved spectroscopic signatures, such as double-peaked emission lines in single-fiber spectra, are excluded from this comparison but retained in the redshift histograms \citep[e.g.,][]{2012ApJS..201...23E,2016MNRAS.462.1603Y,2016MNRAS.463...24L,2013ApJ...763...36B}.

The upper panel of Figure~\ref{fig:dr_z} compares the redshift distributions of DESI and literature samples. The solid red histogram shows the DESI candidates, while the dashed red histogram shows the same sample after applying observational completeness weights (Section~\ref{weights}). The solid black histogram shows the spatially resolved literature sample used in the main panel, and the dashed black histogram shows the full literature compilation, including unresolved spectroscopic candidates.

The DESI-selected dual AGNs highlight several regimes where the survey expands the known population. 
At low redshift, DESI overlaps with a regime already well sampled by SDSS and follow-up studies \citep[e.g.,][]{2011ApJ...737..101L}, but the comparison changes rapidly with redshift.  Below $z\lesssim0.08$, DESI and the literature samples are broadly comparable, with $\approx260$ DESI candidates versus $\approx255$ spatially resolved systems and $\approx350$ systems in the full literature compilation. At $0.08\lesssim z\lesssim0.2$, DESI begins to dominate, identifying $\approx1500$ candidates, a factor of $\sim6.5$ larger than the spatially resolved literature sample and a factor of $\sim2$ larger than the full literature compilation.

The most significant difference appears at intermediate redshifts ($0.2 \lesssim z \lesssim 0.4$), where the DESI distribution of duals peaks, while the literature sample declines.
DESI contains 3410 candidates in this interval, compared to 48 spatially resolved literature systems and 301 systems in the full literature compilation, corresponding to increases of factors of $\sim70$ and $\sim11$, respectively.
A secondary DESI peak appears at $0.7\lesssim z\lesssim1$, where DESI identifies 355 candidates compared to 47 spatially resolved literature systems. 
These two peaks reflect the DESI tracer selection: the lower-redshift peak is dominated by BGS targets, while the higher-redshift peak is driven mainly by ELGs, with additional contributions from LRGs and QSOs.

At higher redshifts, close-pair identification becomes increasingly limited by angular resolution and spectroscopic depth. Nevertheless, DESI identifies 403 dual AGN candidates at $z>1$, roughly three times the 136 spatially resolved literature systems satisfying the same $1.6''$ angular separation criterion.
While \textit{Gaia}, \emph{HST}, and \emph{JWST} probe sub-kpc pairs closer to the gravitationally bound binary phase, DESI is most sensitive to host galaxy-scale duals at separations of a few to tens of kiloparsecs. Relaxing the angular criterion to include all spatially resolved literature systems at $z>1$ yields 245 comparators, a factor of $\sim$1.6 below the DESI count. 

Overall, the DESI DR1 dual AGN catalog contains 7125 systems, compared to 1754 spatially resolved literature systems and $\sim2689$ systems in the complete literature compilation including unresolved spectroscopic candidates, increasing the known sample by factors of $\sim4$ and $\sim2.7$, respectively. DESI fills a previously undersampled intermediate-redshift regime by nearly 2 orders of magnitude.


\begin{figure*}
\centering
\includegraphics[width=0.99\textwidth]{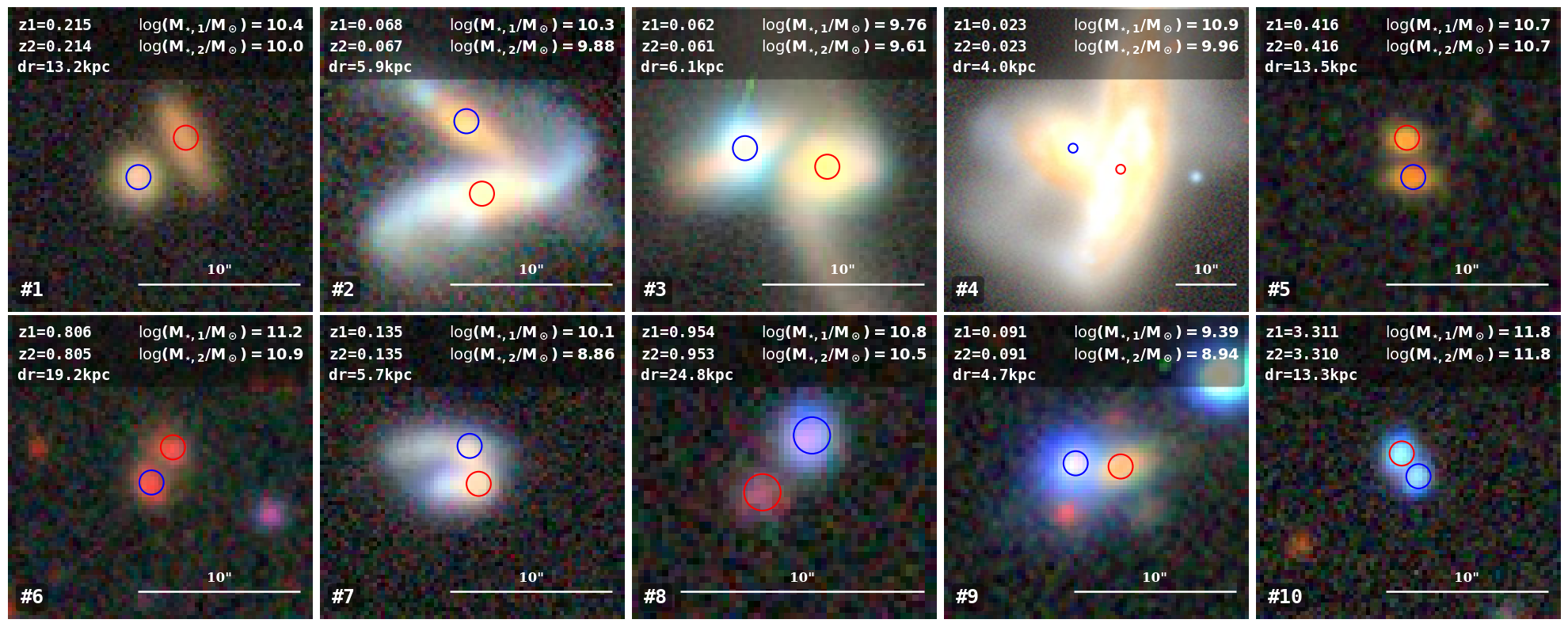}
\caption{Examples of candidate dual AGN systems identified in the sample. Background images are from the DESI Legacy Imaging Surveys. The red and blue circles indicate the positions of the spectroscopic fibers for the primary and secondary targets, respectively. The inset text in each panel lists the spectroscopic redshifts of the two sources ($z_1, z_2$), the projected physical separation ($dr$), and the stellar masses ($\log (M_{\star1,2}/M_\odot)$) of the host galaxies.} 
\label{examples}
\end{figure*}

While Figure~\ref{fig:dr_z} defines the statistical parameter space of the sample, Figure~\ref{examples} illustrates the diversity of host morphologies and interaction stages in the DESI dual AGN catalog. The background images are drawn from the DESI Legacy Imaging Surveys, with red and blue circles marking the spectroscopic-fiber positions of the primary and secondary targets, respectively.

The examples span nearby BGS--BGS systems in the local Universe (Panels~1--4), where the galaxies are well resolved and show a progression from relatively intact pairs at $\sim13~\mathrm{kpc}$ to increasingly disturbed and partially blended systems at $\sim4$--$6~\mathrm{kpc}$. 
Intermediate redshift LRG--LRG pairs (Panels~5--7) show smoother, centrally concentrated light profiles in contrast to the more visibly disturbed BGS systems, spanning mass ratios from near-equal to highly asymmetric. Panel~8 shows a mixed ELG--QSO pair at $z\sim1$. Panels~9 and~10 extend the sample to its mass and redshift extremes, showing a compact dwarf--dwarf pair with $M_\star<10^{9.5}M_\odot$ at $\sim5~\mathrm{kpc}$ and a clearly resolved dual quasar at $z\approx3.3$.

\begin{figure}
\centering
\includegraphics[width=0.45\textwidth]{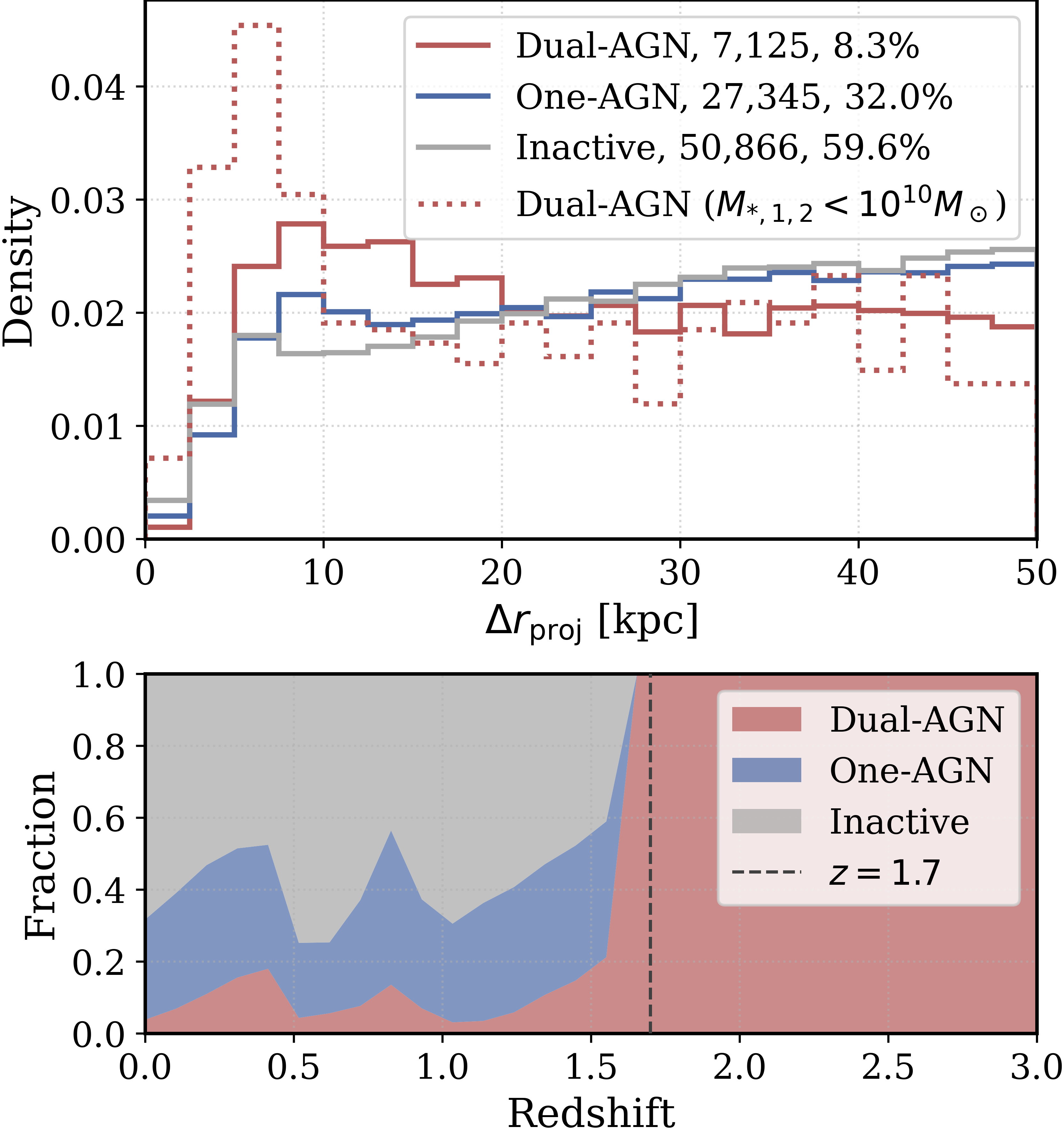}
\caption{
\textbf{Top:} Normalized histogram of projected separations ($\Delta r_{\mathrm{proj}}$) for three populations: dual AGN (red), one-AGN (blue), and inactive (black) galaxy pairs. Percentages indicate the relative fraction of each class in the full sample.
\textbf{Bottom:} Redshift-dependent fractional distribution of the same three populations shown in the top panel, displayed as a stacked area plot. Each bin sums to unity. {The dashed vertical line marks the \texttt{Redrock} classification boundary at $z=1.7$.}
}
\label{fig:dual_one_inactive_fraction}
\end{figure}

Figure \ref{fig:dual_one_inactive_fraction} shows the relative distribution of dual AGN, One-AGN, and Inactive galaxy pairs to isolate the effects of physical separation and redshift on AGN activity. We find 7,125 dual AGNs (8.3\% of the total sample), 27,345  one-AGN pairs (32.0\%), and 50,866 inactive pairs (59.6\%).
The top panel shows the probability density of these three classes as a function of projected separation between pairs. 
Dual AGNs are systematically skewed toward smaller separations relative to the other population. 
The distribution peaks at $\Delta r_{\mathrm{proj}} \sim 5$–$12\mathrm{kpc}$ and declines towards larger separations, indicating that simultaneous SMBH activity occurs preferentially in the later stages of galaxy interactions.
Approximately $41\%$ of dual AGNs lie at $\Delta r_{\mathrm{proj}} \leq 20\mathrm{kpc}$, compared to $\sim32\%$ for one-AGN pairs and $\sim30\%$ for inactive systems. 
In contrast, inactive pairs show a flatter distribution that gradually increases toward larger separations, while one-AGN pairs occupy an intermediate regime. 

The increased concentration of dual AGNs at small separations is consistent with hydrodynamical simulations such as \texttt{ASTRID} \citep{Chen2023} and Horizon-AGN \citep{2022MNRAS.514..640V}, which predict that simultaneous SMBH accretion is most likely during late-stage mergers when tidal torques efficiently drive gas inflows toward both nuclei. Observationally, \citet{2011ApJ...737..101L} first quantified the AGN pair frequency at projected separations of $5$–$100\,\mathrm{kpc}$ in SDSS at $z\sim0.1$, finding a monotonic increase toward smaller separations. Subsequent optical, infrared, and X-ray studies confirmed a significant excess of AGN activity in close pairs \citep{2011MNRAS.418.2043E, 2017ApJ...848..126S, 2023ApJ...951...92B}, establishing enhanced nuclear activity at small separations as a robust signature of merger-driven AGN triggering.

More recently, \citet{2025A&A...699A.330E} extended this picture into the dwarf galaxy regime ($M_\star \leq 10^{10}\,M_\odot$), finding that low-mass pairs show enhanced AGN fractions at small separations and require closer encounters than higher-mass systems. 
DESI enables us to probe this regime over a larger sample, extending to lower stellar masses and higher redshifts, reaching $\sim2.5$--$5$ magnitudes fainter and $\sim1.5$ dex lower in mass than SDSS \citep{2025ApJ...982...10P}.
In the low-mass dual AGN sample (with the same cut of $\log(M_\star/M_\odot)<10$ as in \citealt{2025A&A...699A.330E}), we find a stronger concentration toward small separations than in the full sample, peaking at $\Delta r_{\mathrm{proj}} \sim 6\,\mathrm{kpc}$.

The bottom panel of Figure \ref{fig:dual_one_inactive_fraction} shows the fractional contribution of each class of pairs as a function of redshift. At low redshift ($z \lesssim 0.5$), the sample is dominated by inactive pairs, with one-AGN systems contributing a substantial but smaller fraction and dual AGNs remaining a minority. This regime is primarily populated by DESI BGS targets, where the survey is most complete and uniformly sampled. At intermediate redshifts ($0.5 \lesssim z \lesssim 1.5$), the relative contributions of the three populations become more comparable, reflecting the transition to mixed tracer populations and the increasing AGN duty cycle. 
{At $z\geq1.7$, \texttt{Redrock} imposes an upper redshift limit on the \texttt{GALAXY} spectral class, and only \texttt{QSO} templates are permitted. Consequently, all identified close pairs above this redshift are classified as QSO–QSO systems and enter the dual AGN category.}
The high-redshift dominance of dual AGNs therefore reflects selection effects rather than an intrinsic evolutionary trend.

\begin{figure*}
\centering
\includegraphics[width=0.99\textwidth]{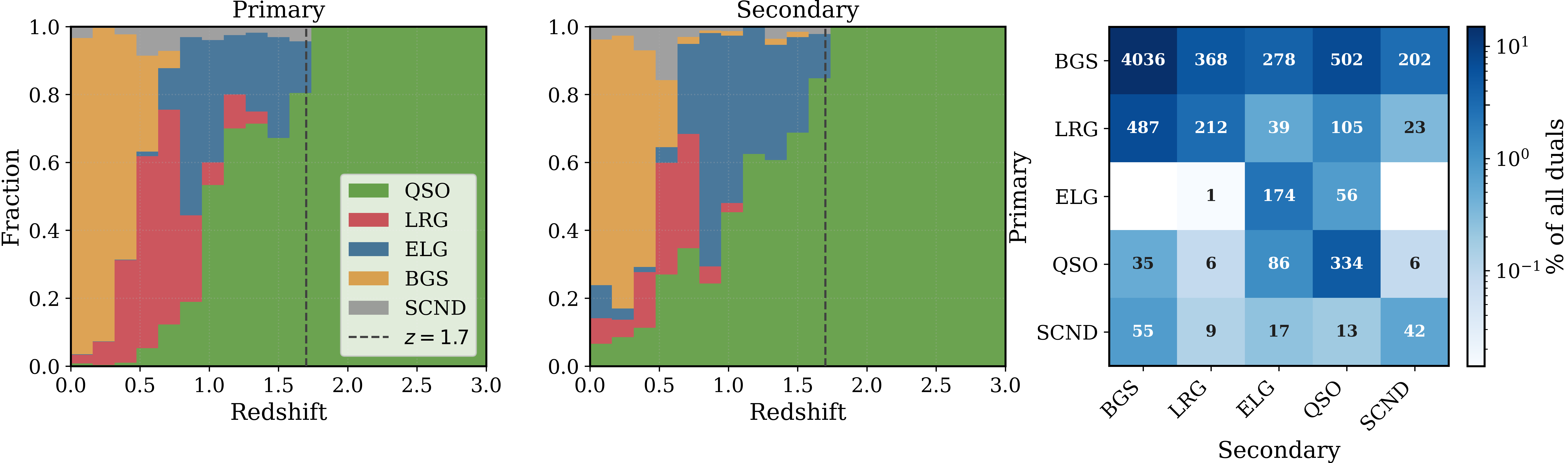}
\caption{
Decomposition of the dual AGN sample by DESI target class. 
\textbf{Left and Middle Panels}: The fractional distribution of target types for the primary (left) and secondary (middle) components as a function of redshift. 
\textbf{Right Panel}: Matrix showing the distribution of pairing combinations across the entire sample. The y-axis represents the target class of the primary object, and the x-axis represents the secondary. 
Numbers give the counts in each combination (total exceeds the number of unique systems), and the color scale indicates the corresponding percentage of all duals.
}
\label{fig:tracer_z}
\end{figure*}
\label{sec:dr_z}

\subsection{Duals across tracers}

Figure~\ref{fig:tracer_z} illustrates how the dual AGN sample spans a wide range of galaxy populations. 
Individual targets can be associated with multiple DESI tracer classes. In this figure, we include all tracer labels attached to each galaxy, so the summed counts exceed the total number of dual AGN systems. For the analysis that follows, we assign each galaxy a single representative tracer (\texttt{BEST\_TRACER}). 
When available, we adopt the tracer class assigned by the DESI LSS catalog, which reflects the survey's targeting and fiber-assignment priorities. If that is not available, we follow the DESI LSS priority ordering (QSO $>$ LRG $>$ ELG $>$ BGS $>$ SCND) \citep{2025JCAP...04..074B}.

The left and middle panels of Figure~\ref{fig:tracer_z} show the fractional contribution of each DESI target class to the primary (more massive) and secondary components of dual AGN systems as a function of redshift, while the right panel summarizes the distribution of primary–secondary tracer combinations across the full sample. {The dashed line marks the $z=1.7$ \texttt{Redrock} boundary discussed above.}
Globally, the dual AGN population is dominated by BGS targets ($70.2\%$ of all components combined).  QSOs account for $6.9\%$ of primaries but $14.4\%$ of secondaries, while LRGs contribute $12.2\%$ of primaries and $8.4\%$ of secondaries. ELGs represent a smaller but non-negligible fraction ($3.4\%$ of primaries and $8.6\%$ of secondaries).

At low redshift ($z \lesssim 0.5$), the dual AGN population is dominated by BGS (Yellow) targets, which account for more than $80\%$ of both primary and secondary components.
In this regime, galaxies are selected primarily by optical brightness rather than nuclear activity, and AGN identification is performed spectroscopically. As a result, the low redshift dual AGN population closely reflects the general population of galaxy mergers, providing a relatively unbiased view of merger-driven nuclear activity in the local Universe.
At intermediate redshifts ($0.5 \lesssim z \lesssim 1.0$), LRG (Red) and ELG (Blue) targets become increasingly prominent. LRGs dominate primaries near $z\sim0.6$–$0.8$, while ELGs become prominent around $z\sim0.8$–$1.0$, particularly among secondaries where they can constitute nearly $70\%$ in some bins.
LRGs trace massive galaxies dominated by older stellar populations, while ELGs represent gas-rich, actively star-forming systems. The presence of pairs that involve ELG in this redshift range is particularly noteworthy, as these systems probe gas-rich interactions and often involve lower-mass companions. 
At $z\gtrsim1.2$, the sample transitions rapidly to QSO-dominated systems, and above $z = 1.7$ it is exclusively QSO–QSO by construction (\ref{pair_sep}).

The rightmost panel of Figure~\ref{fig:tracer_z} quantifies the distribution of tracer–tracer pairings across the entire dual AGN sample. The most common configuration is BGS–BGS, comprising $56.6\%$ of all dual systems. In total, $67.4\%$ of dual AGNs involve same-tracer pairs (diagonal elements of the matrix). This is partly driven by the survey selection itself: DESI tracers probe distinct redshift ranges and galaxy populations, so pairs drawn from the same tracer are more likely to be observed together within a given redshift slice. 
Mixed tracer combinations remain significant, however. BGS–QSO systems account for $7.0\%$ of all duals, while LRG–BGS and BGS–LRG pairs contribute $\sim6\%$ each. QSO–QSO pairs represent $4.7\%$ of the total sample and dominate the high-redshift regime. ELG–ELG systems comprise $2.5\%$, and additional ELG-involving pairs contribute further diversity.

Although ELG–ELG and ELG-involving pairs constitute a relatively small fraction of the total sample, these systems are of particular interest. ELGs trace lower-mass, gas-rich galaxies that are expected to participate in minor mergers, which theoretical models predict to be a dominant channel for supermassive black hole mergers. Such systems are expected to experience prolonged inspiral phases and may contribute significantly to the population of merging black holes detectable by future space-based gravitational wave observatories such as LISA. 

Although SCND appears in more than 200 pairings in the full matrix, most of these objects are also associated with primary DESI tracers. After applying the \texttt{BEST\_TRACER} assignment, SCND contributes only a small fraction of the final sample (8 primaries and 40 secondaries).

\subsection{Host Galaxy Properties} \label{host_agn_properties}

In this section, we compare the host galaxy properties of three populations: 
(i) dual AGN candidates, divided into primary and secondary according to stellar mass, with $q_M\equiv M_{\star,2}/M_{\star,1}\leq 1$ by construction; (ii) one-AGN galaxy pairs, divided into the AGN host and its inactive companion; and (iii) the full isolated AGN hosts comparison sample from the DESI AGN/QSO VAC. 
Stellar masses and star-formation rates are taken from the \texttt{CIGALE} SED fits described in Section~\ref{data}. 
For the quantitative host galaxy analysis, we additionally require reliable \texttt{CIGALE} stellar-mass and SFR estimates, 
\footnote{Following the recommended \texttt{CIGALE} VAC quality cuts, we require $0.2<\texttt{FLAG\_MASSPDF}<5$ and $0.2<\texttt{FLAG\_SFRPDF}<5$,  corresponding to best-fit to Bayesian ratios for $M_\star$ and SFR.}
retaining approximately $65\%$ of the total dual AGN candidate sample.

{Two caveats are important for the SFR measurements used in this work. First, the SFRs are derived from broadband SED fitting.
For BGS and ELG hosts, where the optical--near-IR SED is typically dominated by stellar emission, relative SFR comparisons are the most robust. 
For LRGs, absolute SFRs are noisier because most systems lie in the nearly quiescent regime, where age--dust degeneracies are strong. 
For QSOs the uncertainties are even larger. In luminous type-1 systems, the accretion-disk continuum can dominate the optical--near-IR SED, making the AGN--host decomposition degenerate and producing host SFR uncertainties of order $\sim40$--$50\%$ even with full UV-to-sub-mm photometry \citep{2015A&A...576A..10C}. As a result, we treat QSO SFR trends as qualitative throughout. }
{Second, integrated SED-based SFRs trace the global galaxy light rather than the nuclear regions where tidally driven gas inflows are expected to be strongest. Spatially resolved IFU observations of CALIFA interacting galaxies show that the sSFR enhancement in pairs is concentrated in the central regions: within $\sim0.3\,R_{\rm eff}$, paired galaxies show a factor of $\sim2.5$ enhancement relative to matched controls, while the sSFR averaged over $\sim2.5\,R_{\rm eff}$ is consistent with, or slightly below, the control value \citep{2015A&A...579A..45B}. Any enhancement we recover from our integrated \texttt{CIGALE} SFRs is therefore diluted by the less affected disk light.}

\subsubsection{Star Formation Activity and Main-Sequence Offsets} \label{ssfr_mstar}
\begin{figure*}
\centering
\includegraphics[width=0.99\textwidth]{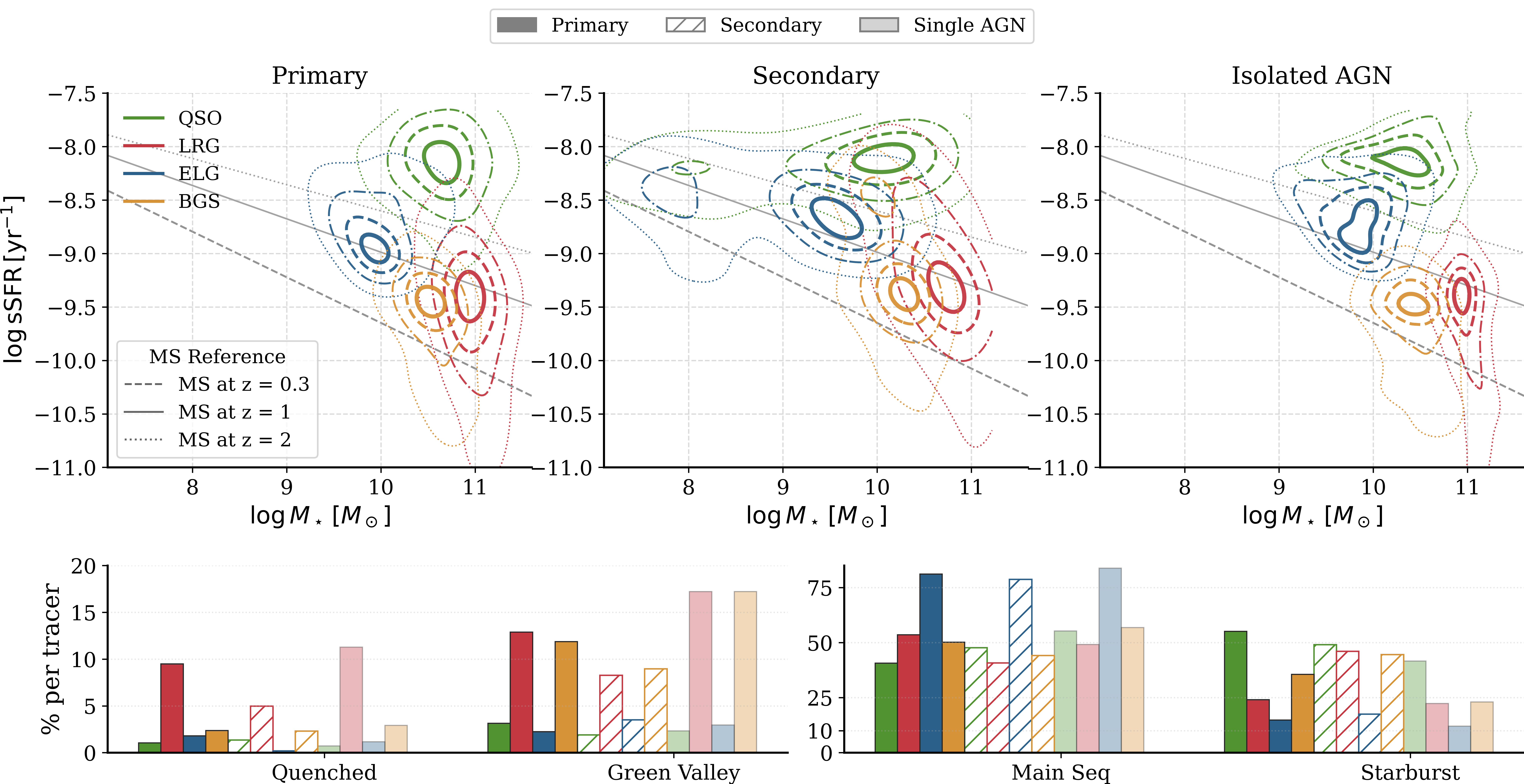}
\caption{
Host galaxy properties of the dual AGN sample compared to the isolated AGN population. 
\textbf{Top Row}: The distribution of specific star formation rate versus stellar mass for the primary (left) and secondary (center) hosts in dual systems, and the isolated AGN comparison sample (right). Colored contours represent each DESI tracer class. KDE contours mark iso-density levels at $20\%, 45\%, 70\%,$ and $90\%$ of the peak density. 
\textbf{Bottom Row}: The fractional distribution of galaxies across four evolutionary stages. Solid colored bars correspond to primary hosts, hatched bars to secondary hosts, and faint bars to the isolated AGN sample.
Dashed, solid, and dotted lines show the redshift-dependent star-forming main sequence, $\log \mathrm{sSFR}_{\rm MS}(M\star, z)
= \log \mathrm{SFR}_{\rm MS}(M\star,z) - \log M_\star$
at $z=0.3, 1,$ and $2$, respectively, computed following \citet{2014ApJS..214...15S}.
}
\label{fig:sfr_Mstar}
\end{figure*}

Figure~\ref{fig:sfr_Mstar} compares the distributions of dual AGN hosts and isolated AGN control samples in the specific star formation rate versus stellar mass plane (sSFR--$M_\star$). The top panels show the two-dimensional distributions for primary galaxies (left), secondary galaxies (center), and isolated AGN hosts (right). Kernel density contours mark iso-density levels at $20\%, 45\%, 70\%,$ and $90\%$ of the peak density, with colors indicating the DESI tracer classes: ELGs (blue), LRGs (red), QSOs (green), and BGSs (yellow). Each tracer occupies a distinct region of the plane due to target-selection differences: ELGs and QSOs lie near the star-forming main sequence, LRGs occupy the massive and quiescent regime, and BGSs span intermediate masses and sSFRs. These selection effects are important to keep in mind when interpreting tracer‐dependent trends.

Dual primaries are consistently massive, but the degree of mass asymmetry varies across tracers. QSO and ELG pairs have the largest mass gaps, with secondaries extending to $\log (M_\star/M_\odot)<9$, though these pairs are relatively sparse. LRG and BGS pairs, which dominate the sample numerically, show more modest mass differences. 
Isolated AGN hosts typically lie between primaries and secondaries in mass.

{Relative to the isolated AGN sample, dual systems are systematically shifted toward higher star-forming activity at fixed stellar mass, with the strongest effect in the lower-mass secondary hosts. Among BGS galaxies, dual secondaries have median $\log({\rm sSFR}/{\rm yr}^{-1})$ values $\sim0.1$--$0.3$ dex above the tracer-matched isolated AGN control, while dual primaries show a weaker excess at the $\lesssim0.1$ dex level. ELG and QSO systems lie closer to their corresponding isolated AGN baselines, possibly because these control populations are already strongly star-forming or AGN-dominated. LRGs show a more mixed behavior: many are gas-poor or environmentally quenched, so even if tidal torques are present, the available cold-gas reservoir may be insufficient to produce a strong SFR response \citep{2019MNRAS.490.2139R,2020MNRAS.493.3716H}.}

To quantify the star-formation state of each host, we compute its offset from the star-forming main sequence,
\begin{equation}
\Delta_{\rm MS} \equiv
\log{\rm SFR}_{\rm galaxy} - \log{\rm SFR}_{\rm MS}(M_\star,z),
\end{equation}
using the redshift-dependent parameterization of \citet{2014ApJS..214...15S},
$
\log{\rm SFR}_{\rm MS}(M_\star,z) =
\left(0.84-0.026\,t\right)
\log\!\left(\frac{M_\star}{M_\odot}\right)
-
\left(6.51-0.11\,t\right),
$
where $t\equiv t(z)$ is the age of the Universe in Gyr. For visualization we divide the population into four regimes: quenched ($\Delta_{\rm MS}\leq -1.0$), green valley ($-1.0<\Delta_{\rm MS}\leq -0.3$), main sequence ($-0.3<\Delta_{\rm MS}\leq 0.6$), and starburst ($\Delta_{\rm MS}>0.6$). These boundaries are intended as descriptive partitions of the distribution rather than strict physical phase divisions.

{The bottom panel of Figure~\ref{fig:sfr_Mstar} shows the resulting fractions: each bar gives the percentage of galaxies within a given tracer and AGN class falling into each $\Delta_{\rm MS}$ regime.}
Most hosts lie on or above the main sequence, with only a minority in the green-valley or quenched tails. ELG hosts  are concentrated near the main sequence ($\sim$80\%) with little difference between duals and singles. QSOs show the highest starburst fractions overall ($\sim$50\% of dual QSOs).
In the BGS sample, the starburst fraction increases from $\sim$23\% in the isolated AGN control to $\sim$36\% in dual primaries and $\sim$45\% in dual secondaries. A similar but smaller trend is present for LRG hosts, where the starburst fraction rises from $\sim$22\% in singles to $\sim$45\% in dual secondaries. 

Smaller fractions of systems populate the green valley and quenched regimes.
Among those, LRGs - consistent with their nature as massive, evolved galaxies with little ongoing star formation -  show the largest fractions of low star formation systems, with quenched fractions decreasing from $\sim11\%$ in the isolated AGN control to $\sim10\%$ in dual primaries and $\sim5\%$ in dual secondaries, alongside a similar decline in the green-valley population ($17\%$, $13\%$, and $8.3\%$, respectively).
BGS systems show a broader distribution of star-formation states: while quenched fractions are low ($\sim2$--$3\%$), the green-valley contribution is still substantial in the control sample ($17.2\%$) and declines in dual systems ($11.9\%$ for primaries and $9.0\%$ for secondaries).
{A notable trend is that the green-valley fraction declines from isolated AGNs to dual primaries to dual secondaries across all tracers, consistent with a merger sequence in which dual AGN activity coincides with the starburst peak, while the green-valley transition generally occurs later, on longer timescales, after AGN feedback has expelled the nuclear gas and beyond the dual phase \citep{DiMatteo2005_BH_model, 2008ApJS..175..356H}.}
ELG and QSO hosts are almost entirely star-forming, with only minimal contributions from the quenched or green-valley regimes.

This behavior is consistent with a picture in which tidal torques drive gas toward the central regions during close interactions, enhancing star formation most efficiently in the lower-mass companion. Low-redshift pair studies find enhanced star formation at small projected separations, with the response often strongest in gas-rich or lower-mass companions \citep{2008AJ....135.1877E,2010ApJ...713..330X,2013MNRAS.430.3128E}. 
{As discussed above, spatially resolved studies show that pair-induced star formation is centrally concentrated; therefore, the $\sim0.1$--$0.3$ dex offsets measured from integrated \texttt{CIGALE} SFRs likely underestimate the nuclear response.}
{Cosmological simulations show qualitatively similar behavior: interacting galaxies in IllustrisTNG and SIMBA are enhanced in sSFR by factors of $\sim2$--$3$ relative to matched controls, with the amplitude depending on gas fraction, stellar mass, and merger stage \citep{2019MNRAS.490.2139R,2020MNRAS.494.4969P,2020MNRAS.493.3716H}. }

\subsubsection{Evolution with Redshift and Projected Separation}
\label{redshift_evolution}

\begin{figure*}
\centering
\includegraphics[width=0.99\textwidth]{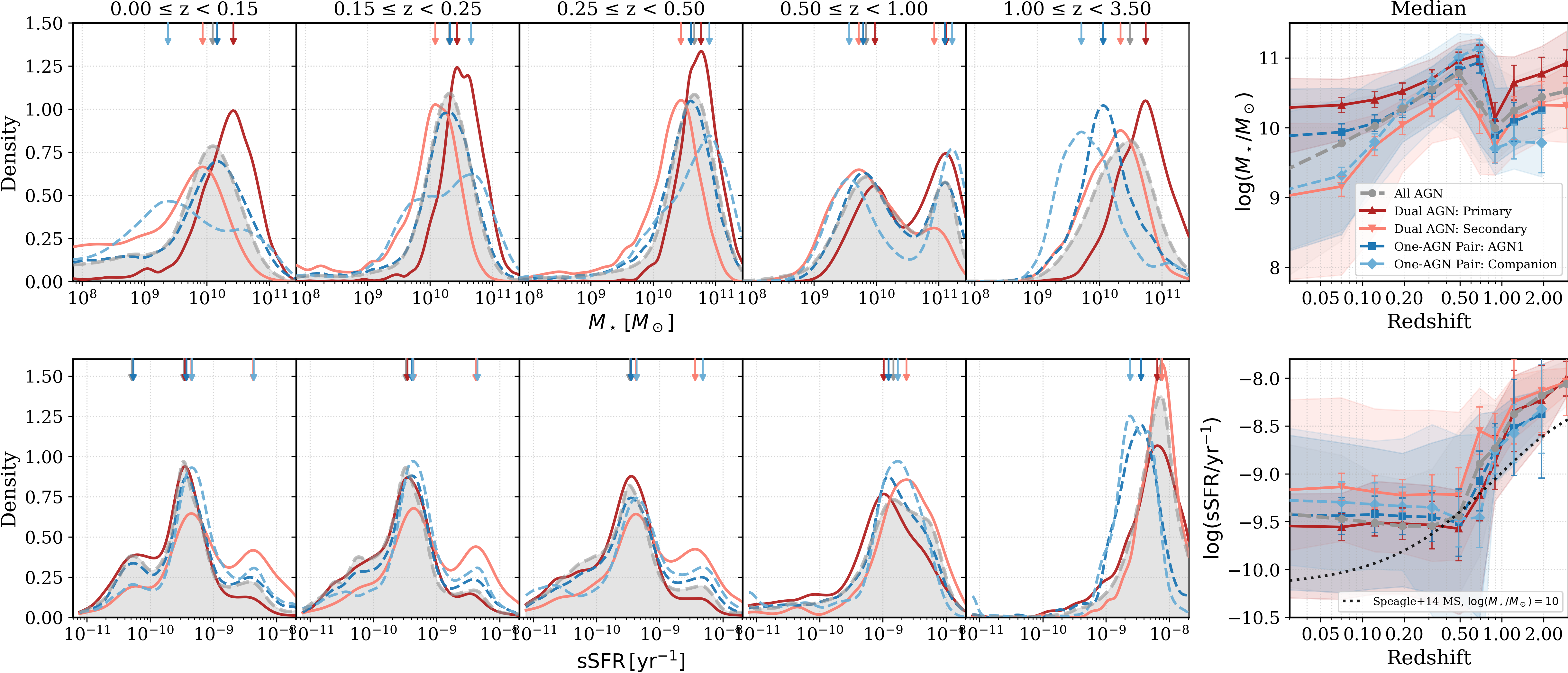}
\caption{
Redshift evolution of host galaxy properties (stellar mass in the top row and specific star-formation rate in the bottom row) for dual AGNs, one-AGN pairs, and the isolated AGN control population. \textbf{Left}: distributions in five redshift bins. \textbf{Right}: Median evolution with redshift. Shaded regions mark the $16^{\rm th}$--$84^{\rm th}$ percentile range.
}
\label{fig:hist}
\end{figure*}

\begin{figure}
\centering
\includegraphics[width=0.39\textwidth]{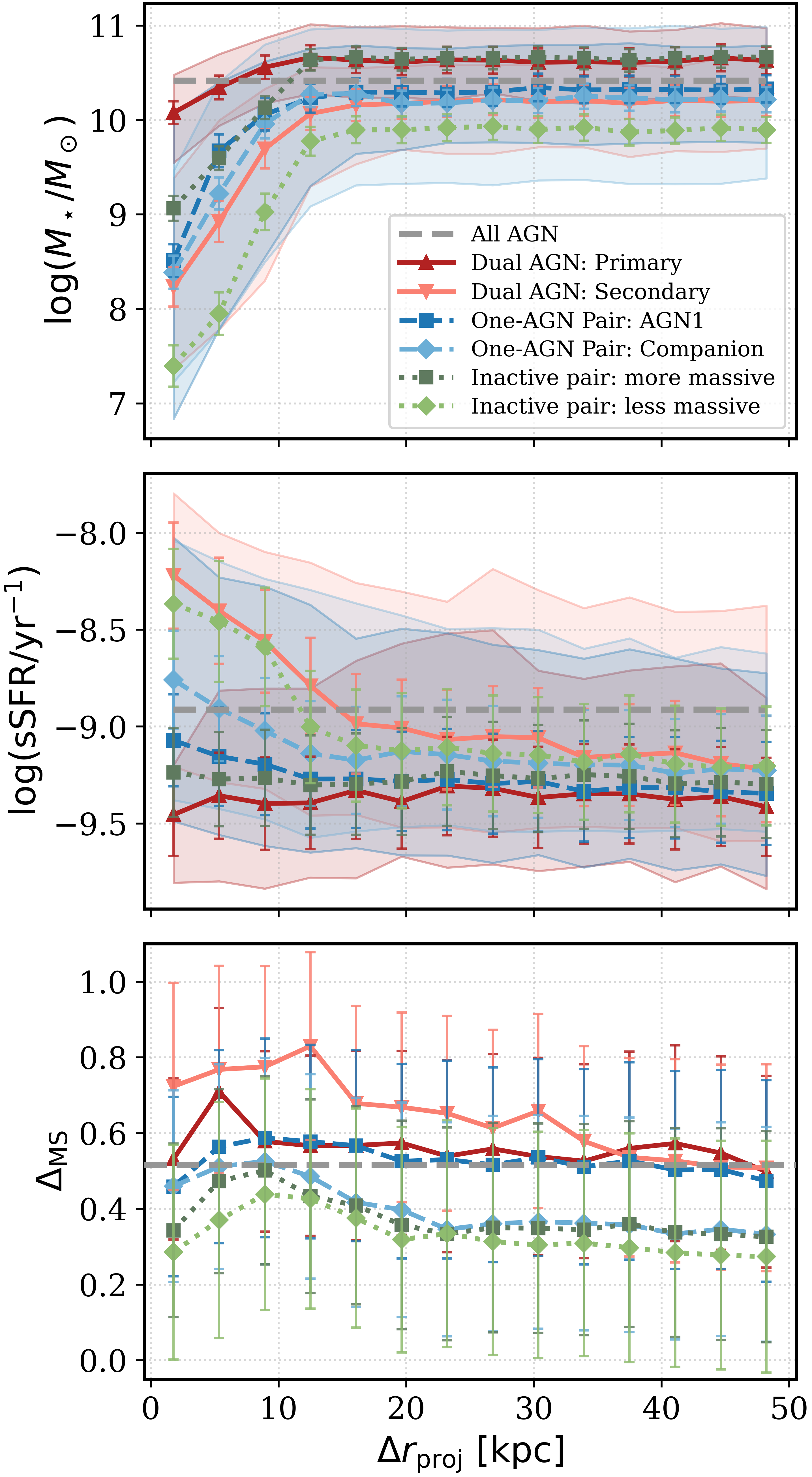}
\caption{Median host stellar mass (top), specific star-formation rate (middle), and offset from the star-forming main sequence, $\Delta_{\rm MS}$ (bottom), as a function of projected separation. Dual AGN primaries and secondaries are shown in red and orange, respectively; one-AGN pairs are separated into the AGN host and inactive companion, shown in dark blue and light blue, respectively; inactive pairs are separated into the more massive and less massive members, shown in dark green and light green; and isolated AGNs are shown as a gray dashed line. Shaded regions indicate the $16^{\rm th}$--$84^{\rm th}$ percentile range in each separation bin, and error bars show the mean \texttt{CIGALE} uncertainty for the objects contributing to that bin.}
\label{fig:parameters_dr}
\end{figure}

{Figure~\ref{fig:hist} shows the observed redshift dependence of $M_\star$ and sSFR for dual AGNs, one-AGN pairs, and the isolated AGN comparison sample. These trends should not be interpreted as evolutionary tracks for a fixed galaxy population. 
The DESI tracer mix changes strongly with redshift (Figure~\ref{fig:tracer_z}), and different tracer classes occupy different regions of the $M_\star$--sSFR plane (Figure~\ref{fig:sfr_Mstar}), so the medians combine any physical evolution with the changing tracer composition.  We therefore use this figure mainly to compare the populations within the same redshift bins, where they share a similar redshift dependent selection. }
Offsets ($\Delta$) denote differences in median logarithmic quantities, in dex, relative to the isolated AGN population, and we characterize the distribution width by the half-width of the central $68\%$ interval, $\sigma_{16\text{--}84}\equiv(P_{84}-P_{16})/2$.

The median stellar mass of dual primaries follows a broad rise--dip--rise pattern with redshift, which largely reflects the changing tracer mix. It increases from $\log(M_\star/M_\odot)\simeq10.18$ at $z<0.05$ to $\simeq10.95$ by $z\sim0.5$, where massive BGS and LRG hosts dominate, declines to $\simeq10.1$ at $0.8\lesssim z\lesssim1.0$, where ELG and QSO targets become more important, and rises again to $\log(M_\star/M_\odot)\simeq10.9$ by $z\gtrsim2$. The other populations show similar behavior, but with systematic offsets. Dual primaries are the most massive hosts at nearly all redshifts, exceeding the isolated AGN median by $\sim0.2$--$0.7$ dex, while dual secondaries lie $\sim0.2$--$0.6$ dex below it. One-AGN hosts occupy an intermediate regime, and their inactive companions broadly follow the lower-mass secondary population.

{The sSFR trend with redshift reflects both the evolving normalization of the star-forming main sequence and the changing tracer composition. For reference, the bottom-right panel of Figure~\ref{fig:hist} shows the \citet{2014ApJS..214...15S} main-sequence normalization at fixed $\log(M_\star/M_\odot)=10$.}
The median dual primary sSFR is near $\log({\rm sSFR}/{\rm yr}^{-1})\simeq-9.4$ to $z\sim0.5$, then rises to $\sim-8.9$ at $z\sim0.9$, $\sim-8.2$ at $z\sim1.2$, and $\sim-8.0$ by $z\sim3$, closely tracking the isolated AGN population. Dual secondaries are enhanced relative to this baseline at $z\lesssim1$ by $\sim0.3$--$0.8$ dex, while the primary--secondary offset is smaller in the higher-redshift bins.
{Part of this behavior may reflect the higher gas fractions and sSFRs of high redshift galaxies, for which the relative enhancement from an interaction is expected to be smaller.} 
In COSMOS/CANDELS pair samples, for example, the SFR excess in major pairs decreases from $\sim1.5\times$ at $0.5<z<1.6$ to $\sim1.1\times$ at $z>1.6$ \citep{2022ApJ...940....4S}.
{This same redshift range is also dominated by ELG and QSO targets in DESI, so tracer composition contributes to the observed offset as well.}

{Figure~\ref{fig:parameters_dr} shows the same asymmetry at different projected separations. {As with the redshift trends, each bin contains a different mix of redshift and tracer, so the curves compare the populations observed at a given separation and are not a sequence followed by individual merging systems.}
At all projected separations, dual AGN primaries are relatively massive, with $\log(M_\star/M_\odot)\simeq10.1$--$10.7$, and have nearly flat median $\log({\rm sSFR}/{\rm yr}^{-1})\simeq-9.4$. The strongest primary--secondary contrast is seen among the closest observed pairs. At $\Delta r_{\rm proj}\lesssim5$~kpc, secondaries have low median mass, $\log(M_\star/M_\odot)\simeq8.2$, and elevated median sSFR, $\log({\rm sSFR}/{\rm yr}^{-1})\simeq-8.2$, whereas at $\Delta r_{\rm proj}\sim50$~kpc the observed secondaries are more massive, $\log(M_\star/M_\odot)\simeq10.2$, and less strongly star-forming, $\log({\rm sSFR}/{\rm yr}^{-1})\simeq-9.2$.
{The factor of $\sim100$ span in the median secondary mass between the closest and widest bins should be interpreted with caution. Simulations find that stars are relatively resilient to stripping and substantial stellar mass loss occurs near final coalescence, whereas the pairs in our sample are still separated by several kpc \citep{2022MNRAS.514..640V}.}
{Here, relevant comparison samples are not explicitly matched in stellar mass, redshift, color, or tracer type. These effects can influence the observed mass-ratio and sSFR trends, and some of this contrast may be driven by selection rather than by the pair interaction alone.} 

{In particular, the smallest physical separations are probed only at low redshift because of the $1.6''$ minimum angular separation cut. At those redshifts, the magnitude limit reaches the lowest stellar masses (see Appendix~\ref{app:tracer_mass_limits}). The closest bins are therefore filled with low-mass secondaries by construction. At higher redshift, low-mass companions fall below the magnitude limit and the $1.6''$ cut excludes the closest pairs, so the surviving systems are wide pairs with more massive secondaries.
Inactive pairs (shown in Figure \ref{fig:parameters_dr} with green dotted lines, with the more massive member shown in darker green and the less massive member in lighter green) trace nearly the same mass and sSFR behavior as the dual secondaries, indicating that the trend is not specific to dual AGNs. Additionally, restricting the sample to $z<0.05$, where the mass limit varies the least, flattens the mass trend; however, it is still present to a smaller extent.}

{The limit is also tracer dependent: BGS and LRG reach $\log(M_\star/M_\odot)\lesssim9$ only at $z\lesssim0.1$, whereas ELG and QSO reach such masses to $z\sim1$ \citep[see also Appendix~\ref{app:tracer_mass_limits}]{2025ApJ...982...10P}. 
We examined the separation dependent trends split by DESI tracer. This comparison shows that the qualitative mass gap and sSFR trends are also present within the BGS subset. Thus, the trend does not appear to be solely an artifact of ELG/QSO low mass selection, but its amplitude and interpretation are certainly affected by the tracer-dependent mass limits. Overall, we conclude that the trend is present in the BGS/ELG samples at $z<0.4$.}

{We also caution for measurement effects. At $\Delta r_{\rm proj}\lesssim3.6$ kpc, the \texttt{CIGALE} uncertainties are relatively large ($\sim0.10$ dex in $\log M_\star$ and $\sim0.19$ dex in $\log{\rm sSFR}$) and this bin also contains fewer systems than the wider separation bins. In addition, at small separations, deblending effects can make the secondary appear artificially faint or low-mass. This would lower the inferred mass ratio and raise the inferred sSFR.}

{A more robust quantity is the offset from the star-forming main sequence, $\Delta_{\rm MS}$, shown in the bottom panel of Figure~\ref{fig:parameters_dr}. Dual secondaries remain elevated, with $\Delta_{\rm MS}\simeq+0.7$--$0.8$ dex among the closest pairs and $\simeq+0.5$ dex in the widest bins. They also lie above both the less massive inactive members and the one-AGN companions at the same projected separation by $\sim0.3$ dex. This residual is modest and comparable to interaction-driven enhancements found in matched close pair studies, where the SFR excess reaches a factor of $\sim2$--$2.5$ at the smallest separations \citep{2022ApJ...940....4S}.
Note that \texttt{CIGALE} and \texttt{FastSpecFit} stellar masses agree to $\lesssim0.2$ dex, and although their absolute sSFR scales differ by $\sim1.5$ dex, the dual-secondary $\Delta_{\rm MS}$ excess over the inactive control is reproduced in both ($+0.31$ and $+0.46$ dex).
We therefore find evidence of a real but modest enhancement of star formation in the lower mass member of dual systems. 
A modest asymmetric tidal response is a natural explanation. In unequal mass pairs, the lower mass galaxy is more easily perturbed, with gravitational torques driving gas inward and raising the central star-formation rate, while the more massive primary, embedded in a deeper potential well, responds more weakly \citep{2008AJ....135.1877E,2015A&A...579A..45B,2020MNRAS.494.4969P,2020MNRAS.493.3716H}.}

One-AGN pairs occupy an intermediate regime. Their active hosts and inactive companions follow the same selection-driven mass behavior as the dual secondaries. In $\Delta_{\rm MS}$, the one-AGN companions lie $\sim0.25$ dex below the dual secondaries. Among the closest pairs, the companions are mildly more elevated than their AGN hosts, typically by $\lesssim0.2$ dex. This likely shows that tidal torques and gas redistribution are already operating in one-AGN pairs, while simultaneous dual activity may require an additional condition, such as a sufficiently large and centrally concentrated gas supply to fuel both black holes. Simulations in which SMBH accretion during mergers is stochastic and often asynchronous support this picture, since enhanced star formation need not imply simultaneous triggering of both nuclei \citep{2019MNRAS.483.2712R,2022MNRAS.514..640V}.

\subsubsection{Merger mass ratios}\label{merger_rates}
\begin{figure}
\centering
\includegraphics[width=0.48\textwidth]{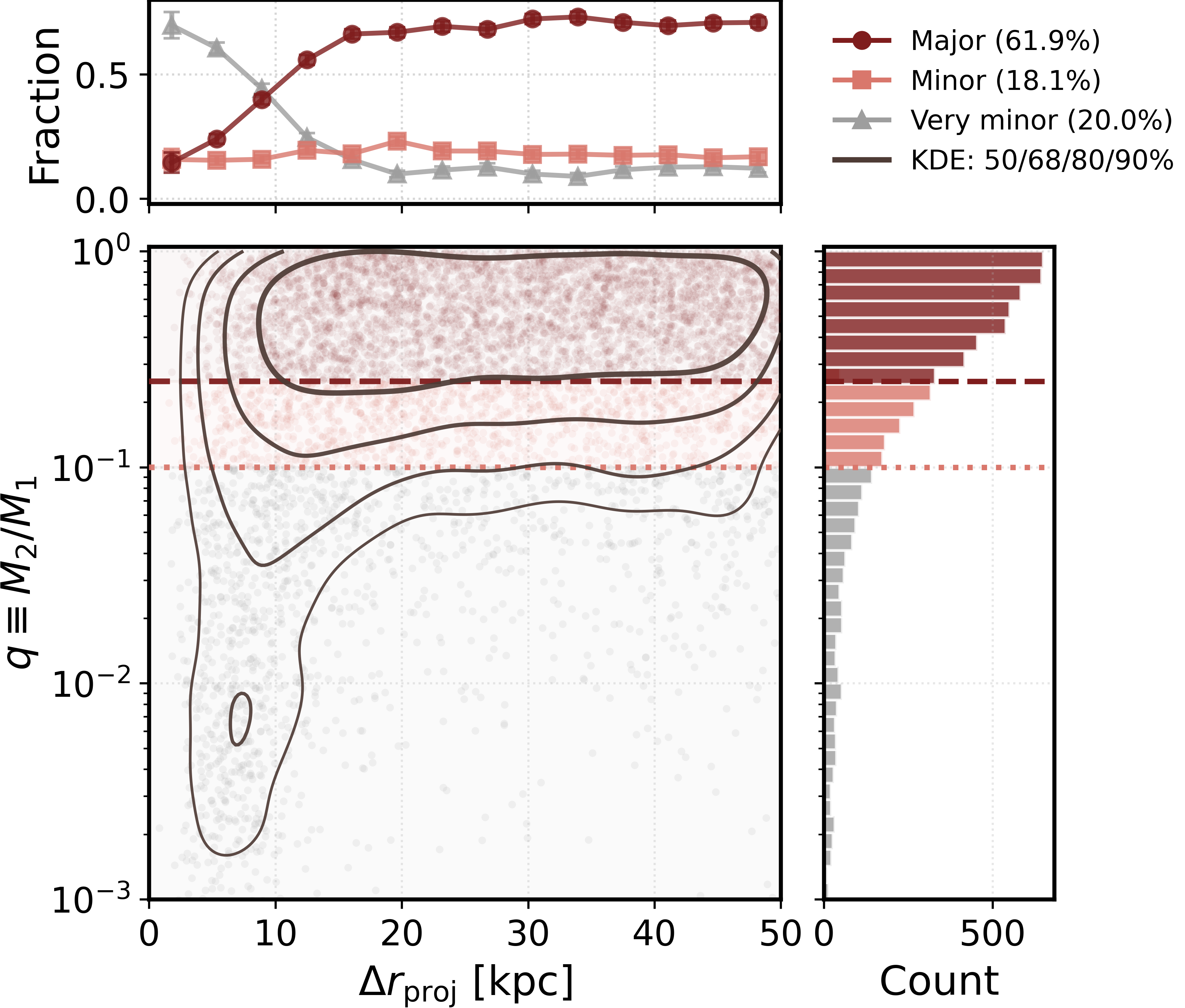}
\caption{
Merger mass ratios for the dual AGN sample. The main panel shows $q_M=M_{\star,2}/M_{\star,1}$ versus projected separation, $\Delta r_{\rm proj}$, with points colored by merger class: major ($q_M\geq0.25$), minor ($0.10\leq q_M<0.25$), and very minor ($q_M<0.10$). Horizontal lines mark the class boundaries, and black contours show the 50, 68, 80, and 90 percent KDE levels. The top panel shows the merger-class fractions as a function of $\Delta r_{\rm proj}$, and the right panel shows the corresponding $q_M$ distribution. Percentages in the legend give the total sample fraction in each class.
}
\label{fig:major_minor}
\end{figure}

Figure~\ref{fig:major_minor} characterizes the observed merger mass-ratio distribution. 
We define
\begin{equation}
q_M \equiv \frac{M_{\star,2}}{M_{\star,1}},
\qquad 0<q_M\leq1,
\end{equation}
and classify systems as major mergers ($q_M\geq0.25$), minor mergers ($0.10\leq q_M<0.25$), and very minor mergers ($q_M<0.10$), following standard conventions \citep[e.g.,][]{2011ApJ...742..103L,2015MNRAS.449...49R}. The dual AGN distribution is broad: $61.9\%$ of systems are major mergers, $18.1\%$ are minor mergers, and $20.0\%$ are very minor mergers. Although major mergers form the largest single class, a substantial fraction of DESI dual AGNs occur in unequal-mass configurations, consistent with expectations from cosmological simulations \citep[e.g. Horizon-AGN;][]{2022MNRAS.514..640V}.

The merger class mix differs strongly between projected-separation bins. The closest bin ($0$--$3.6$~kpc) is dominated by very minor systems ($f_{\rm vminor}\simeq70\%$), while the major-merger fraction rises above $\sim50\%$ beyond $\sim10$--$15$~kpc and reaches $\sim0.68$--$0.72$ in the widest bins.
The KDE contours are concentrated at high $q_M$ over most separations and extend to low $q_M$ mainly among the closest pairs. This pattern follows directly from the selection effects described above. 

\subsection{AGN Luminosity}
\label{L_bol}

\begin{figure}
\centering
\includegraphics[width=0.40\textwidth]{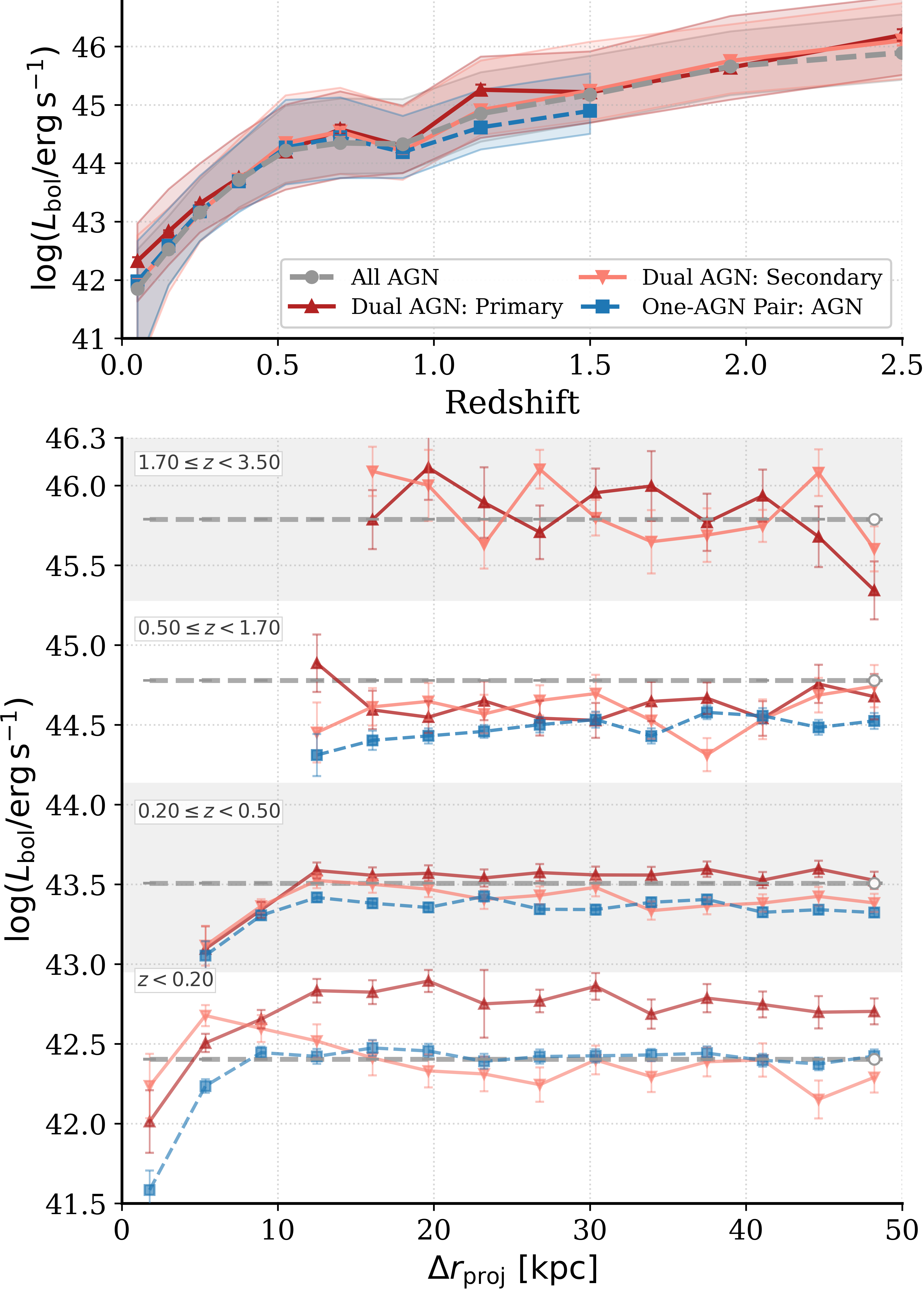}
\caption{
Median AGN bolometric luminosity as a function of redshift (\textbf{top}) and projected separation (\textbf{bottom}). In the bottom panel the sample is split into four redshift bins, where the grey dashed line marks the median of the full AGN sample in that same bin. Dual AGNs are shown as solid red curves, split into primaries and secondaries; the AGN in one-AGN pairs is shown as a dashed blue curve. Shaded regions on the top panel mark the $16^{\rm th}$--$84^{\rm th}$ percentile range.
}
\label{fig:Lbol}
\end{figure}

Figure~\ref{fig:Lbol} shows the median AGN bolometric luminosity as a function of redshift (top) and projected separation (bottom). The overall AGN population in DESI shows a strong variation of $L_{\rm bol}$ with redshift, with the median rising by nearly five orders of magnitude, from $L_{\rm bol}\sim10^{41.2}\,{\rm erg\,s^{-1}}$ at $z\sim0.03$ to $\sim10^{46}\,{\rm erg\,s^{-1}}$ by $z\sim3$. This rise also sets the reference level in the bottom panel: the all AGN median increases across the four redshift bins, from $10^{42.4}\,{\rm erg\,s^{-1}}$ at $z<0.2$ to $10^{43.5}$, $10^{44.8}$, and $10^{45.8}\,{\rm erg\,s^{-1}}$, and the dual and one-AGN-pair populations shift upward with it. This trend is driven partly by the flux limit of the survey, which allows only the most luminous AGN to be detected at high redshift and biases the median upward, and partly by the cosmic evolution of SMBH accretion, in which the typical AGN luminosity and space density peak at $z\sim1$--$3$. We therefore treat the redshift dependence as a varying baseline rather than as intrinsic evolution.

Against this varying baseline, the luminosity offset of the dual components is largest at low redshift and shrinks with $z$. Within each redshift bin the medians are flat with separation beyond $r_p\sim10$ kpc. At $z<0.2$, dual primaries lie $\sim0.3$--$0.4$ dex above the isolated AGN median and $\sim0.3$ dex above the AGN in one-AGN pairs, while dual secondaries are consistent with the median.
The offset falls to $\approx0.05$ dex by $0.2\le z<0.5$, and by $z\ge0.5$ all three populations lie at or slightly below the median. The excess of dual primaries over the AGN in one-AGN pairs likewise weakens with redshift, from $\sim0.3$ dex at $z<0.2$ to $\lesssim0.1$ dex by $z\sim1$. This suggests that, when cold gas is scarce, activating both nuclei at once requires a stronger gas inflow than activating one, so requiring a dual preferentially selects the more luminous interacting systems, while at high redshift, where abundant cold gas can keep both nuclei active without such a trigger, the preference fades.

Overall, the offsets are modest and there is no large excess at high redshift. The bolometric luminosity distributions of dual and single AGN are similar, which indicates that any merger-driven enhancement of the accretion rate is small. At high redshift, gas-rich galaxies can sustain accretion through internal instabilities even without a companion, so a merger raises the probability of gas inflow but adds little to the luminosity of systems already fueled by abundant cold gas \citep[e.g.,][]{2011ApJ...741L..33B,2012ApJ...757...81B,2025ApJ...986..101L}. More efficient AGN feedback at high redshift may further cap any merger-induced enhancement.

At the smallest separations, the medians decrease by a few tenths of a dex toward the smallest $r_p$. At $z<0.2$ the primary--secondary luminosity ordering also reverses, with secondaries slightly more luminous than primaries at $r_p\lesssim10$ kpc ($10^{42.7}$ versus $10^{42.5}\,{\rm erg\,s^{-1}}$ at $r_p\approx5$ kpc) and fainter at larger separations. Such a reversal would be consistent with stochastic, asymmetric accretion during close passages, in line with hard X-ray results in which dual incidence and nuclear activity increase toward small separations \citep{2012ApJ...746L..22K}, and with simulations in which the initially smaller black hole frequently becomes the brighter AGN during the merger \citep{Chen2023}. Because these smallest separations are also where photometric deblending of the close pair is most uncertain, we regard the reversal as suggestive.

\subsection{Comparison with simulation predictions}\label{simulation}

To connect the observed DESI dual AGN sample to the evolution of massive black hole binaries, we use the study of \citet{2025arXiv251216844C}, in which mock dual AGN catalogs are constructed from large cosmological simulation \texttt{ASTRID} to match the selection and properties of the DESI DR1 sample. In particular, the mock catalogs reproduce the DESI targeting strategy across different tracers and redshifts, ensuring that the simulated AGN population reflects the observed distribution of galaxy types. 
{We note that different cosmological parameters adopted by \texttt{CIGALE} and \texttt{ASTRID} shift the DESI–\texttt{ASTRID} stellar-mass comparison by only $\Delta \log M_\star \simeq 0.02$–$0.03$ dex, so we do not apply an additional correction.}
The resulting DESI-mock dual AGN sample provides a framework for connecting observed dual AGN candidates to their underlying MBH merger population and for predicting gravitational wave sources detectable by future experiments such as LISA and \textit{PTAs}.

\subsubsection{Dual Fraction}\label{dual_fraction}

\begin{figure}
\centering
\includegraphics[width=0.48\textwidth]{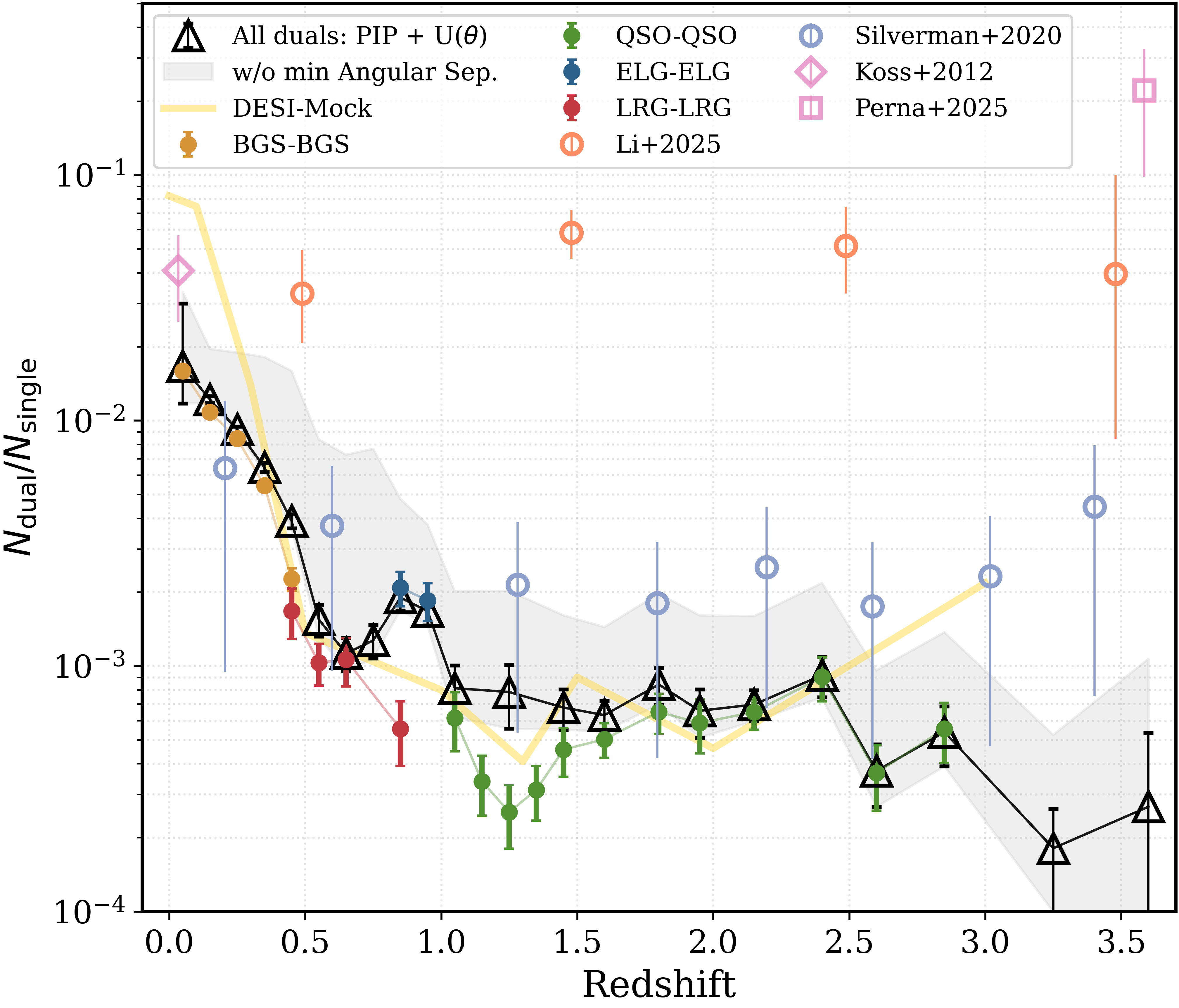}
\caption{
Redshift evolution of the dual AGN fraction, defined as $N_{\rm dual}/N_{\rm single}$, for the DESI sample. Hollow black triangles show the full dual AGN population measured with PIP+$U(\theta)$ weighting, while colored points indicate the same-tracer subsamples. Error bars show Poisson uncertainties propagated with the applied weights.
The gray shaded region shows the \texttt{ASTRID}-based correction when removing the $1.6''$ minimum angular-separation cut.
The yellow line shows the \texttt{ASTRID}-based DESI mock prediction for the dual AGN fraction. 
Literature measurements from \citet{2025ApJ...986..101L}, \citet{2020ApJ...899..154S}, \citet{2012ApJ...746L..22K}, and \citet{2025A&A...696A..59P} are overplotted for comparison.
}
\label{fig:fracdual}
\end{figure}

Figure~\ref{fig:fracdual} shows the redshift evolution of the dual-to-single AGN ratio, $N_{\rm dual}/N_{\rm single}$, for the DESI DR1 sample. The full dual population, measured with PIP+$U(\theta)$ weighting, is shown by hollow black triangles, while the yellow curve shows the DESI-mock reference prediction. 
The observed ratio declines from $\sim2\times10^{-2}$ at $z=0.05$ to $\sim4\times10^{-3}$ by $z=0.45$, and remains at the $\sim(1.10$--$1.92)\times10^{-3}$ level over $0.65\lesssim z\lesssim0.95$. At higher redshift the ratio stays low, typically $(4$--$8)\times10^{-4}$ over $1.05\lesssim z\lesssim2.4$, before becoming increasingly uncertain above $z\gtrsim2.5$, as the raw number of dual systems per bin falls below a few tens.
Relative to the DESI-mock reference curve, the observed DESI ratio is lower by factors of $\sim2$--5 for $z<0.3$, although with large uncertainties, {driven in part by close separation blending affecting nearby, low-mass companions.} At higher redshift, the two converge to within $\lesssim0.3$ dex over $0.4\lesssim z\lesssim2.3$.

The DESI dual AGN fraction is broadly consistent with the \cite{2020ApJ...899..154S} (SDSS+HSC) measurements, generally within $\lesssim1\sigma$ over $0.2\lesssim z\lesssim3.4$ (A recent SDSS quasar study \citep{2026ApJ...998...80J} reports a lower dual fraction and suggests that the measurements in \cite{2020ApJ...899..154S} may be affected by stellar contamination.), but lies below the \cite{2025ApJ...986..101L} (COSMOS-Web) points and also below the high-redshift \cite{2025A&A...696A..59P} (COSMOS sample at $z \sim 3.5$) estimate. 
This offset likely reflects differences in selection, including AGN luminosity cuts, pair definitions, and tracer completeness, rather than a purely statistical discrepancy.
\citet{2025ApJ...986..101L} searched for kpc-scale dual and offset AGN around an X-ray-selected sample using deep HST and JWST NIRCam imaging.
This sample includes heavily obscured AGN, which dominate the merger-triggered population. Dual fraction among such obscured systems is nearly two orders of magnitude higher than among luminous unobscured quasar pairs \citep{2025ApJ...986..101L}, 
{whereas the high-redshift DESI population is biased toward luminous, relatively unobscured type 1 AGN through QSO targeting. At lower redshift, however, the DESI sample includes many galaxy-targeted systems identified through optical emission-line and mid-infrared diagnostics, so the selection differences are broader than a simple obscuration bias.}
Additionally, the JWST NIRCam's resolution allows detailed SED decomposition at sub-arcsecond scales for $z>1$, capturing companions that ground-based surveys might miss.
The high redshift JWST/NIRSpec estimate of \citet{2025A&A...696A..59P}, based on 3--5 multiple systems among 16 AGN at $2<z<6$ (i.e., $\sim$20--30\%),  is also above our DESI measurements. {This offset likely reflects differences in spatial resolution, since IFU observations are sensitive to close pairs that would be unresolved or blended in single fiber surveys like DESI. However, given the small parent sample, this comparison should be interpreted with caution.}

Swift/BAT measurement from \citet{2012ApJ...746L..22K} is consistent within uncertainties with the first corrected DESI bin measurement, with a reported dual AGN frequency of $10\%$ (16/167) at $<100$ kpc separations.

We also note that the observed decline in dual fraction with redshift is partly driven by angular-resolution incompleteness from the DESI $1.6''$ fiber scale, which prevents recovery of close physical pairs. 
{To estimate this effect, we use the \texttt{ASTRID} mock to compute, in each redshift bin, the fraction of duals without this separation cut over the number of duals with the cut
$f_{\rm corr}(z)=N_{\rm dual}^{\rm no\,cut}/N_{\rm dual}^{1.6''\,{\rm cut}}$, 
and multiply the observed weighted DESI dual fraction by this factor.
The gray shaded region  therefore represents an empirical angular-resolution correction derived from the \texttt{ASTRID} mock.
The correction is strongest at intermediate redshift, reaching factors of $\sim4.7-5.6$ at $z\sim0.5-0.8$, and is more modest at both low redshift ($\sim1.1-2$ below $z\sim0.3$) and high redshift ($\sim2$ at $z\sim1.5-2.3$, where the mock counts are small).
The non-monotonic redshift dependence likely reflects both the changing physical scale subtended by the minimum angular separation cut, and redshift-dependent changes in the tracer-matched mock parent population \citep[see][]{Chen2023}. }

The colored points show the same-tracer contributions to the dual AGN fraction, each occupying distinct redshift ranges consistent with DESI target selection. BGS--BGS pairs (orange) dominate at low redshift, contributing $\sim90\%$ of the weighted dual counts at $z<0.3$ and remaining important to $z\sim0.5$. LRG--LRG pairs (red) contribute most strongly at $z\sim0.4$--$0.6$, while ELG--ELG pairs (blue) dominate near $z\sim0.8$--$0.9$, accounting for $\sim64\%$ of the weighted dual counts.
{Their higher abundance relative to LRG--LRG pairs at similar redshifts is expected. LRGs tend to be massive central galaxies, making close LRG–LRG pairs intrinsically rare. ELGs, by contrast, more commonly trace star-forming satellites and field galaxies, which can be more frequently found at small separations.}
At $z\gtrsim1$, the QSO--QSO pairs (green) become the dominant same-tracer contribution, increasing from $\sim39\%$ at $z\simeq1.05$ to $\sim61\%$ at $z\simeq1.6$. 
The incomplete sum of same-tracer points at intermediate redshift reflects the importance of mixed-tracer pairs, with same-tracer systems accounting for only $\sim41\%$--$57\%$ of the weighted counts over $0.5\lesssim z\lesssim0.8$.

\subsubsection{Merger rates and LISA predictions using \texttt{ASTRID}} \label{lisa_merger_rates}

\begin{figure*}
\centering
\includegraphics[width=\textwidth]{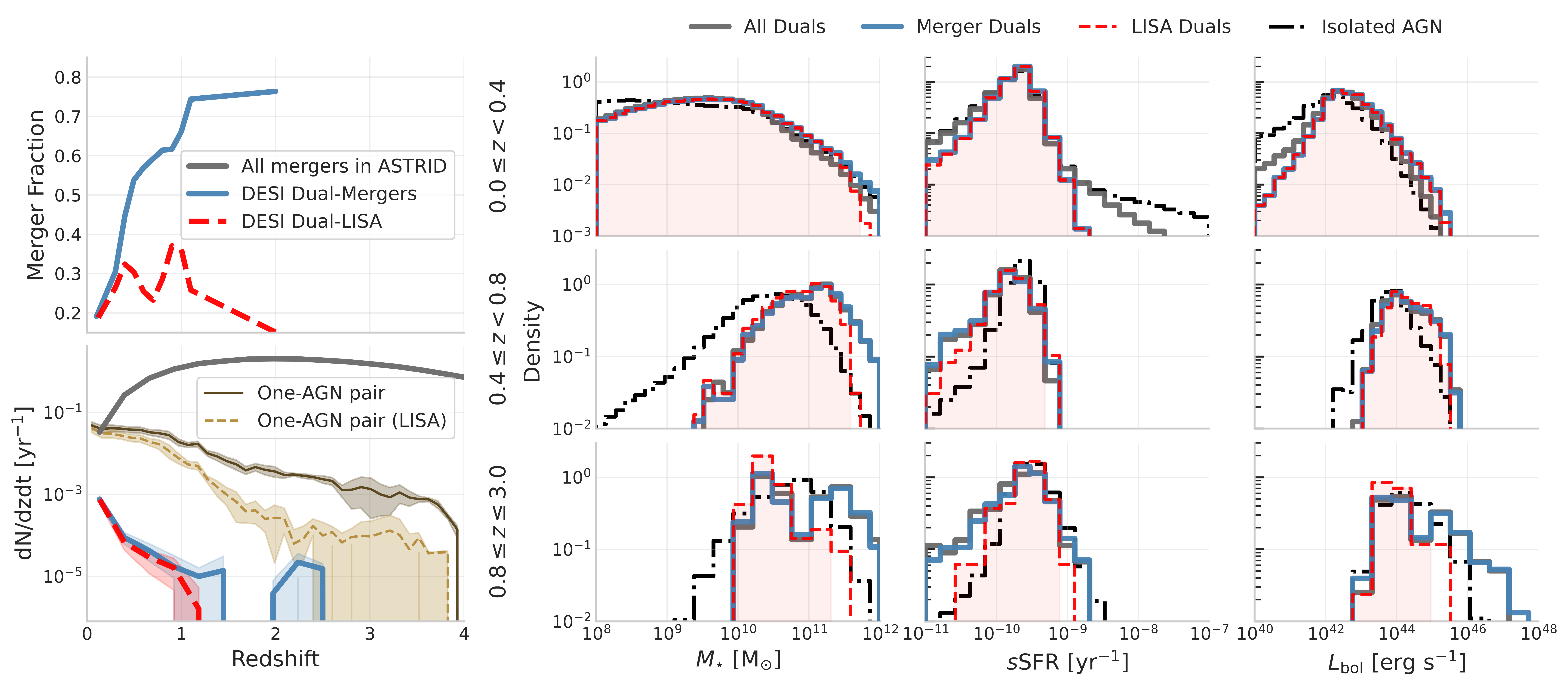}
\caption{
\textbf{Left:} Redshift evolution of the merger fraction for DESI-selected dual AGN in the \texttt{ASTRID} mock (blue solid line) and for the subset expected to be detectable by LISA with $\mathrm{SNR}_{\rm LISA}>10$ (red dashed line). The lower panel shows the corresponding merger rate, $\mathrm{d}N/\mathrm{d}z\mathrm{d}t$, for all intrinsic mergers in \texttt{ASTRID} (gray), for mergers associated with the DESI-like one-AGN pair sample (brown solid), for the LISA-detectable subset of those one-AGN pair mergers (gold dashed), for DESI Dual-Mergers (blue), and for DESI Dual-LISA systems (red dashed). Shaded bands show the $5$th--$95$th percentile range. 
\textbf{Right:} Distributions of host galaxy stellar mass, specific star-formation rate, and AGN bolometric luminosity for mock-selected DESI-like systems are shown in the left, middle, and right columns, respectively. The three rows correspond to different redshift intervals. 
In each panel, the black dash-dotted curve shows the DESI-mock isolated AGN reference sample. The gray curve shows all DESI-mock dual AGN in the mock, the blue curve shows the subset associated with mergers, and the red dashed curve shows the subset expected to be LISA-detectable. 
}
\label{fig:fracmerger}
\end{figure*}

Figure~\ref{fig:fracmerger} summarizes how DESI dual AGNs in the \texttt{ASTRID} mock sample evolve into merging black hole binaries, and which subset would be detectable by LISA. The merger identification and LISA detectability criteria are described in Section~\ref{methods}, with full details in \citet{2025arXiv251216844C}.

In the upper-left panel of Figure~\ref{fig:fracmerger}, the merger fraction of dual AGNs (blue solid line, defined as the fraction of duals in each redshift bin that will merge) increases with redshift and reaches a maximum of $\sim 0.76$ at $z\sim2$. 
We caution that the apparent flattening of the merger fraction at $z\gtrsim1$ reflects limited statistics in the simulation rather than a genuine saturation.
While mergers become more common at high redshift, only a narrow window in redshift and binary parameter space satisfies the LISA sensitivity requirements.
The fraction of LISA detectable systems (red dashed line) broadly tracks the merger fraction at $z\lesssim0.3$, declines at intermediate redshift, and peaks at $z\sim0.9$ reaching $\sim0.37$, before decreasing again toward higher redshift.
The characteristic delay between the dual AGN phase and the final coalescence is of the order $\sim1\,\mathrm{Gyr}$ at all redshifts.

{The lower-left panel of Figure~\ref{fig:fracmerger} compares the merger-rate histories of five populations. The full \texttt{ASTRID} merger population peaks at $z\simeq2$ with $\mathrm{d}N/\mathrm{d}z\mathrm{d}t \simeq 2.0\,\mathrm{yr^{-1}}$, with $\sim68\%$ of its integrated rate 
arising from $z\geq0.8$ and only $\sim0.3\%$ from $ z\leq0.3$. 
In contrast, the one-AGN pair merger rate peaks at $z\simeq0.05$ with $\mathrm{d}N/\mathrm{d}z\mathrm{d}t \simeq 4.8\times10^{-2}\,\mathrm{yr^{-1}}$ and then declines steadily with redshift,  with its integrated rate distributed more evenly across redshift intervals. 
The LISA-detectable subset accounts for $\simeq78\%$ of the integrated one-AGN pair merger rate at $z\leq0.3$, but only $\simeq27\%$ at $z\geq0.8$, showing LISA's diminishing sensitivity at higher redshift and larger separations.}

The DESI-selected dual samples are much rarer. Both the DESI Dual-Merger and DESI Dual-LISA rates are highest in the lowest-redshift bin and decline rapidly with redshift, with $\mathrm{d}N/\mathrm{d}z\mathrm{d}t \simeq 7.6\times10^{-4}\,\mathrm{yr^{-1}}$ in the first bin, and $\sim60$--$70\%$ of their integrated rates arising from $z\leq0.3$. 
Even in this lowest-redshift interval, the integrated DESI Dual-Merger rate is only $\simeq0.5\%$ of the intrinsic \texttt{ASTRID} merger rate, dropping to $\simeq6\times10^{-4}\%$ at $z\geq0.8$. DESI-selected duals therefore trace only a small and biased subset of the total merger population.

The right panels of Figure~\ref{fig:fracmerger} compare the distributions of host galaxy stellar mass, sSFR, and AGN bolometric luminosity for the isolated AGN reference, all dual AGNs, the merger subset, and the LISA-detectable subset.

At low redshift ($0.0 \leq z \leq 0.3$), the dominant effect is a mass-dependent selection: more massive black holes sink and merge on shorter dynamical-friction timescales \citep{2017ApJ...840...31D}, so the duals that coalesce within the $\lesssim3$\,Gyr available are preferentially drawn from more massive hosts and, through the $M_{\rm BH}$--$M_\star$ relation \citep{2015ApJ...813...82R}, contain more massive and more luminous AGN. In Figure~\ref{fig:fracmerger} relative to the isolated AGN reference, dual AGNs have median host stellar mass $\sim7\times$ higher ($\sim3\times10^{8}$ to $\sim2\times10^{9}\,M_{\odot}$) and median bolometric luminosity $\sim3\times$ higher, and the merger subset is further concentrated at the top end: the fraction above $10^{10}\,M_{\odot}$ rises from $\sim20\%$ for all duals to $25\%$, with a similar trend in $L_{\rm bol}$ ($22\%$ vs.\ $28\%$ above $10^{43}\,\mathrm{erg\,s^{-1}}$). 
The merger and LISA subsets are nearly indistinguishable because at these redshifts LISA is sensitive across essentially the full mass range of DESI dual mergers ($M_{\rm tot}\sim10^{5}$--$10^{7}\,M_{\odot}$; \citealt{2025ApJ...993..199W}); only the very top of the mass function is filtered out ($11\%$ of mergers vs.\ $8\%$ of LISA systems above $10^{11}\,M_{\odot}$), where $M_{\rm BH}\gtrsim10^{8}\,M_{\odot}$ binaries radiate below the LISA band and into the PTA regime. 
The sSFR distributions are broadly similar across all populations; the slightly broader high-sSFR tail of the full dual sample ($12\%$ vs.\ $10\%$ above $10^{-9.5}\,\mathrm{yr^{-1}}$) comes from low-mass, gas-rich systems whose black holes are too light to merge by $z=0$ or to produce detectable signals.

At intermediate redshift ($0.4 \leq z \leq 0.7$), the LISA mass window begins to cut into the population from above. The full dual and merger populations are nearly identical and both sit well above the isolated AGN reference, with median host masses ($\sim10^{11}\,M_{\odot}$) roughly $4\times$ higher and median luminosities ($\simeq1.5$--$1.7\times10^{44}\,\mathrm{erg\,s^{-1}}$) about $2\times$ higher. The LISA subsample, by contrast, is shifted to slightly lower masses (median $M_{\star}\sim8\times10^{10}\,M_{\odot}$) with a strongly suppressed tail above $10^{11.5}\,M_{\odot}$: these hosts contain black holes with $M_{\rm BH}\gtrsim10^{8}$--$10^{9}\,M_{\odot}$, whose gravitational wave frequencies fall below the LISA band. In sSFR, all dual populations are slightly skewed low relative to the isolated reference (median $\sim1.5\times10^{-10}$ vs.\ $\sim2.3\times10^{-10}\,\mathrm{yr^{-1}}$), so duals at $z\sim0.5$ preferentially inhabit massive but less actively star-forming hosts. Within them, the LISA subset leans toward higher sSFR ($11.6\%$ vs.\ $7.9\%$ above $10^{-9.5}\,\mathrm{yr^{-1}}$), consistent with its lower-mass, more gas-rich hosts.

The merger and LISA populations diverge most strongly at $0.8 \leq z \leq 2.2$, where LISA's V-shaped sensitivity curve, combined with a DESI selection that increasingly favors massive, luminous systems, narrows the detectable window to intermediate masses. Relative to the isolated AGN reference, the dual and merger populations are no longer simply shifted upward: their medians are comparable or slightly lower, but their high-mass and high-luminosity tails are strongly enhanced ($44\%$ of mergers vs.\ $27\%$ of isolated AGNs above $10^{11}\,M_{\odot}$; $32\%$ vs.\ $16\%$ above $10^{45}\,\mathrm{erg\,s^{-1}}$). The LISA subset filters these tails away, retaining only $\sim8\%$ above $M_\star\sim10^{11}\,M_{\odot}$ and $L_{\rm bol}\sim10^{45}\,\mathrm{erg\,s^{-1}}$: at $z\gtrsim1$, mergers with $M_{\rm tot}\gtrsim10^{7}$--$10^{8}\,M_{\odot}$, precisely the systems in the most massive and luminous hosts, radiate on the low-frequency side of LISA's peak sensitivity and become undetectable \citep{2025ApJ...993..199W}. What remains are intermediate-mass binaries ($M_{\rm tot}\sim10^{5}$--$10^{7}\,M_{\odot}$) in moderately massive hosts ($M_{\star}\sim2$--$5\times10^{10}\,M_{\odot}$). The same filtering explains why the LISA sample's sSFR distribution is closer to the isolated AGN reference than to the broader dual samples: removing the massive-host tail leaves a moderate-mass subset with higher typical sSFR.

\subsection{Extreme Regimes: High-z Duals and Dwarf Pairs} \label{Case Studies}

\subsubsection{Dual AGNs at Cosmic Noon ($z \gtrsim 2$)}\label{high_z}

\begin{figure*}
\centering
\includegraphics[width=0.99\textwidth]{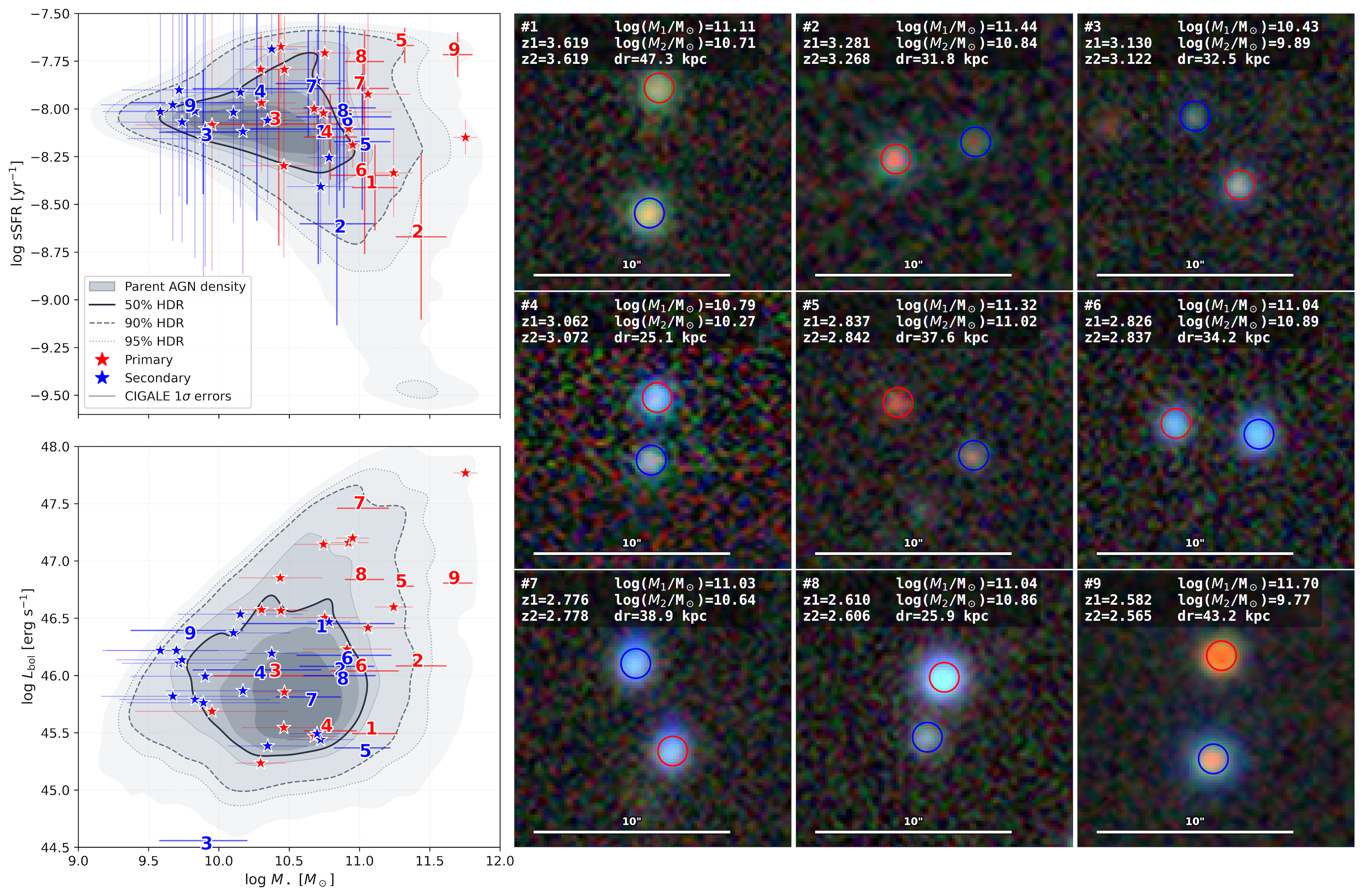}
\caption{
\textbf{Left panels}: Distribution of sSFR (Top) and $L_\text{bol}$ (Bottom) as a function of stellar mass for the full high-z ($z>2.5$) AGN population, shown as shaded contours. Red and blue stars indicate the primary and secondary components of high-redshift dual AGN systems, respectively. Numbers label selected systems highlighted in the right panels.
\textbf{Right panels}: Legacy Survey image cutouts of the numbered systems. Colored circles mark the positions of the primary (red) and secondary (blue) AGN. The top-left corner of each panel lists the redshifts, stellar masses, and projected separations of the two components.}
\label{fig:special_highz}
\end{figure*}

Studies of dual AGN beyond $z>2$ are scarce. Deep Chandra imaging of 62 X-ray selected AGNs at $2.5<z<3.5$ found no confirmed dual systems, placing an upper limit of $\sim5\%$ on the dual fraction in that sample \citep{2024ApJ...974..121S}. 
High-resolution JWST and HST imaging \citep{2025ApJ...986..101L} identified tens of dual/offset AGNs among moderate-luminosity, obscured AGNs up to $z\sim4.5$, with a pair fraction rising to $\sim23\%$ by $z\sim4.5$.
Individual high-$z$ examples include a lensed dual quasar at $z=2.37$ (PS J1721+8842) with $\sim6.5$ kpc separation \citep{2021MNRAS.508L..64M}, as well as a candidate dual AGN at $z\simeq5.4$ identified with JWST \citep{2025MNRAS.544.4160L}. 
{A literature comparison sample at $z>2$ comprises 80 systems from more than 20 studies in the BigMAC Catalog \citep{2025ApJS..281...25P}, as well as 24 COSMOS-Web systems and 5 HST systems. After applying a common $1.6''$ resolution-equivalent cut, this comparison sample is reduced to 40 systems. By comparison, DESI DR1 dual catalog has 132 dual AGNs at $z>2$ (34 at $z>2.5$).}
A complementary study by \citet{2026ApJ..1000..311J} identified DESI DR1 QSO--QSO pairs using broader selection cuts of $r_p \leq 110\,{\rm kpc}$ and $|\Delta V_r| \leq 2000\,{\rm km\,s^{-1}}$. At $z>2$, their catalog adds $\sim60$ systems below our $1.6''$ angular-separation cut and $\sim220$ beyond our $50\,{\rm kpc}$ projected-separation limit, highlighting the sensitivity of the inferred high-$z$ dual AGN census to the adopted selection. 
{We further cross-match our high-redshift candidates with the lens classifications from \citet{2026ApJ..1000..311J}. This identifies 4 previously reported lensed quasars and 7 additional visually classified lensed quasar candidates, which we exclude from the main sample. Excluding these objects leaves 121 primary dual AGN candidates at $z>2$, including 28 at $z>2.5$. We note that eight additional systems have lens-candidate labels in the previous literature, but are visually classified by \citet{2026ApJ..1000..311J} as quasar pairs or quasar pair candidates. We retain these systems with caution flags.}

To illustrate the properties of these systems, Figure~\ref{fig:special_highz} shows representative examples of the highest redshift dual AGNs ($z_{1,2}>2.5$). The left panels present the star formation rates (sSFR) of the host galaxies (top) and the bolometric luminosities of the AGNs (bottom) as functions of the host galaxy stellar mass for the entire AGN population at redshifts ($z \geq 2.5$) (shown as KDE contours). The components of the dual AGNs are overplotted as red (primary) and blue (secondary) stars. Numbered points denote selected systems in the right panels, which include cutouts with component redshifts, stellar masses, and projected separations, while circles indicate the primary AGN in red and the secondary AGN in blue.

Of the 28 primary dual AGN candidates at $z>2.5$, 25 have reliable \texttt{CIGALE} stellar-mass estimates for both components and 24 also have reliable star-formation-rate estimates.
In the sSFR$-M_\star$ plane, $33\%$ of primaries and $54\%$ of secondaries fall within the densest $50\%$ of the parent AGN distribution, while 4 primaries and no secondaries lie outside the densest $90\%$ contours, indicating that most dual components have broadly typical star-formation properties at fixed stellar mass.  
In the $L_{\rm bol}-M_{\star}$ plane, the duals are similarly well embedded, with
four primaries and two secondaries lying outside the densest $90\%$ contours. The primaries tend to occupy more underpopulated regions of the AGN population at the same redshift and show a clear tendency toward the massive, high-luminosity wing. We caution, however, that the inferred host properties for these high-redshift quasar hosts are uncertain, as the \texttt{CIGALE} fits are sensitive to AGN--host decomposition. 

{The nine highlighted systems span $z=2.52$--$3.62$ and projected separations of $r_p\simeq25$--$48$ kpc. They cover a wide range of host and AGN properties, with stellar mass ratios $q_{M_\star}\simeq0.01$--$0.71$ and bolometric-luminosity ratios $q_{L_{\rm bol}}\simeq0.02$--$0.99$. 
The highest-redshift pair in this set (\#1; $z=3.62$), was previously reported as a dual AGN candidate in BigMAC and is a wide pair with $r_p\simeq47$ kpc and a very small line-of-sight velocity offset, $|\Delta v|\simeq5~{\rm km\,s^{-1}}$.
These examples illustrate the diversity of the high-redshift DESI candidate population. System~\#4 is the most compact highlighted pair ($r_p\simeq25$ kpc), system~\#6 has the most comparable stellar masses ($q_{M_\star}\simeq0.71$), and system~\#2 has nearly equal inferred bolometric luminosities ($q_{L_{\rm bol}}\simeq0.99$). 
The sample also includes highly asymmetric systems: system~\#7 hosts the most luminous AGN in the highlighted set but has $q_{L_{\rm bol}}\simeq0.02$, while system~\#9 is an extreme mass-ratio candidate ($q_{M_\star}\simeq0.01$) with a large velocity offset ($|\Delta v|\simeq1472~{\rm km\,s^{-1}}$), and should therefore be treated cautiously as a physical dual system.}

\subsubsection{Dwarf Dual AGNs as potential LISA Sources} \label{dwarfs}

\begin{figure*}
\centering
\includegraphics[width=0.99\textwidth]{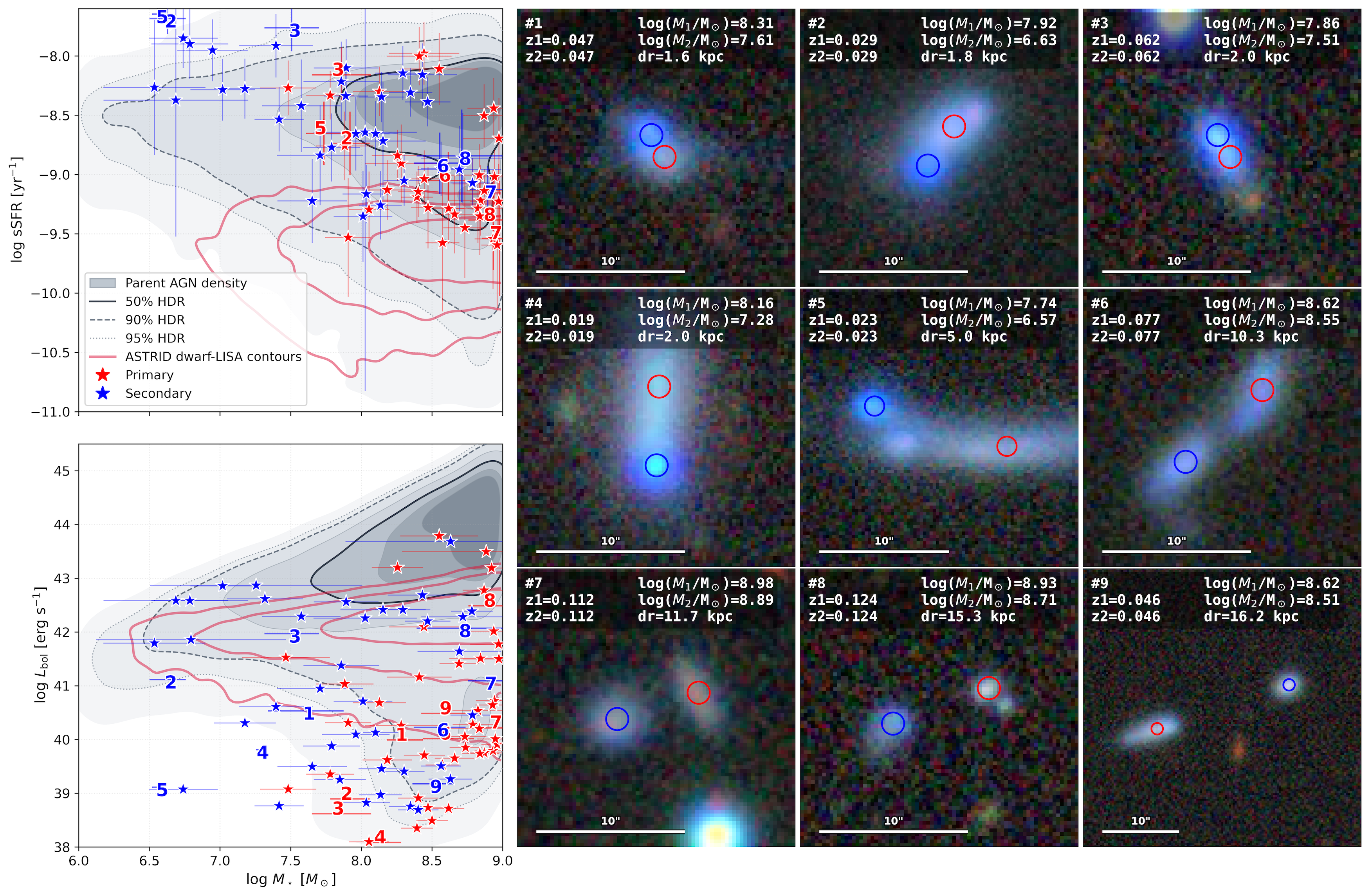}
\caption{
Same as Figure~\ref{fig:special_highz}, but for dwarf dual AGN systems ($M_{\star,1,2} < 10^9\,M_\odot$). Red contours show the distribution of individual nuclei in \texttt{ASTRID} dwarf--dwarf dual systems whose black holes are expected to enter the LISA sensitivity regime, enclosing 50\%, 80\%, and 95\% of the simulated dwarf-LISA population.
}
\label{fig:special_dwarfs}
\end{figure*}


{
Dwarf galaxies ($M_\star \lesssim 10^9\,M_\odot$) are the most numerous galaxy class in the Universe and constitute the building blocks from which more massive systems assemble through hierarchical merging \citep[e.g.,][]{2010A&ARv..18..279V, 2020ARA&A..58..257G}.
Identifying AGN in dwarfs is difficult, since their black holes are expected to have low masses and to be weakly accreting, making AGN intrinsically faint. $\sim83\%$ of observed dwarf AGN have bolometric luminosities below $10^{42}\,{\rm erg\,s}^{-1}$ \citep{2024MNRAS.528.5252M,2025MNRAS.536..295M}. 
In single-fiber surveys such as SDSS, this faint nuclear signal is further diluted because the $3''$ fiber integrates emission from both the nucleus and the surrounding star-forming disk, weakening the AGN signature in the integrated spectrum. In the MaNGA IFU survey, spaxel-by-spaxel analysis showed that SDSS single-fiber spectra missed $\sim80\%$ of dwarf AGN in the same objects \citep{2024MNRAS.528.5252M}. MaNGA is nonetheless limited in volume and redshift reach ($z<0.15$). 
DESI's smaller $1.6''$ fiber reduces this dilution relative to SDSS and enables dwarf-AGN searches over a much larger spectroscopic sample. 
Using the DESI EDR sample \citet{2025ApJ...982...10P} identified $2,444$ dwarf AGN candidates with $\log(M_\star/M_\odot)\leq9.5$, corresponding to an AGN fraction of $\approx2.1\%$, nearly four times higher than previous single-fiber estimates. This fraction is a lower limit because the [N II] BPT diagnostic is incomplete for low-metallicity dwarf galaxies \citep{2024MNRAS.528.5252M}. The full DESI main survey will expand this sample by more than an order of magnitude. 
Here we use DESI DR1 to extend this search to dwarf galaxies in close pairs and dual AGN systems.
}

{Galaxy mergers are well established as triggers of enhanced AGN activity in massive galaxies, but their role at low stellar masses is less well characterized. 
\cite{2025A&A...699A.330E} found that galaxy mergers  enhance AGN activity in dwarfs at projected separations $r_\mathrm{p} \lesssim 20$kpc, 
1.4 times higher than in non-interacting dwarfs of comparable mass and redshift, although $\approx91\%$ of dwarf AGN reside in non-interacting systems. 
}

Confirmed dual AGN with both nuclei in dwarf-mass galaxies are extremely rare.  
\citet{2023ApJ...944..160M} reported two strong candidates for dual AGN in dwarf--dwarf mergers based on dual X-ray sources, infrared AGN-like colors, and tidal features, but the physical association of all components is not confirmed through spectroscopic redshifts. 
Earlier studies also found Mrk~709, a blue compact dwarf pair at $z = 0.052$ with component masses $M_\star \sim 2.5 \times 10^9$ and $1.1 \times 10^9\,M_\odot$, hosts an AGN in its southern nucleus confirmed by X-ray and radio emission \citep{2014ApJ...787L..30R, 2021ApJ...912...89K} 
but no AGN has been detected in the northern companion. In the minor merger regime, an AGN has been identified in a dwarf satellite of the massive galaxy NGC~3341 \citep{2013MNRAS.435.2335B}, and \citet{2024OJAp....7E...3M} confirmed dual AGN via Chandra X-ray observations of SDSS~J125417.98+274004.6, a $\sim$11:1 minor merger in which the secondary component has $M_\star \approx 2.2 \times 10^9\,M_\odot$.
Recent Euclid Q1 work has identified nine dual AGN candidates in low-mass galaxy pairs with $\log(M_\star/M_\odot)\leq10$, using DESI spectroscopic redshifts and multiwavelength AGN diagnostics \citep{2026arXiv260413170M}; however, these systems extend above the dwarf mass threshold adopted here, and most remain candidate rather than robust dual AGNs. 

These systems are particularly important because the black hole masses expected in dwarf galaxies ($\sim10^{4}$–$10^{6}\,M_\odot$) lie near the peak sensitivity of LISA, making them promising gravitational wave progenitors. However, their intrinsic faintness and the low AGN occupation fraction in dwarfs have so far limited systematic searches \citep{Miller2015, 2018MNRAS.478.2576M, 2020MNRAS.492.2268B}.

After applying stellar mass cuts ($M_{\star,1,2} < 10^{9}\,M_\odot$) and visual inspection, our catalog contains $\sim50$ dwarf dual AGN candidates. 
This sample is dominated by BGS-selected pairs and spans $z=0.02$--$0.47$, projected separations of $1.6$--$46$~kpc, and mass ratios $q=0.005$--$0.957$ (25 major, 12 minor, and 13 very minor mergers). Only $11\%$ satisfy explicit BPT-based AGN cuts, while $72\%$ carry the \texttt{OPT\_OTHER\_AGN} flag and $32\%$ the \texttt{WISE\_ANY\_AGN} criterion, confirming that the [N II]-BPT diagnostic alone misses the majority of AGN in low-metallicity dwarfs \citep{2024MNRAS.528.5252M}.

Figure~\ref{fig:special_dwarfs} compares these systems to isolated AGNs with the same stellar mass cuts. The dwarf dual AGN candidates have median stellar masses of $\log(M_{\star,1}/M_\odot)=8.6$ and $\log(M_{\star,2}/M_\odot)=7.9$.
For the 39 pairs with reliable \textsc{cigale} SFR quality flags for both nuclei, all primaries and $\sim80\%$ of secondaries fall within the $90\%$ contour of the comparison AGN population in the sSFR--$M_\star$ plane.
However, the primaries preferentially occupy the high-mass, low-sSFR edge of the dwarf AGN distribution, while the few secondaries outside the contours are shifted toward enhanced sSFR at fixed stellar mass.
}
A stronger offset appears in the $M_\star-L_{\rm bol}$ plane. The median bolometric luminosities of primaries and secondaries are $10^{40.1}$ and $10^{40.7}\,\mathrm{erg\,s^{-1}}$, respectively, far below the comparison AGN median of $10^{43.0}\,\mathrm{erg\,s^{-1}}$. Thus, these systems primarily occupy the faint tail of the AGN luminosity distribution. Accordingly, 7 primaries and 11 secondaries lie outside the $95\%$ KDE contour. 


Red contours in Figure~\ref{fig:special_dwarfs} show the $50\%$, $80\%$, and $95\%$ highest-density KDE regions of the low-redshift \texttt{ASTRID} LISA population. The strongest overlap with such LISA population occurs in the $\log M_\star$--$\log L_{\rm bol}$ plane, where $47\%$ of DESI dwarf dual AGN candidates fall within the $95\%$ contour,
most of the remaining DESI dwarf duals are underluminous for their stellar mass. 
In sSFR, the DESI dwarf dual AGN candidates lie systematically above the \texttt{ASTRID} LISA contours. This offset may be partly physical, since our AGN selection favors gas-rich interacting dwarfs, but it is likely also affected by SED-fitting systematics in low-mass AGN hosts. 
In low mass AGN hosts the nuclear continuum contributes a disproportionately large fraction of the UV--optical SED, causing the AGN--host decomposition to become strongly degenerate at low $f_{\rm AGN}$; unmodeled nuclear light can then leak into the young stellar component and artificially inflate the recovered SFR \citep{2015A&A...576A..10C}.
{A direct comparison with \texttt{FastSpecFit}, which jointly fits DESI spectrophotometry and broadband photometry with stellar population and emission-line templates but does not include the same AGN--host decomposition adopted by \texttt{CIGALE}, shows consistent stellar masses. However, \texttt{CIGALE} sSFRs are higher by a median of $\sim1.4$ dex, with the offset increasing toward lower masses (Appendix~\ref{app:fsf_vs_cigale}). 
We therefore treat the two estimates as bracketing the plausible sSFR range.}
The nine systems shown as cutouts in Figure~\ref{fig:special_dwarfs} highlight both compact, nearly equal-mass systems and more asymmetric minor or very minor pairs.


\subsubsection{Radio and X-ray counterparts}

{We cross-match low mass ($\log(M_\star/M_\odot) < 10$) dual AGN candidate sample against LoTSS DR3 \citep{2026A&A...707A.198S}, FIRST \citep{2015ApJ...801...26H}, and VLASS QL Epoch~1 \citep{2020PASP..132c5001L} in the radio, and eROSITA/eRASS1 \citep{2024A&A...682A..34M}, Chandra CSC~2.1 \citep{2024ApJS..274...22E, 2010ApJS..189...37E}, and XMM-Newton 4XMM-DR13 \citep{2020A&A...641A.136W} in the X-ray. We adopt matching radii of $5\arcsec$ for the radio catalogs, $10\arcsec$ for eROSITA, $5\arcsec$ for XMM, and $3\arcsec$ for Chandra. For pairs whose separation is smaller than the survey beam or PSF, a source associated with both components is classified as a shared pair-level detection rather than evidence for two distinct nuclei. LoTSS dominates the radio matches because of its sensitivity to faint 144~MHz synchrotron emission, though at the stellar masses of interest such emission can trace star formation as well as AGN activity; FIRST and VLASS provide higher-resolution checks at 1.4 and 3~GHz. X-ray associations are rarer: eROSITA is shallow for low-luminosity AGN at $z \sim 0.1$--$0.3$, and Chandra and XMM-Newton coverage is limited to sparse pointed fields.}

At $\log(M_\star/M_\odot) < 10$, 145 pairs (${\sim}23\%$) are radio-associated: 63 one-component, 75 shared-source, and 7 distinct-source matches. 9 pairs (${\sim}1\%$) have X-ray associations, including 4 one-component and 5 shared-source matches. No pair is resolved into two distinct X-ray nuclei.
At $\log(M_\star/M_\odot) < 9.5$, 15 pairs (${\sim}10\%$) have radio associations and one has an X-ray association. 
For the strict dwarf subset, with both components satisfying $\log(M_\star/M_\odot)<9$, we find no X-ray associations and one radio association (MW8 in Table~\ref{tab:multiwavelength_candidates}), in which both components coincide with the same LoTSS source (ILTJ141725.63$+$323801.6) at a separation $\Delta\theta = 5.98''$, comparable to the ${\sim}6\arcsec$ LoTSS beam, with a peak $S/N = 2.8$; we treat this as a tentative unresolved pair-level association.

Of the 7 distinct-source radio pairs, 4 are robust. MW1 is the clearest case, with independent FIRST ($S/N = 21.6$, $13.6$) and VLASS ($S/N = 8.1$, $9.1$) detections at both components. 
In MW2, Target~1 is the highest-significance radio counterpart in the sample (FIRST $S/N = 62.7$, VLASS $S/N = 24.4$), while Target~2 has a separate FIRST detection ($S/N = 13.4$). MW3 and MW4 each have distinct FIRST counterparts at both components, with VLASS confirmation for Target~1. The remaining three cases, MW5--MW7, rely on LoTSS entries with $S/N <2$ and are tentative.
Details of the X-ray-associated systems, together with cutouts and DESI spectra for representative examples, are given in Appendix~\ref{app:radio} (Table~\ref{tab:multiwavelength_candidates}, Figure~\ref{fig:mw_examples}).

%% file: Sec5_Conclusion.tex
\section{Conclusions}\label{conclusions}

In this study, we have constructed a DESI DR1 census of candidate dual and offset AGNs selected from spectroscopic galaxy/QSO pairs. AGNs are identified using the DESI AGN/QSO value-added catalog together with optical emission-line, broad-line, and WISE mid-infrared diagnostics, and the pair statistics are corrected for data quality, targeting, and fiber assignment systematics. The sample spans the DESI BGS, LRG, ELG, and QSO tracer classes, enabling a multi-tracer census of dual AGN activity across quiescent massive galaxies, star-forming systems, and luminous quasars.

Our final catalog contains 7,125 dual AGN candidates ($8.3\%$ of all pairs), 27,347 one-AGN pairs ($32.0\%$), and 50,866 inactive pairs ($59.6\%$) over $0 \lesssim z \lesssim 3.6$. Within the dual sample, $31\%$ of systems have both nuclei supported by two or more diagnostic families, and $6.7\%$ by multiple independent channels, defining conservative high-purity subsets. 
Relative to previous compilations \citep{2025ApJS..281...25P}, DESI increases the known dual AGN sample by a factor of $\sim4$ against the $1,754$ spatially resolved literature systems.
DESI exceeds the resolved sample by a factor of $\sim6.5$ at $z\sim0.1$--$0.2$ ($\approx1500$ candidates), and by $\sim70$ ($\sim11\times$ the full compilation, including candidates through unresolved spectroscopic diagnostics) at $0.2\lesssim z\lesssim0.4$, where DESI contributes 3,410 candidates against 48 resolved literature systems, filling a previously undersampled intermediate redshift regime by nearly two orders of magnitude. 

Dual AGN candidates are preferentially found at small projected separations relative to one-AGN and inactive pairs:
their distribution peaks at $\Delta r_\mathrm{proj}\sim5$--$12~\mathrm{kpc}$, and ${\sim}40\%$ of dual systems lie at $\Delta r_\mathrm{proj}\leq20~\mathrm{kpc}$, consistent with simulations in which simultaneous SMBH activation peaks in late-stage mergers \citep{Chen2023, 2022MNRAS.514..640V}. The DESI fiber diameter (${\sim}1.6''$) sets an effective resolution floor, so this census mainly probes the kpc-scale dual AGN phase rather than the sub-kpc regime accessed by high-resolution imaging with {HST}, {JWST}, Euclid, or {Gaia} \citep{2025ApJ...986..101L, 2025A&A...696A..59P}.

Host galaxy properties of dual AGN systems differ systematically between the two members of a pair. Dual primaries exceed the isolated AGN median stellar mass by $\sim0.2$--$0.7$ dex at all redshifts, while secondaries lie $\sim0.2$--$0.6$ dex below it and reach $\log(M_\star/M_\odot)\simeq8.2$ in the closest pairs.
The mass-ratio distribution is broad ($61.9\%$ major, $18.1\%$ minor, $20.0\%$ very minor mergers), with the very-minor excess concentrated at the smallest separations. 
Part of this contrast is selection: as discussed in \ref{redshift_evolution}, the smallest separations are probed only at low redshift, where low-mass companions are detectable, and deblending at $\Delta r_{\rm proj}\lesssim5~\mathrm{kpc}$ may bias secondary masses low.

In the main-sequence offset $\Delta_{\rm MS}$, which is robust to the mass-limit coupling, dual secondaries lie ${\sim}0.3$ dex above both the less massive inactive-pair members and the one-AGN companions at the same separation, which is evidence for a moderate enhancement of star formation in the lower-mass member of dual systems.
One-AGN companions show a similar, weaker excess, suggesting that tidal torques operate in both one-AGN and dual pairs, while simultaneous accretion requires an additional condition, such as a larger and more centrally concentrated gas supply \citep{2019MNRAS.483.2712R}. 
Because our SFRs are integrated over the whole galaxy, these offsets are lower limits on the nuclear response (\S\ref{host_agn_properties}).

{AGN bolometric luminosities show only modest evidence for merger-related enhancement. The full DESI AGN population rises by nearly five orders of magnitude in median $L_{\rm bol}$ from $z\sim0.03$ to $z\sim3$, reflecting both the survey flux limit and the cosmic evolution of SMBH accretion. Against this evolving baseline, dual primaries are brighter than isolated AGN by $\sim0.3$--$0.4$ dex at $z<0.2$, an offset that falls to $\sim0.05$ dex by $0.2\leq z<0.5$ and vanishes at higher redshift, where gas-rich galaxies sustain accretion without a merger trigger 
\citep{2011ApJ...741L..33B, 2012ApJ...757...81B}.}
We also find that the dual-to-single AGN ratio declines from $\sim2\times10^{-2}$ at $z\simeq0.05$ to $\sim4\times10^{-3}$ at $z\simeq0.45$, and mostly remains at the level of $\sim10^{-3}$ or below over $1\lesssim z\lesssim2.4$, in agreement (within ${\sim}1\sigma$) with SDSS-based measurements
\citep{2020ApJ...899..154S}. 
Part of this decline is the angular-resolution limit, which excludes an increasing fraction of close physical pairs at higher redshift.

Forward modeling with the DESI-matched \texttt{ASTRID} mock \citep{2025arXiv251216844C} shows that the fraction of DESI duals whose black holes merge by $z=0$ increases with redshift, reaching ${\sim}76\%$ at $z\sim2$ in the mock.
The DESI-selected duals nevertheless trace only a small and biased subset of the total merger population: their merger rate peaks at $z\simeq0.13$ with $dN/dz\,dt\simeq7.6\times10^{-4}~\mathrm{yr^{-1}}$, and most of the integrated dual-merger and dual-LISA rates arise from $z\leq0.3$. 
The LISA-detectable fraction of DESI dual mergers falls from $\simeq0.97$ at $z\leq0.3$ to $\simeq0.22$ at $z\geq0.8$, because high redshift DESI mergers are increasingly dominated by $M_{\rm BH}\gtrsim10^8$--$10^9\,M_\odot$ black holes whose emission falls below the LISA band and into the PTA regime; the systems LISA retains inhabit intermediate-mass, relatively gas-rich hosts.

We identify ${\sim}50$ dwarf dual AGN candidates ($M_\star<10^9\,M_\odot$), expanding a population in which only a
handful of systems were previously known
\citep{2014ApJ...787L..30R, 2023ApJ...944..160M}. With expected black hole masses of ${\sim}10^4$--$10^6\,M_\odot$, these systems lie near peak sensitivity of LISA and represent promising gravitational wave progenitors. 
{A multiwavelength cross match of the low mass dual AGN yields radio associations more often than X-ray. No pair is yet resolved into two distinct X-ray nuclei, marking these systems as targets for pointed \textit{Chandra} and sub-arcsecond radio follow-up.} 
At high redshift, we identify 121 dual AGN candidates at $z>2$, more than three times the systems retained from literature samples at the same separations. 

DESI's wide area, dense spectroscopic sampling, and multi-tracer targeting strategy enable the construction of the largest uniformly selected, single-survey spectroscopic census of kpc-scale dual and offset AGN candidates, spanning $0 \lesssim z \lesssim 3.6$ and a broad range of host galaxy properties and AGN luminosities.
Combined with \texttt{ASTRID}-based forward modeling, it links the observed galaxy-scale dual-AGN phase to the later massive black hole mergers that produce low-frequency gravitational waves detectable by LISA.

{Looking ahead, the full DESI survey is expected to cover ${\sim}17{,}000~\mathrm{deg}^2$ and obtain spectra for ${\sim}63$ million unique galaxies and quasars. A simple scaling of the DR1 yield suggests that the dual AGN catalog could grow by factors of ${\sim}2$ in DR2 and ${\sim}4$ in the final data release, reaching ${\sim}14{,}000$ and ${\sim}28{,}000$ candidates, respectively. In reality, more survey passes will enable a more complete sampling of small separation objects, yielding potentially greater numbers.
These expanded datasets will allow the dual AGN population to be mapped with greater statistical power, providing a more complete view of how its demographics evolve across cosmic time.}

%% file: Sec6_Appendix.tex
\section{Comparison of \texttt{CIGALE} and \texttt{FastSpecFit} stellar masses and SFRs}
\label{app:fsf_vs_cigale}

\begin{figure*}
\centering
\includegraphics[width=0.89\textwidth]{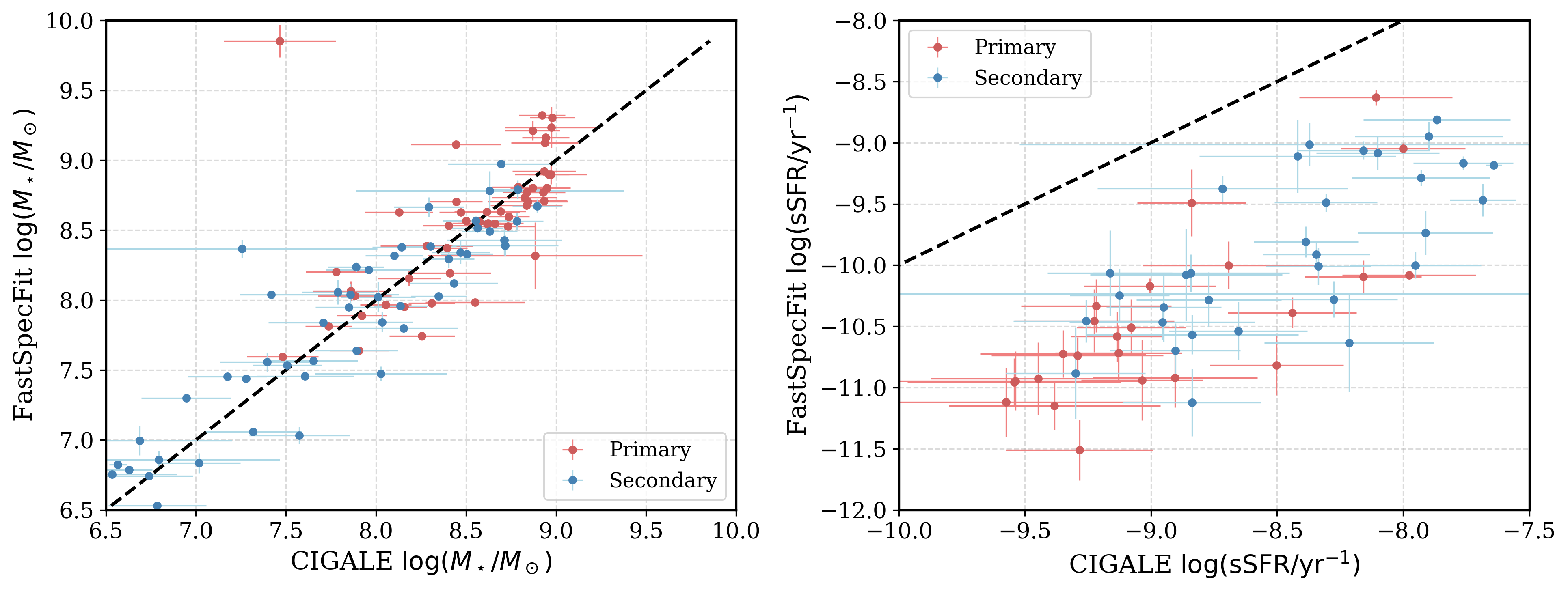}
\caption{
Comparison of host galaxy properties derived from \texttt{CIGALE} and \texttt{FastSpecFit} for the dwarf dual AGN subsample. 
The top and bottom panels show $\log(M_\star/M_\odot)$ and $\log(\mathrm{sSFR}/\mathrm{yr}^{-1})$, respectively, with primaries and secondaries shown as red and blue points. 
Error bars indicate the reported uncertainties from each pipeline, and the dashed black line denotes the one-to-one relation. 
Only components with reliable \texttt{CIGALE} posterior quality flags are shown.}
\label{fig:cigale_fsf}
\end{figure*}

To study a possible \texttt{CIGALE} SFR bias as motivated by \S \ref{dwarfs}, we cross-match our dwarf dual AGN catalog with the DESI \texttt{FastSpecFit} value-added catalog \citep{2023ascl.soft08005M}. 
{\texttt{FastSpecFit} jointly fits stellar population synthesis templates and nebular emission lines to the DESI spectrum and the same \textit{grz}+WISE photometry used by \texttt{CIGALE}, without an explicit AGN component. Its reported SFR comes from this stellar fit, not from a nebular emission-line luminosity.}
Figure~\ref{fig:cigale_fsf} compares the two methods for all dwarf dual AGN components with good quality fits in both pipelines.
Stellar masses are consistent: the distribution is tightly centered on the $1\!:\!1$ relation, with a median offset of $\sim-0.01$~dex.
Specific star formation rates, however, show a systematic discrepancy, with a median sSFR excess of $\sim1.4$~dex for \texttt{CIGALE} with respect to \texttt{FastSpecFit}. The offset is mass-dependent, increasing from $\sim1.3$~dex at $\log(M_\star/M_\odot)\sim8.5$ to $\sim1.6$~dex at $\log(M_\star/M_\odot)<7.5$.
{Neither code fully separates AGN and host light in this regime: \texttt{CIGALE}'s AGN fraction relies on WISE photometry that is often marginal for dwarf hosts, and \texttt{FastSpecFit} has no AGN component at all. In both cases residual AGN continuum can be absorbed into the young stellar component, inflating the recovered SFR, plausibly more so for \texttt{CIGALE}.}
{However, we do not investigate the origin of the discrepancy further here.}



\section{Stellar Mass limits for DESI pair tracers}
\label{app:tracer_mass_limits}
\begin{figure}
\centering
\includegraphics[width=0.39\textwidth]{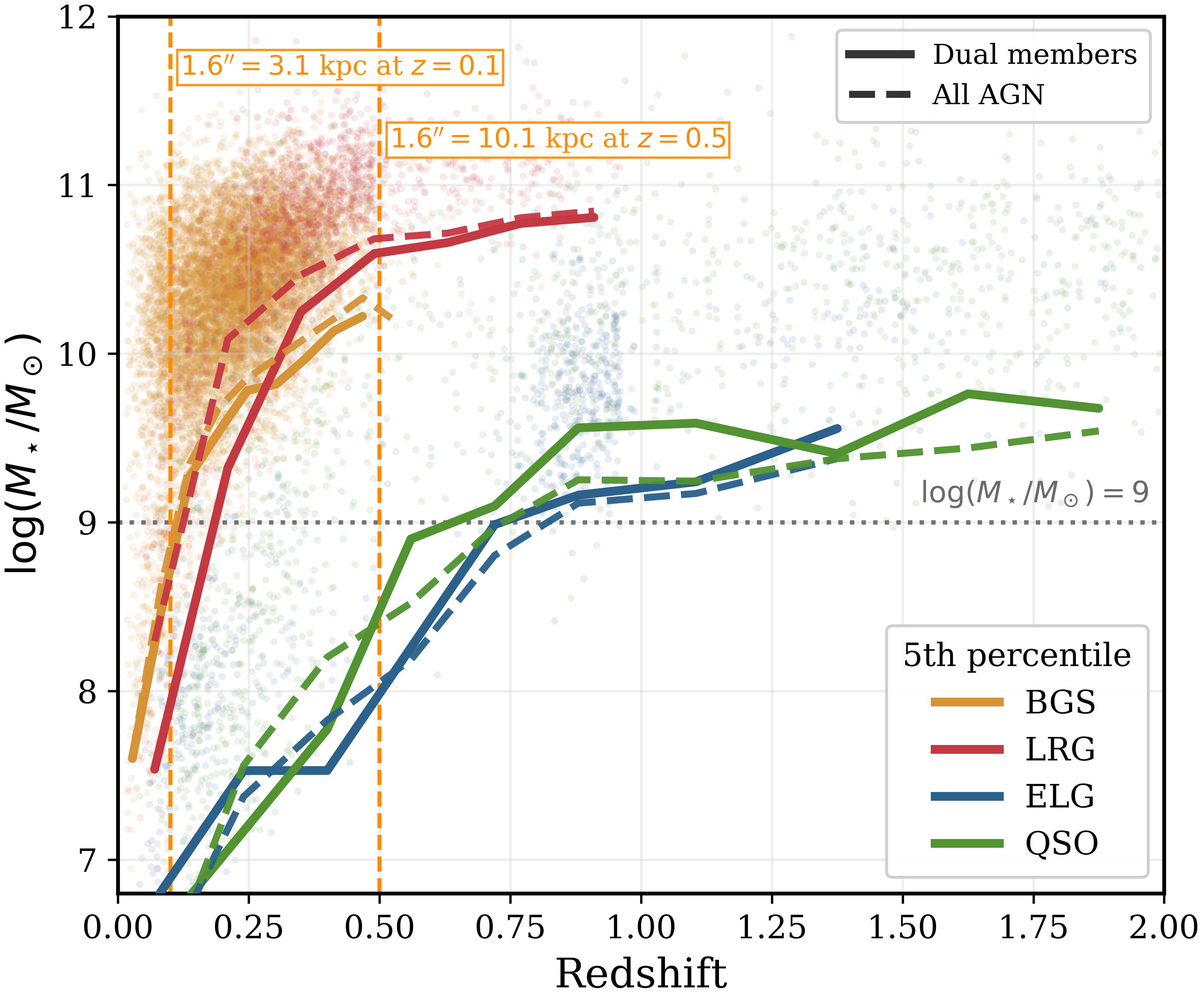}
\caption{
{Stellar mass as a function of redshift for dual-AGN members, separated by DESI tracer class. Points show individual dual-AGN host galaxies, and solid curves show the 5th-percentile stellar-mass envelope of the dual AGN members in redshift bins. Dashed curves show the corresponding 5th-percentile envelope for the isolated AGN comparison sample. The horizontal dotted line marks $\log(M_\star/M_\odot)=9$. Vertical dashed lines mark the redshifts at which the DESI minimum angular separation scale of $1.6^{\prime\prime}$ corresponds to 3.1 kpc and 10.1 kpc.}
}
\label{fig:tracer_mass_limits}
\end{figure}

Figure~\ref{fig:tracer_mass_limits} shows the redshift and tracer-dependent stellar mass coverage of the selected dual AGN sample. The 5th-percentile envelopes for dual members are shown as solid curves, and the corresponding isolated-AGN envelopes are shown as dashed curves. 
These envelopes are not formal completeness limits, but they show that low-mass dual members are preferentially present at low redshift: objects with $\log(M_\star/M_\odot)<9$ make up $27\%$ of dual members at $z<0.1$, $5.7\%$ at $0.1<z<0.5$, and only $1.1\%$ at $z>0.5$. At $z<0.1$, most low-mass dual members are BGS targets ($59\%$), while at $0.1<z<0.5$ they are dominated by QSO and ELG targets. This coupling between redshift, tracer class, and stellar-mass coverage underlies the selection effects discussed in \S\ref{redshift_evolution}.

\section{Radio and X-ray matches}
\label{app:radio}

{Table~\ref{tab:multiwavelength_candidates} summarizes the systems selected for detailed counterpart inspection. For the radio search, we include 4 strongest distinct-source candidates, 3 lower significance LoTSS-only distinct source associations, and the sole radio-associated pair in the strict dwarf subset. For the X-ray search, we include all nine associations. We distinguish one-component associations from shared-source detections because the latter do not resolve the two optical nuclei. Shared-source and single-component associations identify useful follow-up targets, but do not by themselves establish dual accretion.}

MW9 has the cleanest single-component X-ray detection: a Chandra source (2CXO~J125828.0$+$280550) lies only $0.12''$ from Target~1, with no cataloged counterpart at Target~2. MW13 is the only X-ray-associated system with both components at $\log(M_\star/M_\odot) < 9.5$; its eROSITA source is unresolved at the $2.30''$ pair separation, while the single Chandra source is substantially closer to Target~2 ($0.67''$) than to Target~1 ($2.95''$), supporting at most one X-ray-emitting component rather than a resolved X-ray dual. MW10 contains an eROSITA source $0.17''$ from Target~2, which also carries broad-line and optical/IR AGN diagnostics, while Target~1 has no X-ray counterpart. MW11 is a notable multiwavelength single-component case: Target~1 has both an eROSITA association and significant FIRST emission ($S/N = 11.2$). 
The remaining associations are lower priority: MW14 contains one XMM source whose centroid favors Target~2 but is unresolved at the pair scale; MW15 is a shared eROSITA association with only weak supporting LoTSS emission ($S/N = 1.3$); MW12 has an eROSITA source offset $7.33''$ from Target~1; and MW16 and MW17 are broad-line QSO--QSO systems with shared X-ray associations.

\begin{figure*}
\centering
\includegraphics[width=0.89\textwidth]{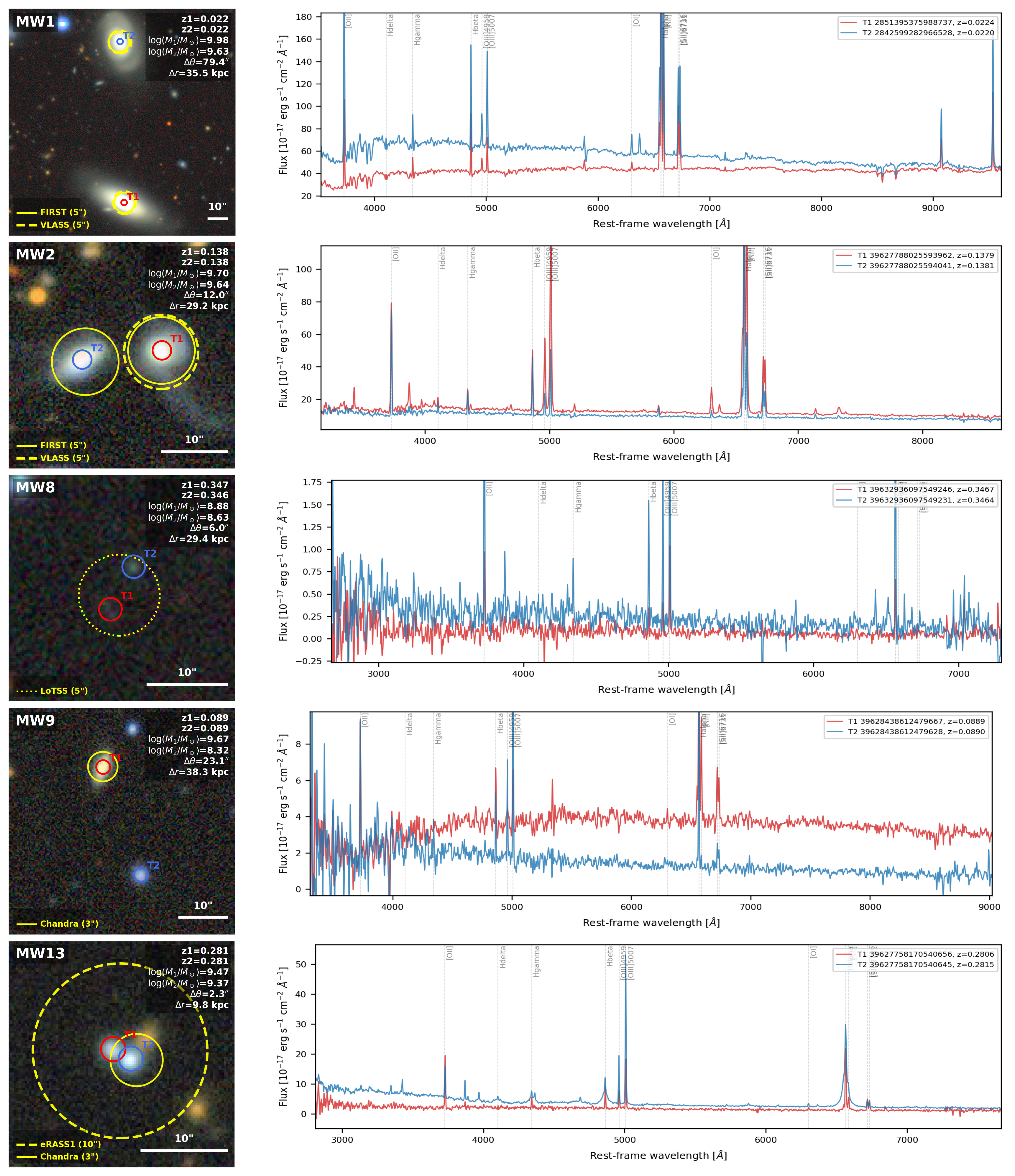}
\caption{
Examples from Table~\ref{tab:multiwavelength_candidates}. 
For each system, the left panel shows the Legacy Survey optical cutout with DESI component positions marked in red (T1) and blue (T2), and radio/X-ray catalog matching apertures shown in yellow. 
Solid yellow circles mark FIRST or Chandra matching radii, dashed circles mark VLASS or eROSITA/eRASS1, and dotted circles mark LoTSS; the radius corresponds to the catalog matching radius adopted in the cross-match. 
The right panel shows the DESI rest-frame spectra of the two components.
}
\label{fig:mw_examples}
\end{figure*}

Figure~\ref{fig:mw_examples} shows representative examples: the two strongest radio cases (MW1, MW2), the strict-dwarf association (MW8), and the two most informative X-ray cases (MW9, MW13).

\begin{table*}
\centering
\caption{Selected radio and X-ray counterpart associations among low-mass dual AGN candidates.}
\label{tab:multiwavelength_candidates}
\footnotesize
\setlength{\tabcolsep}{3pt}
\renewcommand{\arraystretch}{1.08}

\begin{tabular}{lllllll}
\hline
Class & ID & Comp. & TARGETID &
$\log(M_\star/M_\odot)$ &
$\Delta\theta$ ($r_p$) &
\parbox[t]{5.2cm}{Counterpart evidence} \\
\hline

Radio distinct & MW1 & T1 & 2851395375988737 & 9.98 & $79.43''$ &
\parbox[t]{5.2cm}{FIRST S/N=21.6; VLASS S/N=8.1} \\
               &     & T2 & 2842599282966528 & 9.63 & 35.5 kpc &
\parbox[t]{5.2cm}{FIRST S/N=13.6; VLASS S/N=9.1} \\[1pt]

Radio distinct & MW2 & T1 & 39627788025593962 & 9.70 & $12.01''$ &
\parbox[t]{5.2cm}{FIRST S/N=62.7; VLASS S/N=24.4} \\
               &     & T2 & 39627788025594041 & 9.64 & 29.2 kpc &
\parbox[t]{5.2cm}{FIRST S/N=13.4} \\[1pt]

Radio distinct & MW3 & T1 & 39627591962854470 & 9.89 & $36.21''$ &
\parbox[t]{5.2cm}{FIRST S/N=24.8; VLASS S/N=15.8} \\
               &     & T2 & 39627591962854301 & 9.14 & 34.6 kpc &
\parbox[t]{5.2cm}{FIRST S/N=6.0} \\[1pt]

Radio distinct & MW4 & T1 & 39628121955110675 & 9.88 & $35.90''$ &
\parbox[t]{5.2cm}{FIRST S/N=14.0; VLASS S/N=5.2} \\
               &     & T2 & 39628121955110903 & 9.17 & 26.8 kpc &
\parbox[t]{5.2cm}{FIRST S/N=11.7} \\

\hline
\multicolumn{7}{l}{\textit{Tentative radio associations}} \\
\hline

Radio distinct? & MW5 & T1 & 39633100497488122 & 9.73 & $16.43''$ &
\parbox[t]{5.2cm}{LoTSS S/N=0.7} \\
                &     & T2 & 39633100497488167 & 9.61 & 29.7 kpc &
\parbox[t]{5.2cm}{LoTSS S/N=1.1} \\[1pt]

Radio distinct? & MW6 & T1 & 39632936898663958 & 9.98 & $17.54''$ &
\parbox[t]{5.2cm}{LoTSS S/N=1.0} \\
                &     & T2 & 39632936898663959 & 9.91 & 36.4 kpc &
\parbox[t]{5.2cm}{LoTSS S/N=1.4} \\[1pt]

Radio distinct? & MW7 & T1 & 39628488327566297 & 9.71 & $21.06''$ &
\parbox[t]{5.2cm}{LoTSS S/N=1.1} \\
                &     & T2 & 39628488327566285 & 9.53 & 49.9 kpc &
\parbox[t]{5.2cm}{LoTSS S/N=1.2} \\[1pt]

Radio shared & MW8 & T1 & 39632936097549246 & 8.88 & $5.98''$ &
\parbox[t]{5.2cm}{Both components match one LoTSS source with S/N=2.8.} \\
             &     & T2 & 39632936097549231 & 8.63 & 29.4 kpc &
\parbox[t]{5.2cm}{} \\

\hline
\multicolumn{7}{l}{\textit{X-ray associations}} \\
\hline

X-ray single & MW9 & T1 & 39628438612479667 & 9.67 & $23.12''$ &
\parbox[t]{5.2cm}{Chandra source at $0.12''$} \\
              &     & T2 & 39628438612479628 & 8.32 & 38.3 kpc &
\parbox[t]{5.2cm}{No cataloged X-ray counterpart} \\[1pt]

X-ray single & MW10 & T1 & 39627787715218976 & 9.67 & $12.93''$ &
\parbox[t]{5.2cm}{No cataloged X-ray counterpart} \\
              &      & T2 & 2305843037487506435 & 9.60 & 29.5 kpc &
\parbox[t]{5.2cm}{eROSITA source at $0.17''$} \\[1pt]

X-ray single & MW11 & T1 & 39627666516611990 & 9.93 & $15.47''$ &
\parbox[t]{5.2cm}{eROSITA source at $2.92''$; FIRST S/N=11.2} \\
              &      & T2 & 39627666516612111 & 9.88 & 42.0 kpc &
\parbox[t]{5.2cm}{No cataloged X-ray counterpart} \\[1pt]

X-ray single & MW12 & T1 & 39627533108384848 & 9.99 & $10.79''$ &
\parbox[t]{5.2cm}{eROSITA source at $7.33''$} \\
              &      & T2 & 39627533108384956 & 9.31 & 34.3 kpc &
\parbox[t]{5.2cm}{No cataloged X-ray counterpart} \\[1pt]

X-ray shared & MW13 & T1 & 39627758170540656 & 9.47 & $2.30''$ &
\parbox[t]{5.2cm}{Shared eROSITA association} \\
              &      & T2 & 39627758170540645 & 9.37 & 9.8 kpc &
\parbox[t]{5.2cm}{Chandra source at $0.67''$ from T2 ($2.95''$ from T1)} \\[1pt]

X-ray shared & MW14 & T1 & 39633443121791179 & 9.98 & $4.47''$ &
\parbox[t]{5.2cm}{One XMM source matches both components; centroid favors T2.} \\
              &      & T2 & 39633443121791172 & 9.50 & 12.9 kpc &
\parbox[t]{5.2cm}{} \\[1pt]

X-ray shared & MW15 & T1 & 39627853309936517 & 9.86 & $5.78''$ &
\parbox[t]{5.2cm}{Shared eROSITA source; also shared weak LoTSS association with S/N=1.3.} \\
              &      & T2 & 39627853309936489 & 9.76 & 27.9 kpc &
\parbox[t]{5.2cm}{} \\[1pt]

X-ray shared & MW16 & T1 & 39627933198844858 & 9.88 & $5.31''$ &
\parbox[t]{5.2cm}{Shared eROSITA association; broad-line QSO--QSO system.} \\
              &      & T2 & 39627933198844841 & 9.81 & 44.4 kpc &
\parbox[t]{5.2cm}{} \\[1pt]

X-ray shared & MW17 & T1 & 39627842245366926 & 9.79 & $5.00''$ &
\parbox[t]{5.2cm}{Shared XMM association; broad-line QSO--QSO system.} \\
              &      & T2 & 39627842245366900 & 9.43 & 43.0 kpc &
\parbox[t]{5.2cm}{} \\

\hline
\end{tabular}
\end{table*}

